# Breakup of SUSY Quantum Mechanics in the Limit-Circle Region of the Reflective Kratzer Oscillator

G. Natanson
ai-solutions Inc.
2232 Blue Valley Dr.
Silver Spring MD 20904
U.S.A.
greg_natanson@yahoo.com

The paper studies violation of conventional rules of SUSY quantum mechanics for the inverse-square-at-origin (IS@O) radial potential V(r) within the limit-circle (LC) range. A special attention is given to transformation properties of the Titchmarsh-Weyl *m*-function under Darboux deformations of the reflective Kratzer oscillator: centrifugal Kepler-Coulomb (KC) potential plus a Taylor series in r. Since our analysis is based on Fulton's representation of a regular-at-infinity (R@∞) solution [Math. Nachr. **281**, 1418 (2008)] as a superposition of two Frobenius solutions at the origin, we refer to the appropriate expressions as the Titchmarsh-Weyl-Fulton (TWF) functions. Explicit transformation relations are derived for partner TWF functions associated with SUSY pairs of IS@O potentials. It is shown that these relations have a completely different form for Darboux transformations (DTs) keeping the potential within the LC range.

As an illustration, we use regular nodeless Frobenius solutions to construct SUSY partners of the radial *r*- and *c*-Gauss-reference (GRef) potentials solvable via hypergeometric and confluent hypergeometric functions, respectively. We explicitly demonstrate existence of *non-isospectral* partners of both radial potentials in the LC region and obtain their discrete energy spectra using the derived closed-form expressions for the TWF functions. The general transformation relations for the TWF function have been verified taking advantage of form-invariance of the radial GRef potentials under double-step DTs with the so-called 'basic' seed solutions (SSs). Similarly we directly ratify that TWF functions for three shape-invariant reflective potentials on the half-line -- hyperbolic Pöschl-Teller (h-PT), Eckart/Manning-Rosen (E/MR), and centrifugal KC potentials – do retain their form under basic DTs.

## 1. Introduction

In the early nineties Gangopadhyaya, Panigrahi and Sukhatme [1] pointed to some non-trivial features of SUSY partners for inverse-square-at-origin (IS@O) radial potentials within the so-called 'limit-circle' (LC) range [2-5] of the exponent difference at the singular point r=0 (ExpDiff@O). In particular, they realized that the conventional rules of SUSY quantum mechanics on the line [6-9] can be automatically extended to singular radial potentials only if the ExpDiff@O is restricted to the limit-point (LP) region [2-5]. Unfortunately this remarkable work has been completely disregarded in the literature. To our knowledge, the introductory course written by Gangopadhyaya et al [10] is the only quantum-mechanical textbook which correctly approaches pitfalls of SUSY quantum mechanics within the LC range. (As elaborated in more details in Section 3, the breakdown of SUSY quantum mechanics for non-singular radial potentials has been also noticed by Baye [11], but again without any significant impact on following developments in the field.)

The purpose of this study is to draw attention to an anomalous behavior of regular solutions of the Schrödinger equation with an IS@O radial potential under Darboux transformations (DTs) keeping the ExpDiff@O within the LC range. It is explicitly demonstrated that these DTs violate the well-established rules of SUSY quantum mechanics [12-14]. This breakup of the conventional rules was overlooked by Gangopadhyaya et al. [1, 10] since they limited their analysis only to LC-to-LP and LP-to-LC pairs of the DTs.

We restrict our analysis solely to IS@O potentials because we are mostly interested in SUSY partners of the so-called [15] 'regular and confluent Gauss-reference' (*r*-GRef and *c*-GRef) potentials exactly solvable via hypergeometric and confluent (*c*-) hypergeometric functions [16-18], respectively, both families of radial potentials from this class as well as all their SUSY partners have the IS@O singularity. As discussed in this paper these 'transition' [6] potentials inherit many conventional features of non-singular potentials. For singularities of a different type (see, i. g., [19] for nontrivial complications arising in those cases) violations of the conventional rules of SUSY quantum mechanics might be even more severe. However those potentials simply lie beyond our current interests.

The paper is organized as follows. In Section 2 we present a simplified version of Kemble' proof [20] that the lowest-energy eigenfunction of the Schrödinger equation with an IS@O radial potential as well as regular-at-origin (R@O) and regular-at-infinity (R@∞) solutions lying below this 'ground' energy level do not have nodes on the positive semi-axis regardless of the range of potential parameters.

As shown in Section 3, the DT with a factorization function (FF) irregular at the origin converts any R@O solution (i. e., the one with a characteristic exponent at the origin (ChExp@O) between 1 and $3/2$) into the one with the ChExp@O between $1/2$ and 1. This implies that the appropriate SUSY pairs are formed by one regular and one irregular solutions, contrary to the conventional rule. We re-examine our original arguments [21] for transformation properties of the Jost function under Darboux deformations of IS@O potentials with exponential tails (later re-discovered by Sukumar [22]) and show that our initial derivation is indeed inapplicable to DTs keeping the ExpDiff@O within the LC range. We also extend our analysis to IS@O potentials with Coulomb tails and briefly compare our results with those presented by Sparenberg and Baye [23, 24] in the late nineties.

In Section 4 we analyze transformation properties of the Titchmarsh-Weyl (TW) *m*-function [2, 25] under Darboux deformations of the Kratzer oscillator [26] using nodeless Frobenius solutions at the origin as the FFs. Since our analysis is based on Fulton's representation [27] of a R@∞ solution as a superposition of two Frobenius solutions at the origin, we refer to the appropriate formulas as the Titchmarsh-Weyl-Fulton (TWF) functions. It is directly shown that the partner TWF functions are interrelated in an anomalous way if the Darboux transformation keeps the potential within the LC region.

In subsections 5.1 and 5.2 of Section 5 the results of the general analysis are applied to the radial *r*-GRef and *c*-GRef potentials. We derive closed-form expressions for the TWF functions associated with these potentials and in particular confirm that the functions do have single poles at the bound-state energies as expected.

Section 6 uses the results of Section 4 to obtain closed-form expressions for the partner TWF functions in particular cases of DTs using regular nodeless Frobenius solutions near the origin as the FFs. Again the analysis is split into two subsections dealing separately with the *r*-GRef and *c*-GRef potentials.

Section 7 takes advantage of the fact that both radial potentials discussed in subsections 5.1 and 5.2 retain their form under double-step DTs using the so-called [15] 'basic' seed solutions (SSs) for the Crum Wronskians [28]. It is explicitly confirmed that the derived closed-form expressions for the TWF functions of the GRef radial potential are invariant under the double-step DTs of this type. A similar analysis is performed for SUSY pairs of the appropriate Jost functions.

As initially found in [21], the hyperbolic Pöschl-Teller (h-PT) potential [29] retains its form under the DTs using four 'basic'[15] solutions as the FFs. The common remarkable feature of these basic wavefunctions is that the underlying 'almost-every holomorphic' (AEH) solutions [15] remain finite at any regular point of the appropriate Sturm-Liouville (SL) equation extended into the complex plane. As a direct consequence of this observation, the DTs in question do not lead to new singular points and the resultant SL equation is again solvable via hypergeometric and *c*-hypergeometric functions for the *r*-GRef and *c*-GRef potentials, respectively [15]. In particular, since the eigenfuction f the ground-energy state in any 'shape-invariant'[x)] GRef potential [30] turns out to be a basic solution the DT using the latter as the FF preserves the potential representation in terms of elementary functions.

In subsection 8.1 of Section 8 we use the form-invariance of the h-PT potential under four DTs with basic FFs to explicitly confirm that the partner TWF functions are indeed interrelated in the predicted anomalous way if the DTs keep the ExpDiff@O within the LC range. The second shape-invariant radial potential solvable via hypergeometric functions -- the Eckart/Manning-Rosen (E/MR) potential [32, 33] as we refer to it here, has only two basic solutions, respectively, regular and irregular at the

---

[x)] It should be mentioned that in the original (Russian) versions of Gendenshtein's papers [30, 31], he uses the Russian equivalent of the English term 'form-invariance', retained in the translation of his joint paper with Krive [31]. The broadly accepted term: 'shape-invariance' first appearing in the English translation of [30] is thus nothing but a translation slip.

origin. Again, it is explicitly ratified that the TWF function within the LC region is transformed in an anomalous way when the second basic solution is used as the FF.

In subsection 8.2 of Section 8 we extend our analysis to the generic reflective radial potential exactly solvable via *c*-hypergeometric functions [17, 18]. The anomalous behavior of the TWF function under action of the DT keeping the potential within the LC region is then confirmed for the shape-invariant centrifugal Kepler-Coulomb (KC) potential. (We add the epithet ‘centrifugal’ to the potential name to stress that the parameter range covers the region with a repulsive IS@O term.)

Finally, Section 9 outlines possible directions for future developments, with a special emphasis on possible re-formulation of our analysis as the spectral problem for a self-adjoint operator in the LC and LP regions [3].

The paper is accompanied by three Appendices. In Appendix A we re-derive the expression [17] for the Jost function of the radial *c*-GRef potential with a Coulomb tail to point out to some details essential for the current analysis. Appendix B presents a list of auxiliary formulas used to prove form-invariance of the TMF function for the h-PT potential in the particular case of DTs with basic FFs.

Originally the primary purpose of Appendix C was just to relate our approach to rational SUSY partners of the E/MR potentials to Quesne’s study [34] who quantizes these potentials by means of polynomials in the reciprocal argument, as against the ’Gauss-seed (GS) Heine’ [15] polynomials used here. However, during this analysis we revealed a very close relationship between the radial *r*-GRef potential of our current interest and the so-called ‘Linear-Tangent-Polynomial’ (LTP) potential on the line [35]. Some common features between the E/MR and Rosen-Morse (RM) [36] potentials revealed by Quesne [34] turned out to be thus the very particular case of the dualism between the appropriate ‘PF beams’ in mutually reciprocal arguments.

## 2. Definition of regular solutions of the Schrödinger equation within the LC range

In this series of publications we are only interested in the radial Schrödinger equation for the Kratzer oscillator

$$V(r;\Diamond_o) \equiv V(r;\Diamond_o,q_1,\tau) = \frac{\Diamond_o^2 - 1/4}{r^2} + \frac{q_1}{r} + \sum_{k=0}^{\infty} q_{k+2}(\tau) r^k \qquad (2.1)$$

having the second-order pole at the origin:

$$\lim_{r\to 0} [r^2 V(r;\Diamond_o)] = q_0 \equiv \Diamond_o^2 - 1/4 . \qquad (2.2)$$

By definition the 'singularity index' (SI) $\Diamond_o > 0$ (as we refer to it) differs from the larger ChExp@O, $s_o$, by $1/2$:

$$s_o = \Diamond_o + 1/2 \qquad (2.3)$$

and therefore coincides with one half of the ExpDiff@O. It is assumed that the potential has a LP singularity at $+\infty$. In following [1, 10], we only consider the region with a real SI $\Diamond_o$ ($q_0 > -1/4$ or $s_o > 1/2$ ), where the indicial equation at the origin has two distinct real roots. (Otherwise the energy of wave packets has no lower bound so that they will be simply "sucked into" the singularity [10].)

As initially pointed out by Scarf [37], the region $-1/4 < q_0 < 3/4$ ($1/2 < s_o < 3/2$ or $0 < \Diamond_o < 1$) with the continuous spectrum represents an example of the LC problem [3]. In the latter case a regular solution at the origin is defined as the one with a larger characteristic exponent. For any centrifugal-barrier potential ($q_0 > 0$ or $s_o > 1$) this definition of a regular solution is equivalent to the Direchlet boundary condition

$$\psi_a(0;\varepsilon;\Diamond_o) = 0 \qquad (2.4)$$

at an arbitrary energy $\varepsilon$. In the quantum-mechanical applications the boundary condition based on a larger value of the ChExp ('the less singular function' in terms of [10]) is sometimes [6, 10] referred to as an 'ad hoc constraint', however this is a natural approach in the spectral theory of the singular self-adjoint operators [3-5]. In particular, this constraint is equivalent to solving the self-adjoint SL equation

$$\left[\frac{d}{dr} r \frac{d}{dr} + r\varepsilon - \frac{\Diamond_o^2}{r} - q_1 - \sum_{k=0}^{\infty} q_{k+2}(\tau) r^{k+1}\right] \chi_{\mathsf{a}}(r;\varepsilon;\Diamond_o) = 0 \qquad (2.5)$$

under the Direchlet boundary condition

$$\chi_{\mathsf{a}}(0;\varepsilon;\Diamond_o) = 0, \qquad (2.6)$$

where we put

$$\chi_{\mathsf{a}}[r;\varepsilon;\Diamond_o] = \frac{1}{\sqrt{r}} \psi_{\mathsf{a}}[r;\varepsilon;\Diamond_o]. \qquad (2.7)$$

An extension of the conventional Sturm-Liouville theory [38] to the so-called ‘singular-point boundary conditions’ (s.p.b.c.) has been done by Kemble [20] at the early stage of quantum mechanics. It seems useful to present a simplified version of his analysis taking advantage of some remarkable feature of the Wronskians of different pairs of solutions for the generic IS@O potential. In following our original study [21] on DTs of IS@O potentials we use the labels $\mathsf{a}$, $\mathsf{b}$, $\mathsf{c}$, and $\mathsf{d}$ to specify solutions: R@O, R@∞, square integrable, and irregular at both ends, accordingly. In Sukumar’s [22] notation for solution types

$$\mathsf{a} = T_3,\ \mathsf{b} = T_4,\ \mathsf{c} = T_1,\ \mathsf{d} = T_2. \qquad (2.8)$$

By analyzing properties of the appropriate Jost functions [39], it was proven that DTs with FFs of type $\mathsf{a}$ and $\mathsf{b}$ result in isospectral potentials if the FFs in question do not have nodes on the positive semi-axis. It was taken for granted that regular solutions lying below the ground energy level are nodeless as it indeed takes place for the non-singular case with boundary conditions at the ends of a finite interval [38].

It seems necessary to re-examine mathematical grounds of the latter assumption more carefully to better understand differences between the LP and LC regions. Let $r_{\mathsf{a}}$ and $r_{\mathsf{b}}$ be zeros of R@O and, respectively, R@∞ solutions $\psi_{\mathsf{a}}(r;\varepsilon;\Diamond_o)$ and $\psi_{\mathsf{b}}(r;\varepsilon;\Diamond_o)$ of the Schrödinger equation with the potential $V(r;\Diamond_o, q_1, \tau)$ at an arbitrary energy $\varepsilon$ below the ground energy level $\varepsilon_{c;0}$. In following the conventional arguments for the finite interval [38] and taking into account that

$$W\{\psi_a(r;\varepsilon;\Diamond_o), \psi_a(r;\varepsilon+\delta\varepsilon;\Diamond_o)\}\Big|_{r=0} = 0 \tag{2.9}$$

one finds

$$-\frac{dr_a}{d\varepsilon} \,|\, \psi'_a(r_a;\varepsilon;\Diamond_o)\,|^2 \approx \frac{1}{\delta\varepsilon} W\{\psi_a(r;\varepsilon+\delta\varepsilon;\Diamond_o), \psi_a(r;\varepsilon;\Diamond_o)\}\Big|_{r=r_a}$$

$$= \int_0^{r_a} \psi_a(r';\varepsilon;\Diamond_o)\; \psi_a(r';\varepsilon+\delta\varepsilon;\Diamond_o) dr' > 0 \tag{2.10a}$$

and

$$-\frac{dr_b}{d\varepsilon} \,|\, \psi'_b(r_b;\varepsilon;\beta_o)\,|^2 \approx \frac{1}{\delta\varepsilon} W\Big\{\psi_b(r;\varepsilon+\delta\varepsilon;\beta_o), \psi_b(r;\varepsilon;\beta_o)\Big\}\Big|_{r=r_b}$$

$$= -\int_{r_b}^{\infty} \psi_b(r';\varepsilon;\beta_o), \psi_b(r';\varepsilon+\delta\varepsilon;\beta_o) dr'. \tag{2.10b}$$

In the particular case of a non-singular radial potential ($s_o$ =1 or $\Diamond_o = 1/2$) the derivative of the function $\psi_a(r;\varepsilon;1/2)$ at r = 0 can be simply set to 1. Otherwise it either vanishes ($s_o$ >1) or infinitely large ($s_o$ <1). One can easily verify that

$$\frac{dr_a}{d\varepsilon} \approx -\frac{r_a}{2\Diamond_o} < 0\,. \tag{2.11}$$

It then directly follows from (2.10a) and (2.10b) that the node of the solution regular at one end moves toward another end as the energy decreases. In particular, if the ground eigenfunction is nodeless this confirms that both R@O and R@∞ solutions lying below the ground energy level are necessarily nodeless regardless of the value of $s_o$.

It is worth to stress that condition (2.9), with $\varepsilon = \varepsilon_v$, $\varepsilon + \delta\varepsilon = \varepsilon_{v'}$ $(v \neq v')$, and a similar relation for the Wronskian at other end are sufficient for eigenfunctions with eigenvalues $\varepsilon_v$ and $\varepsilon_{v'}$ to be mutually orthogonal. Likewise the Wronskian of any two R@∞ solutions (other than eigenfunctions) also vanishes at the singular point representing the infinitely deep well if $\frac{1}{2} < s_o < 1$ which means that these solutions must be also mutually orthogonal.

To prove that the ground eigenfunction may not have nodes, first note that changing r for the energy-dependent variable $\varsigma \equiv 2\sqrt{-\varepsilon}\, r$ and making $\varepsilon \to -\infty$ turns the Schrödinger equation into the Whittaker equation [40]

$$\left[\frac{d^2}{d\varsigma^2} - \frac{\Diamond_o^2 - \frac{1}{4}}{\varsigma^2} - \frac{1}{4}\right] M_{0,\Diamond_o}[\varsigma] = 0, \tag{2.12}$$

where

$$M_{0,\Diamond_o}[\varsigma] = \lim_{\varepsilon \to -\infty} \psi_a(\varsigma/\sqrt{-\varepsilon}; \varepsilon; \Diamond_o + \tfrac{1}{2}) \tag{2.13}$$

is nothing but the Whittaker function

$$M_{\kappa,\Diamond_o}[\varsigma] = \varsigma^{\frac{1}{2}+\Diamond_o}\, e^{-\frac{1}{2}\varsigma} F(\Diamond_o + \tfrac{1}{2} - \kappa, 2\Diamond_o + 1; \varsigma). \tag{2.14}$$

Since both parameters a= $\Diamond_o + \frac{1}{2}$ and c= $2\Diamond_o + 1$ are positive this is also true for each term of the Taylor series $F(a, c; \varsigma)$ in $\varsigma$ so that solution (2.13) must be nodeless which concludes the proof.

To conclude, let us notice that significance of using square integrability of regular solutions as the boundary condition within the LP range of the ExpDiff@O was exaggerated by the author in [15]. As seen from the above analysis the parameter range is completely nonessential for the proof that regular solutions below the ground energy levels are nodeless at least for the potentials of our current interest. The crucial factor is that the given definition of the R@O solutions assures that the Wronskian of any two of them vanishes at the lower end so that their nodes move in the regular way specified by the Sturm-Liouville theory [38].

## 3. Breakup of SUSY quantum mechanics in the LC region

Since function

$$^{*}\psi_{\dagger}(r;\varepsilon_{\dagger};\lozenge_{o;\dagger}\,|\,\dagger,\varepsilon_{\dagger}) \equiv \psi_{\dagger}^{-1}(r;\varepsilon_{\dagger};\lozenge_{o}) \qquad (3.1)$$

is a solution of the Schrödinger equation with the potential

$$^{*}V(r;\lozenge_{o;\dagger}\,|\,\dagger,\varepsilon_{\dagger}) = V(r;\lozenge_{o}) - 2\frac{d^2}{dr^2}\,ln\,\psi_{\dagger}(r;\varepsilon_{\dagger};\lozenge_{o}) \qquad (3.2)$$

The very specific feature of IS@O potentials usually disregarded in the literature is that transformation properties of R@O solutions under action of DTs with irregular FFs generally depend on the values of the SIs $\lozenge_o$ and $\lozenge_{o;\dagger}$:

$$\lozenge_{o;\dagger} = \begin{cases} \lozenge_o + 1 & \text{for } \dagger = \dagger_+ = \mathsf{a} \text{ or } \mathsf{c} \text{ (LP, LC)} \\ \quad \text{or} & \\ \lozenge_o - 1 > 1/2 & \text{for } \dagger = \dagger_- = \mathsf{b} \text{ or } \mathsf{d} \text{ (LP).} \end{cases} \qquad (3.3)$$

While (3.3) does not make any distinction between the LP and LC regions for DTs with R@O FFs the LC range of the SI requires a special attention for FFs of types b and d, namely,

$$.\ 0 < \lozenge_{o;\dagger_-} = 1 - \lozenge_o < 1 \quad (\dagger_- = \mathsf{b} \text{ or } \mathsf{d}) \qquad (3.3^*)$$

if $1/2 < s_o < 3/2$. We thus conclude that DTs with FFs irregular at the origin keep parameters of the Schrödinger equation within the LC range. The crucial difference compared with the LP case is that the SUSY partner of any eigenfunction regular at the origin behaves at small r as $r^{1-s_{o;\dagger_-}}$ -- a distinctive feature of the general (irregular) solution.

If $1 < \lozenge_o < 2$ then any DT with a FF irregular at the origin converts the given potential into the one with the SI $\lozenge_{o;\dagger_-} = \lozenge_o - 1$ lying within the LC range. Nonetheless, the resultant SUSY partner of $\psi_{\mathsf{a}}(r;\varepsilon;\lozenge_o)$ still behaves as $r^{s_{o;\dagger_-}}$ at small r and therefore remains regular, in agreement with the conventional rules of SUSY quantum mechanics. Similarly, if the larger ChExp@O lies within the LC range then any DT with a R@O FF brings the potential parameters into the limit point (LP) range with the SI $\lozenge_{o;\dagger_{+}} = \lozenge_o + 1$ [1, 10] and the SUSY partner of the function $\psi_{\mathrm{a}}(r;\varepsilon;\lozenge_o)$ vanishes at the origin as required by the boundary condition imposed on regular solutions in the LP region. It was Gangopadhyaya et al. [1] who first drew attention to this nontrivial problem. By using the exactly-solvable h-PT potential as an illustrative example these authors (see also [10]) explicitly confirmed validity of SUSY quantum mechanics for DTs between the LP and LC regions.

However, the conventional rules of SUSY quantum mechanics become invalid in case of DTs with FFs irregular at the origin when the latter are applied in the LC region. In fact, since the reverse of the FF $\psi_{\dagger_{-}}(r;\varepsilon;\lozenge_o)$ for $\dagger$ = $\mathrm{b}$ or $\mathrm{d}$ approaches zero as $r^{s_o - 1}$ the larger ChExp@O for the SUSY partner also lies within the LC range:

$$\tfrac{1}{2} < s_{o;\dagger_{-}} = 2 - s_o < \tfrac{3}{2}. \tag{3.4}$$

We thus conclude that DTs with FFs irregular at the origin keep parameters of the Schrödinger equation within the LC range. *The crucial difference between these Darboux deformations of the generic IS@O potential and all others is that they convert any R@O solution into an irregular one. The appropriate pairs of SUSY partners are thus formed by potentials with two completely different sets of eigenvalues.*

It was Baye [11] who first realized that the DT of a regular solution results in an irregular solution for the 'singularity strength' $n = s_o - 1$ equal to 0. He explicitly pointed to the fact there is no simple relation between phase shifts of scattering states in two potentials and concluded that *"the* n=0 *case is therefore useless for the construction of phase-equivalent potentials but is an interesting example of supersymmetric partners whose spectra are not simply linked."* (In his early papers [11, 41] Baye restricted his analysis only to integer values of the so-called [42] 'singularity strength' $\nu = n$ so that $\nu=0$ was the only value of $\nu$ lying within the LC range.)

Keeping in mind this anomalous behavior of the SI within the LC range, let us re-examine more accurately the arguments used in [21] to derive transformation properties of the Jost function under DTs with FFs of four different types:

$$f(k;\Diamond_o+1\,|\,\mathbf{a},\varepsilon_{\mathbf{a}}) = \frac{2\Diamond_o}{\sqrt{-\varepsilon_{\mathbf{a}}}-i\,k} f(k;\Diamond_o)\,, \tag{3.5a}$$

$$f(k;\Diamond_o-1\,|\,\mathbf{b},\varepsilon_{\mathbf{b}}) = \frac{\sqrt{-\varepsilon_{\mathbf{b}}}-i\,k}{2(\Diamond_o-1)} f(k;\Diamond_o) \qquad \text{for } \Diamond_o>1\,, \tag{3.5b}$$

$$f(k;\Diamond_o+1\,|\,\mathbf{c},\varepsilon_{\mathbf{c}}) = \frac{2\Diamond_o}{-\sqrt{-\varepsilon_{\mathbf{c}}}-i\,k} f(k;\Diamond_o)\,, \tag{3.5c}$$

$$f(k;\Diamond_o-1\,|\,\mathbf{d},\varepsilon_{\mathbf{d}}) = \frac{-\sqrt{-\varepsilon_{\mathbf{d}}}-i\,k}{2(\Diamond_o-1)} f(k;\Diamond_o) \qquad \text{for } \Diamond_o>1\,. \tag{3.5d}$$

Note that we define the Jost function via (12.28) in [39], namely, we represent the R@O solution as

$$\varphi_{\mathbf{a}}(r;k;\Diamond_o) = \frac{1}{2i\,k}[f_-(k;\Diamond_o)\,f_+(r;k;\Diamond_o) - f_+(k;\Diamond_o)\,f_-(r;k;\Diamond_o)]\,, \tag{3.6}$$

where

$$\lim_{r\to 0}[r^{-s_o}\,\varphi_{\mathbf{a}}(r;k;\Diamond_o)] = 1 \tag{3.7a}$$

and

$$\lim_{r\to\infty}[e^{\mp i\,k\,r} f_\pm(r;k;\Diamond_o)] = 1\,. \tag{3.7b}$$

The Jost function $f(k;\Diamond_o) \equiv f_+(k;\Diamond_o)$ and its counter-part $f_-(k;s_o)$ are then given by the Wronskians

$$f_\pm(k;\Diamond_o) \equiv W\{f_\pm(r;k;\Diamond_o), \varphi_{\mathbf{a}}(r;k;\Diamond_o)\}\,. \tag{3.8}$$

Alternatively one can use (12.140) in [39]

$$f(k;\Diamond_o) = 2\Diamond_o \lim_{r\to 0}[r^{\Diamond_o - 1/2}\, f_+(r;k;\Diamond_o)] \tag{3.9}$$

which is obtained by evaluating Wronskians (3.8) at r=0 [39]. It should be mentioned that Sukumar [22] apparently uses an alternative definition for the Jost function via (12.142) in [39]:

$$F(k;\nu) = \frac{1}{\Gamma(\nu+3/2)}\left(\frac{k}{2i}\right)^{\nu} f(k;\nu+1)\,. \qquad (3.10)$$

When extended to non-integer values of the singularity strength ν, transformation relations (3.5a)–(3.5d) thus turn into Sukumar's formulas (38), (44), (17), and (32):

$$F(k;\nu+1\,|\,\mathbf{a},\varepsilon_{\mathbf{a}}) = \frac{k}{k+i\sqrt{-\varepsilon_{\mathbf{a}}}}F(k;\nu) \qquad \text{for } \nu > -1/2, \qquad (3.11a)$$

$$F(k;\nu-1\,|\,\mathbf{b},\varepsilon_{\mathbf{b}}) = \frac{k+i\sqrt{-\varepsilon_{\mathbf{b}}}}{k}F(k;\nu) \qquad \text{for } \nu > 1/2, \qquad (3.11b)$$

$$F(k;\nu+1\,|\,\mathbf{c},\varepsilon_{\mathbf{c}}) = \frac{k}{k-i\sqrt{-\varepsilon_{\mathbf{c}}}}F(k;\nu) \qquad \text{for } \nu > -1/2, \qquad (3.11c)$$

$$F(k;\nu-1\,|\,\mathbf{d},\varepsilon_{\mathbf{d}}) = \frac{k-i\sqrt{-\varepsilon_{\mathbf{d}}}}{k}F(k;\nu) \qquad \text{for } \nu > 1/2, \qquad (3.11d)$$

where the lower bounds for ν were added to emphasize an anomalous behavior of the Jost function for the DTs converting the LC region onto itself. Though the definition of the Jost function via (3.10) was more recently used both in [23, 24, 40] and in [42], we prefer to stick with our original formulas [17, 21] because the additional factor seems to be rather artificial for non-integer values of ν and moreover it only complicates the closed-form expressions derived in [21] for the GRef radial potentials. Since the scale factor in transformation relations (3.11a)-(3.11d) is independent of the singularity strength, contrary to our original expressions (3.5a)-(3.5d), one may prefer to deal with Jost function (3.10), instead of (3.8). However, the latter definition looks more natural.

Let us briefly re-examine our original arguments [21] used to derive transformation relations (3.5a)-(3.5d) for IS@O radial potentials with exponential tails at infinity. We thus assume that the factorization function $\psi_{\dagger}(r;\varepsilon_{\dagger};\lozenge_o)$ has an asymptotics

$$\psi_{\dagger}(r;\varepsilon_{\dagger};\lozenge_o) \sim A(\lozenge_o)\,\exp(-\sigma_{1;\dagger}\sqrt{-\varepsilon_{\dagger}}) \quad \text{at large r}, \qquad (3.12)$$

where

$$\sigma_{1;\dagger} = \begin{cases} + \quad \text{for } \dagger = \mathsf{a} \text{ or } \mathsf{c} \\ \text{or} \\ - \quad \text{for } \dagger = \mathsf{b} \text{ or } \mathsf{d}. \end{cases} \tag{3.13}$$

The direct consequence of this assumption is that solutions

$$^{\star}f_{\pm}(r;k;\Diamond_{o;\dagger}\,|\,\dagger;\varepsilon_{\dagger}) = \frac{W\{\psi_{\dagger}(r;\varepsilon_{\dagger};\Diamond_o), f_{\pm}(r;k;\Diamond_o)\}}{(\pm i\,k + \sigma_{1;\dagger}\sqrt{-\varepsilon_{\dagger}})\;\psi_{\dagger}(r;\varepsilon_{\dagger};\Diamond_o)} \tag{3.14}$$

of the Schrödinger equation with potential (3.2) satisfies the required boundary conditions

$$\lim_{r\to\infty}[e^{\mp i\,k\,r}\;{}^{\star}f_{\pm}(r;k;\Diamond_{o;\dagger}\,|\,\dagger;\varepsilon_{\dagger})] = 1 \tag{3.15}$$

so that

$$f_{\pm}(k;\Diamond_{o;\dagger}\,|\,\dagger;\varepsilon_{\dagger}) = \frac{2\Diamond_{o;\dagger}}{\pm i\,k + \sigma_{1;\dagger}\sqrt{-\varepsilon_{\dagger}}} \tag{3.16}$$

$$\times \lim_{r\to 0}[r^{s_{o;\dagger}-1} W\{\psi_{\dagger}(r;\varepsilon_{\dagger};\Diamond_o), f_{\pm}(r;k;\Diamond_o)\}/\psi_{\dagger}(r;\varepsilon_{\dagger};\Diamond_o)].$$

If $\dagger = \dagger_{+} = \mathsf{a}$ or $\mathsf{c}$ then the Wronskian in the right-hand side of (3.14) becomes energy-independent at r = 0:

$$W\{\psi_{\dagger_+}(r;\varepsilon_{\dagger_+};\Diamond_o), \varphi_{\dagger_-}(r;k;\Diamond_o)\}\Big|_{r=0} = -2\Diamond_o \lim_{r\to 0}[r^{-s_o}\,\psi_{\dagger_+}(r;\varepsilon_{\dagger};\Diamond_o)] \tag{3.17}$$

for any solution $\varphi_{\dagger_-}(r;k;\Diamond_o)$ irregular at the origin ($\dagger_- = \mathsf{b}$ or $\mathsf{d}$), including $f_+(r;k;\Diamond_o)$ and $f_-(r;k;\Diamond_o)$, unless k coincides with one of zeros of the functions $f_+(k;\Diamond_o)$ and $f_-(k;\Diamond_o)$, respectively. Therefore

$$f_{\pm}(k;\Diamond_{o;\dagger_+}\,|\,\dagger_+;\varepsilon_{\dagger_+}) = -\frac{2\Diamond_o}{\pm i\,k + \sigma_{1;\dagger_+}\sqrt{-\varepsilon_{\dagger_+}}}\,f_{\pm}(k;\Diamond_o) \tag{3.18}$$

which confirms both (3.5a) and (3.5c). As expected, the appropriate S-matrix elements for potential (3.2) $\dagger = \dagger_+ = \mathsf{a}$ or $\mathsf{c}$ are given by the conventional expressions [44-46]

$$ {}^{*}S(k;\Diamond_{o};\dagger_{+}\,|\,\dagger;\varepsilon_{\dagger_{+}}) = \frac{i\,k + \sigma_{1;\dagger_{+}}\sqrt{-\varepsilon_{\dagger_{+}}}}{-i\,k + \sigma_{1;\dagger_{+}}\sqrt{-\varepsilon_{\dagger_{+}}}}\, {}^{*}S(k;\Diamond_{o}) \qquad (3.19) $$

for potentials with exponential tails (keeping in mind that the appropriate superpotential tends to $\sigma_{1;\dagger_{+}}\sqrt{-\varepsilon_{\dagger_{+}}}$ as $r\rightarrow\infty$).

It should be stressed that both transformation relations (3.5a) and (3.5c) were derived with no restriction imposed on the value of $s_o$ so that they are applicable to both LP and LC regions. In particular, they hold for DTs converting the LC region of the IS@O potential into the LP region of the resultant potential as well as for the inverse transformations – the only cases discussed in the literature so far [1, 10].

Since transformation relations (3.5b) and (3.5d) are simply reverse of (3.5a) and (3.5c) the first impression may be that they are also hold for any values of $s_o$. However, the DT with a FF of type **b** or **d** is the inverse of the DT with a FF of type **a** or **c**, respectively, only for $s_o > 3/2$. In [21] both (3.5b) and (3.5d) were derived by analyzing behavior of the solution

$$ \begin{aligned} {}^{*}\varphi(r;k;\Diamond_{o}\,|\,\dagger;\varepsilon_{\dagger}) = \frac{1}{2i\,k}\Big[ & f_{-}(k;\Diamond_{o})\,W\{f_{+}(r;k;\Diamond_{o}),\psi_{\dagger}(r;\varepsilon_{\dagger};\Diamond_{o})\} \\ & - f_{+}(k;\Diamond_{o})\,W\{f_{-}(r;k;\Diamond_{o}),\psi_{\dagger}(r;\varepsilon_{\dagger};\Diamond_{o})\}\Big] \end{aligned} . \qquad (3.20) $$

near the origin. However, the latter function is regular at the origin only if $\Diamond_{o} > 1$.

Let us now show the derived transformation relations for the Jost function under DTs are also applicable to IS@O radial potentials with a Coulomb tail $q_1\, r^{-1}$ at infinity [23, 24]. First, since

$$ \psi_{\dagger}(r;\varepsilon_{\dagger};\Diamond_{o}) \sim A(\Diamond_{o})\, r^{-\frac{1}{2}\sigma_{1;\dagger}q_{1}/\sqrt{-\varepsilon_{\dagger}}}\ \exp(-\sigma_{1;\dagger}\sqrt{-\varepsilon_{\dagger}}) \quad \text{at large r} \qquad (3.21) $$

its reciprocal (3.1) has the asymptotics:

$$ {}^{*}\psi_{\dagger}(r;\varepsilon_{\dagger};\Diamond_{o}) \sim A^{-1}(\Diamond_{o})\, r^{\frac{1}{2}\sigma_{1;\dagger}q_{1}/\sqrt{-\varepsilon_{\dagger}}}\ \exp(\sigma_{1;\dagger}\sqrt{-\varepsilon_{\dagger}}) \quad \text{at large r,} \qquad (3.21^{*}) $$

so that DTs do not affect the Coulomb tail [23, 24, 43]. (In the particular case of the centrifugal Coulomb potential it was Sukumar [47, 22] who first noticed that the DT eliminating the ground state keeps the Coulomb term unchanged.)

Again, in following Newton's [39] conventions (14.36) and (14.37) for the Coulomb potential, we introduce the Jost function and its counterpart via (3.6), where the functions $f_{\pm}(r;k;s_o)$ are now satisfy the boundary conditions

$$\lim_{r\to\infty} e^{\mp ikr + i\eta(k)\ln(2kr)} f_{\pm}(r;k;\lozenge_o) = 1, \tag{3.22}$$

where the k-dependent phase

$$\eta(k) \equiv \frac{q_1}{2k}, \tag{3.23}$$

matches the one appearing in (A1) in [24]. Within the physical range of real momenta k parameter (3.23) coincides with the parameter n used in (14.37) in [39] to define the Jost function for the Coulomb potential. As demonstrated below, the additional phase shifts $\frac{1}{2}\pi\nu$ appearing in (A1) in [24] only complicates the resultant expression for the transformed Jost function since it depends on the singularity strength ν and therefore changes under action of the DTs:

$$\nu \equiv s_o - 1 \to s_{o;\dagger} - 1 = \nu + \sigma_{o;\dagger}. \tag{3.24}$$

Note also that both (9) in [23] and (A1) in [24] have an additional real factor $\exp[\eta(k)\pi/2]$ which is not affected by DTs and therefore makes no difference for our analysis. Is remarkable that the Wronskian between the functions $f_{\pm}(r;k;s_o)$ still satisfies to (12.27) in [39]:

$$W\{f_+(r;k;\lozenge_o), f_-(r;k;\lozenge_o)\} = -2ik \tag{3.25}$$

despite the change in the boundary condition at infinity. Therefore both Jost function and its counterpart are again given by (3.8). We thus conclude that functions (3.12) satisfy the boundary conditions:

$$\lim_{r\to\infty} e^{\mp[ikr - i\eta(k)]\ln(2kr)} \star f_{\pm}(r;k;\Diamond_{o;\dagger} \,|\, \dagger;\varepsilon_{\dagger}) = 1 \tag{3.26}$$

and therefore SUSY partners of the Jost function for any IS@O radial potential with a Coulomb tail are also related via (3.5a)-(3.5d) for DTs with nodeless FFs of any of four types. In particular, this implies that all the general conclusions made for SUSY pairs of IS@O potentials with exponential tails (including their phase-equivalence) can be automatically extended to potentials with the Coulomb asymptotics at infinity.

## 4. Transformation properties of the Titchmarsh-Weyl-Fulton function under Darboux deformations of the Kratzer oscillator

Let us now consider the transformation properties of the TW function under Darboux transformations of Kratzer oscillator (2.1) using Frobenius solutions at the origin as FFs. By definition the Taylor series in (4.1) is assumed to converge at any finite point on the positive semi-axis. We also require that it converges to zero at infinity so that the potential vanishes in this limit. In following [27, 48], let us consider the superposition

$$\psi_{\mathbf{b}}(r;\varepsilon;\Diamond_o) = \psi_{\mathbf{d}}(r;\varepsilon;\Diamond_o)/C + m(\varepsilon;\Diamond_o)\psi_{\mathbf{a}}(r;\varepsilon;\Diamond_o), \tag{4.1}$$

where

$$\psi_{\mathbf{a}}(r;\varepsilon;\Diamond_o) = r^{s_o}\, f_{\mathbf{a}}(r;\varepsilon;\Diamond_o) \tag{4.2}$$

and

$$\psi_{\mathbf{d}}(r;\varepsilon;\Diamond_o) \equiv r^{1-s_o} f_{\mathbf{d}}(r;\varepsilon;\Diamond_o) \tag{4.2*}$$

are two Frobenius solutions near the origin and C denotes their Wronskian:

$$C \equiv W\{\psi_{\mathbf{a}}(r;\varepsilon;\Diamond_o),\ \psi_{\mathbf{d}}(r;\varepsilon;\Diamond_o)\} = -2\Diamond_o \tag{4.3}$$

The Taylor series

$$f_{\dagger}(r;\varepsilon;\Diamond_o) = 1 + \sum_{k=1}^{\infty} f_{\dagger;k}(\varepsilon;\Diamond_o)\, r^k \tag{4.4}$$

satisfy the differential equations:

$$f''_{a}(r;\varepsilon;\Diamond_o) + 2s_o r^{-1} f'_{a}(r;\varepsilon;\Diamond_o) = [V(r;\Diamond_o) - s_o(s_o - 1)r^{-2} - \varepsilon] f_{a}(r;\varepsilon;\Diamond_o) \quad (4.5)$$

and

$$f''_{d}(r;\varepsilon;\Diamond_o) + 2(1 - s_o)r^{-1} f'_{d}(r;\varepsilon;\Diamond_o) = [V(r;\Diamond_o) - s_o(s_o - 1)r^{-2} - \varepsilon] f_{d}(r;\varepsilon;s_o) \quad (4.5^*)$$

for t = a and d, respectively. The coefficient $C^{-1}$ of $\psi_{d}(r;\varepsilon;\Diamond_o)$ in (4.2) is chosen in such a way that

$$W\{\psi_{a}(r;\varepsilon;\Diamond_o), \psi_{b}(r;\varepsilon;\Diamond_o)\} = 1 \quad (4.6)$$

assuming that ε is not a Hamiltonian eigenvalue $\varepsilon_{c;v}$. If $\varepsilon = \varepsilon_{c;v}$ then the Wronskian is vanishes as prescribed by the general theory of energy spectra of singular Hamiltonians [20]. As mentioned in Introduction, the energy-dependent coefficient *m*($\varepsilon;s_o$) is referred to below as the TWF function to stress that all the results of this Section are obtained within the framework of Fulton's theory [27]. In this paper we only consider the general case of non-integer exponent differences $2\Diamond_o = 2s_o - 1$. Cases 2A and 2B of integer $M = 2\Diamond_o$ in terms of [48] will be analyzed in a separate publication.

As stressed above it is assumed that the potential remains finite in the limit $r \to \infty$ so that the zero-point energy can be chosen in such a way that $V(\infty;q) = 0$. In addition, we only consider cases when the general solution of the Schrödinger equation with the given potential behaves as $r^{\lambda} \exp(-\nu r)$ at large r. As a result, the correction to the potential due to its Darboux deformation also vanishes at infinity so that the zero-point energy remains unchanged. The main consequence of the cited restrictions is that the energy becomes an unambiguously defined parameter for all the SUSY partners of our interest.

One can verify that the linear coefficient in the Tailor series in the right-hand side of (4.4) is energy-independent in both cases t = a and d:

$$2s_o f_{a,1}(\varepsilon;\Diamond_o) = q_1 \quad (4.7)$$

and

$$2(1-s_o)f_{d,1}(\varepsilon;s_o)=q_1. \qquad (4.7^*)$$

Since the first derivatives of functions (4.4) are energy-independent at r=0 it directly follows from (4.5) and (4.5*) that

$$f_{a,2}(\varepsilon;\Diamond_o)=\frac{\varepsilon-V_0}{4(\Diamond_o+1)} \qquad (4.8)$$

and

$$f_{d,2}(\varepsilon;\Diamond_o)=\frac{V_0-\varepsilon}{4(\Diamond-1)} \qquad (4.8^*)$$

and that the Wronskian of two Frobenius of the same type at energies ε+Δε and ε can be approximated as

$$W\{f_{\dagger}(r;\varepsilon+\Delta\varepsilon;\Diamond_o),\, f_{\dagger}(r;\varepsilon;\Diamond_o)\}=2r\,[f_{\dagger;2}(\varepsilon+\Delta\varepsilon;\Diamond_o)-f_{\dagger;2}(\varepsilon;\Diamond_o)]+O(r^2), \qquad (4.9)$$

where

$$f_{a,2}(\varepsilon+\Delta\varepsilon;\Diamond_o)-f_{a,2}(\varepsilon;\Diamond_o)=\frac{\Delta\varepsilon}{4(\Diamond_o+1)} \qquad (4.10)$$

and

$$f_{d,2}(\varepsilon+\Delta\varepsilon)-f_{d,2}(\varepsilon)=-\frac{\Delta\varepsilon}{4(\Diamond_o-1)}. \qquad (4.10^*)$$

**i) LP region**

Let us now consider the SUSY partner of solution (4.1),

$$^*\psi_b[r;\varepsilon;\Diamond_{o;\dagger}\,|\,\dagger;\varepsilon_\dagger]={}^*\psi_d[r;\varepsilon;\Diamond_{o;\dagger}\,|\,\dagger;\varepsilon_\dagger]+{}^*m(\varepsilon;\Diamond_{o;\dagger}\,|\,\dagger;\varepsilon_\dagger)\;{}^*\psi_a[r;\varepsilon;\Diamond_{o;\dagger}\,|\,\dagger;\varepsilon_\dagger],$$
$$(4.11)$$

constructed using nodeless Frobenius solution (4.2) or (4.2*) at the energy $\varepsilon_\dagger$ († = a or d) as the appropriate FF. The scale factor $w(\dagger',\varepsilon\,|\,\dagger;\varepsilon_\dagger)$ in the expression

$$^*\psi_{\dagger'}(r;\varepsilon;\Diamond_{o;\dagger}\,|\,\dagger;\varepsilon_\dagger)=\frac{W\{\psi_\dagger(r;\varepsilon_\dagger;\Diamond_o),\,\psi_{\dagger'}(r;\varepsilon;\Diamond_o)\}}{w_{\dagger'}(\varepsilon;\Diamond_o\,|\,\dagger;\varepsilon_\dagger)\psi_\dagger(r;\varepsilon_\dagger;\Diamond_o)} \qquad (4.12)$$

for the SUSY partner of the Frobenius solution ($\mathsf{t}'$ = $\mathsf{a}$ or $\mathsf{d}$ ) is chosen in such a way that

$$*\psi_{\mathsf{a}}(r;\varepsilon;\Diamond_{o;\mathsf{a}}\,|\,\mathsf{a};\varepsilon_{\mathsf{a}}) = r^{s_o+1}\,*f_{\mathsf{a}}(r;\varepsilon;\Diamond_{o;\mathsf{a}}\,|\,\mathsf{a};\varepsilon_{\mathsf{a}}), \tag{4.13aa}$$

$$*\psi_{\mathsf{a}}(r;\varepsilon;\Diamond_{o;\mathsf{d}}\,|\,\mathsf{d};\varepsilon_{\mathsf{d}}) = r^{s_o-1}\,*f_{\mathsf{a}}(r;\varepsilon;\Diamond_{o;\mathsf{d}}\,|\,\mathsf{d};\varepsilon_{\mathsf{d}}) \quad \text{within LP range}, \tag{4.13ad}$$

$$*\psi_{\mathsf{d}}(r;\varepsilon;\Diamond_{o;\mathsf{a}}\,|\,\mathsf{a};\varepsilon_{\mathsf{a}}) = r^{-s_o}\,*f_{\mathsf{d}}(r;\varepsilon;\Diamond_{o;\mathsf{a}}\,|\,\mathsf{a};\varepsilon_{\mathsf{a}}), \tag{4.13da}$$

$$*\psi_{\mathsf{d}}(r;\varepsilon;\Diamond_{o;\mathsf{d}}\,|\,\mathsf{d};\varepsilon_{\mathsf{d}}) = r^{2-s_o}\,*f_{\mathsf{d}}(r;\varepsilon;\Diamond_{o;\mathsf{d}}\,|\,\mathsf{d};\varepsilon_{\mathsf{d}}) \quad \text{within LP range}, \tag{4.13dd}$$

with

$$*f_{\mathsf{t}'}(r;\varepsilon;\Diamond_{o;\mathsf{t}}\,|\,\mathsf{t};\varepsilon_{\mathsf{t}}) = 1 + \sum_{k=1} *f_{\mathsf{t}';k}(\varepsilon;\Diamond_{o;\mathsf{t}}\,|\,\mathsf{t};\varepsilon_{\mathsf{t}})\,r^k. \tag{4.14}$$

Assuming that $\Diamond_o > 1$ for $\mathsf{t}$ = $\mathsf{d}$, the TWF function for the partner SL problem thus takes the form

$$*m(\varepsilon;\Diamond_{o;\mathsf{t}}\,|\,\mathsf{t};\varepsilon_{\mathsf{t}}) = \frac{\Diamond_o\ w(\mathsf{a},\varepsilon\,|\,\mathsf{t};\varepsilon_{\mathsf{t}})}{\Diamond_{o;\mathsf{t}}\ w(\mathsf{d},\varepsilon\,|\,\mathsf{t};\varepsilon_{\mathsf{t}})}\,m(\varepsilon;\Diamond_o). \tag{4.15}$$

Taking into account that

$$W\{\psi_{\mathsf{d}}(r;\varepsilon_{\mathsf{d}};\Diamond_o),\psi_{\mathsf{a}}(r;\varepsilon;\Diamond_o)\}\big|_{r=0} = -2\Diamond_o = -W\{\psi_{\mathsf{a}}(r;\varepsilon_{\mathsf{a}};\Diamond_o),\psi_{\mathsf{d}}(r;\varepsilon;\Diamond_o)\}\big|_{r=0} \tag{4.16}$$

we conclude that the appropriate scale factors are energy-independent:

$$w(\mathsf{a},\varepsilon\,|\,\mathsf{d};\varepsilon_{\mathsf{d}}) = 2\Diamond_o = -w(\mathsf{d},\varepsilon\,|\,\mathsf{a};\varepsilon_{\mathsf{a}}). \tag{4.17}$$

On other hand, substituting (4.10) and (4.10*) into the Wronskians in the right-hand side of the expressions

$$\frac{W\{\psi_{\mathsf{a}}(r;\varepsilon_{\mathsf{a}};\Diamond_o),\ \psi_{\mathsf{a}}(r;\varepsilon;\Diamond_o)\}}{\psi_{\mathsf{a}}(r;\varepsilon_{\mathsf{a}};\Diamond_o)} = \frac{r^{s_o}\,W\{f_{\mathsf{a}}(r;\varepsilon_{\mathsf{a}};\Diamond_o),\ f_{\mathsf{a}}(r;\varepsilon;\Diamond_o)\}}{f_{\mathsf{a}}(r;\varepsilon_{\mathsf{a}};\Diamond_o)} \tag{4.18}$$

and

$$\frac{W\{\psi_{\mathsf{d}}(r;\varepsilon_{\mathsf{d}};\Diamond_o),\ \psi_{\mathsf{d}}(r;\varepsilon;\Diamond_o)\}}{\psi_{\mathsf{d}}(r;\varepsilon_{\mathsf{d}};\Diamond_o)} = \frac{W\{f_{\mathsf{d}}(r;\varepsilon_{\mathsf{d}};\Diamond_o),\ f_{\mathsf{d}}(r;\varepsilon;\Diamond_o)\}}{r^{s_o-1}f_{\mathsf{d}}(r;\varepsilon_{\mathsf{d}};\Diamond_o)} \tag{4.18*}$$

gives

$$w(\mathbf{a},\varepsilon \mid \mathbf{a};\varepsilon_{\mathbf{a}}) = \frac{\varepsilon - \varepsilon_{\mathbf{a}}}{2(\lozenge_o + 1)} \tag{4.19}$$

and

$$w(\mathbf{d},\varepsilon \mid \mathbf{d};\varepsilon_{\mathbf{d}}) = \frac{\varepsilon_{\mathbf{d}} - \varepsilon}{2(\lozenge_o - 1)}\,. \tag{4.19*}$$

We thus come to the following explicit relations between the TWF functions for two partner SL problems:

$$^{*}m(\varepsilon;\lozenge_o + 1 \mid \mathbf{t}_+, \varepsilon_{\mathbf{t}_+}) = \frac{\varepsilon_{\mathbf{t}_+} - \varepsilon}{4(\lozenge_o + 1)^2}\, m(\varepsilon;\lozenge_o) \quad \text{for } \mathbf{t}_+ = \mathbf{a} \text{ or } \mathbf{c} \tag{4.20}$$

($\mathbf{c} \equiv \mathbf{c}$,0) and

$$^{*}m(\varepsilon;\lozenge_{o;\mathbf{d}} \mid \mathbf{d}, \varepsilon_{\mathbf{d}}) = \frac{4\lozenge_o^2}{\varepsilon_{\mathbf{d}} - \varepsilon}\, m(\varepsilon;\lozenge_o) \quad \text{for } \lozenge_o > 1. \tag{4.20*}$$

As expected, the DT with the FF of type $\mathbf{d}$ adds a pole to the TWF function at $\varepsilon = \varepsilon_{\mathbf{d}}$. Note that TWF function (4.20) for $\mathbf{t}_+ = \mathbf{a}$ must vanish at $\varepsilon = \varepsilon_{\mathbf{a}}$ because the function $m(\varepsilon;s_o)$ may not have a pole at this energy (otherwise the FF would be of type $\mathbf{c}$, not $\mathbf{a}$).

Keeping in mind that

$$^{*}f_{\mathbf{b}}(r;\varepsilon_{\mathbf{a}};\lozenge_{o;\mathbf{a}} \mid \mathbf{a};\varepsilon_{\mathbf{a}}) = {}^{*}f_{\mathbf{d}}(r;\varepsilon_{\mathbf{a}};\lozenge_{o;\mathbf{a}} \mid \mathbf{a};\varepsilon_{\mathbf{a}}) = f_{\mathbf{a}}^{-1}(r;\varepsilon_{\mathbf{a}};\lozenge_o)\,, \tag{4.21}$$

one can represent (4.20) in an alternative form similar to (4.20*):

$$m(\varepsilon;\lozenge_o) = \frac{4\lozenge_{o;\mathbf{a}}^2}{\varepsilon_{\mathbf{a}} - \varepsilon}\, {}^{*}m(\varepsilon;\lozenge_{o;\mathbf{a}} \mid \mathbf{a};\varepsilon_{\mathbf{a}}). \tag{4.22}$$

Note that TWF function (4.22) has a pole at $\varepsilon = \varepsilon_{\mathbf{a}}$ since TWF function (4.20) has a zero at this point.

Similarly, (4.20*) can be represented as

$$^{*}m(\varepsilon;\lozenge_{o;\mathbf{d}} \mid \mathbf{d}, \varepsilon_{\mathbf{d}}) = \frac{4(\lozenge_{o;\mathbf{d}} + 1)^2}{\varepsilon_{\mathbf{d}} - \varepsilon}\, m(\varepsilon;\lozenge_o) \tag{4.22*}$$

which is reverse of (4.21), with $\mathbf{t}_+ = \mathbf{c}$ and $\varepsilon_{\mathbf{c}}$ changed for $\varepsilon_{\mathbf{d}}$ -- the direct consequence of the fact the DT with the FF of type **d** creates the ground-energy state in the new potential.

Since the R@∞ solution turns into Frobenius solution (4.2*) at any zero, $\varepsilon = \varepsilon_{\mathbf{b},m}$, of the function $m(\varepsilon;\Diamond_o)$ the appropriate TWF function is also transformed as

$$*m(\varepsilon;\Diamond_{o;\mathbf{b}} \,|\, \mathbf{b},\varepsilon_{\mathbf{b},m}) = \frac{4\Diamond_o^2}{\varepsilon_{\mathbf{b},m} - \varepsilon}\, m(\varepsilon;\Diamond_o) \quad \text{for } \Diamond_o > 1. \tag{4.23}$$

However, since the TWF function $m(\varepsilon;s_o)$ has the zero at $\varepsilon = \varepsilon_{\mathbf{b},m}$ the DT in question does not add a new pole to its SUSY partner, as expected from the fact [16, 22, 11] that the radial potentials related via this DT are isospectral. Again, we can alternatively represent (4.23) as

$$*m(\varepsilon;\Diamond_{o;\mathbf{b}} \,|\, \mathbf{b},\varepsilon_{\mathbf{b},m}) = \frac{4(\Diamond_{o;\mathbf{b}} + 1)^2}{\varepsilon_{\mathbf{b},m} - \varepsilon}\, m(\varepsilon;\Diamond_o) \quad \text{for } \Diamond_o > 1 \tag{4.23'}$$

which is reverse of (4.20), with $\mathbf{t}_+ = \mathbf{a}$ and $\varepsilon_{\mathbf{a}}$ changed for $\varepsilon_{\mathbf{b},m}$ as a direct consequence of the relation

$$*f_{\mathbf{a}}(r;\varepsilon_{\mathbf{b},m};\Diamond_{o;\mathbf{b}} \,|\, \mathbf{b};\varepsilon_{\mathbf{b},m}) = f_{\mathbf{b}}^{-1}(r;\varepsilon_{\mathbf{b},m};\Diamond_o)\,. \tag{4.24}$$

**ii) LC region**

For $\frac{1}{2} < s_o < \frac{3}{2}$ the DT with the FF of type **d** keeps the potential within the LC region but (contrary to the LP case) changes the solution type. Namely, it turns Frobenius solutions (4.2) and (4.2*) into the Frobenius solutions

$$*\psi_{\mathbf{a}}(r;\varepsilon;\Diamond_{o;\mathbf{d}} \,|\, \mathbf{d};\varepsilon_{\mathbf{d}}) = r^{s_o}\; *f_{\mathbf{a}}(r;\varepsilon;\Diamond_{o;\mathbf{d}} \,|\, \mathbf{d};\varepsilon_{\mathbf{d}}) \tag{4.25}$$

and

$$*\psi_{\mathbf{d}}(r;\varepsilon;\Diamond_{o;\mathbf{d}} \,|\, \mathbf{d};\varepsilon_{\mathbf{d}}) = r^{1-s_o}\; *f_{\mathbf{d}}(r;\varepsilon;\Diamond_{o;\mathbf{d}} \,|\, \mathbf{d};\varepsilon_{\mathbf{d}})\,, \tag{4.25*}$$

respectively, so that

$$^{*}m(\varepsilon;\Diamond_{o;\mathsf{d}}\,|\,\mathsf{d},\varepsilon_{\mathsf{d}}) = \frac{\varepsilon-\varepsilon_{\mathsf{d}}}{16\Diamond_o^2(\Diamond_o-1)^2\; m(\varepsilon;\Diamond_o)} \qquad \text{for } 0<\Diamond_o<1. \tag{4.26}$$

The most important consequence of the derived expression is that function (4.26) has poles at zeros of the TWF function $m(\varepsilon;s_o)$, whereas all the poles of the latter function turn by the DT into zeros of the TWF function (4.26).

Since the R@∞ solution turns into Frobenius solution (4.2*) at any zero $\varepsilon = \varepsilon_{\mathsf{b},m}$ of the function $m(\varepsilon;\Diamond_o)$, the WTF function in question also transforms as

$$^{*}m(\varepsilon;\Diamond_{o;\mathsf{b}}\,|\,\mathsf{b},\varepsilon_{\mathsf{b},m}) = \frac{\varepsilon-\varepsilon_{\mathsf{b},m}}{16\Diamond_o^2(\Diamond_o-1)^2\; m(\varepsilon;\Diamond_o)} \qquad \text{for } 0<\Diamond_o<1. \tag{4.26′}$$

Note that function (4.26′) remains finite at $\varepsilon = \varepsilon_{\mathsf{b},m}$ since the WTF function $m(\varepsilon;s_o)$ has the zero at this energy value. Both derived expressions (4.26) and (4.26′) have fundamental significance because it allows one to obtain the discrete energy spectra for Darboux deformations of exactly solvable potentials in cases when the conventional rules of SUSY quantum mechanics are not valid anymore.

Let us also remind the reader that DTs with FFs of both types $\mathsf{b}$ and $\mathsf{d}$ keep the potential parameters within the LC range while changing the value of the SI from $\Diamond_o$ to $1-\Diamond_o$. The reverse of any DT with the FF of type $\mathsf{b}$ with the factorization energy $\varepsilon = \varepsilon_{\mathsf{b},m}$ has type $\mathsf{d}$. Keeping in mind that the reverse DT has the same factorization energy $\varepsilon = \varepsilon_{\mathsf{b},m}$ and making use of (4.26) one can then directly verify that the latter transformation does bring TMF function (4.26′) back to $m(\varepsilon;\Diamond_o)$ as expected.

Representing (4.1) as

$$\psi_{\mathsf{d}}(r;\varepsilon;\Diamond_o) = 2\Diamond_o[m(\varepsilon;\Diamond_o)\psi_{\mathsf{a}}(r;\varepsilon;\Diamond_o) - \psi_{\mathsf{b}}(r;\varepsilon;\Diamond_o)] \tag{4.27}$$

and taking into account that the R@∞ and R@O solutions defined via (4.1) and (4.2) has an opposite sign at small r we conclude that Frobenius solution (4.24) lying below the ground energy level is nodeless if $m(\varepsilon; \lozenge_o) > 0$.

As originally pointed out by the author in [49], one can always use nodeless superpositions of R@∞ and R@O solutions below the ground energy level as FFs for DTs to construct a continuous family of isospectral IS@O potentials. As we understand now our conclusion that the appropriate DTs insert a new bound state at the factorization energy is only correct if the SI for the original potential lies within the LP range.

Approximately at the same time with our work [49] Deift [50] came up with similar results for the Schrödinger equation with a nonsingular radial potential which was solved with the Direchlet boundary condition. The energy spectrum of the potential with the inserted bound energy state was unambiguously determined by analyzing transformed reflection and transmission coefficients, instead of the transformed Jost function studied in our works [21, 49]. Compared with non-singular potentials, the main complication arising in problems of our interest comes from the fact that one has to find the particular R@∞ solutions of the Schrödinger equation with the given potential which are converted by the given DT into bound eigenfunctions of the deformed potential. If the TWF function for the deformed potential is known then its poles unambiguously determine the eigenvalues sought for *including the LC region*. The DTs of the R@∞ solutions at those energies then give the appropriate eigenfunctions.

The very specific feature of superposition (4.21) is that this is a Frobenius solution. Therefore we can write down a closed form expression for the TWF function associated with the deformed potential if the latter is known for the original IS@O potential and as a result obtain the energy spectrum for the SUSY partner. We give examples of such exactly-solvable pairs of *non-isospectral* SUSY partners in Section 6 below.

## 5. Closed-form expressions for TWF functions of exactly-solvable reflective potentials on the half-line

### 5.1 Exactly-solvable radial potential with an exponential tail at infinity

Let us apply the results of the previous Section to the *r*-GRef radial potential [17, 18] exactly solvable via hypergeometric functions by means of the change of variable z(r) such that

$$z' = {}_1\wp_G^{-1/2}[z;a,b] \equiv 2(1-z)\sqrt{\frac{z}{az+b}}\,, \qquad (5.1.1)$$

where prime stands for the first derivative with respect to r. When expressed in terms of z, the potential takes the form

$$V[z\,|\,{}_1G^{K\Im\aleph}] \equiv {}_1V[z;\lambda_o,\mu_o;a,b] = -\,{}_1\wp^{-1}[z;a,b]\;\; {}_1I_G^o[z;\lambda_o,\mu_o] - \tfrac{1}{2}\{z,x\}\,, \qquad (5.1.2)$$

where

$${}_1I_G^o[z;\lambda_o,\mu_o] \equiv {}_1I^o[z;\lambda_o,0,\mu_o] \qquad (5.1.3)$$

and $\{z,x\}$ denote the so-called [15] 'reference polynomial fraction' (Ref-PF) and 'Schwartz derivative', respectively [51, 52]. Compared with [16-18], the original expression

$${}_1I^o[z;\lambda,\nu,\mu] = \frac{1-\lambda^2}{4z^2} + \frac{1-\nu^2}{4(1-z)^2} + \frac{\mu^2-\lambda^2-\nu^2-1}{4z(1-z)} \qquad (5.1.4)$$

for the Schwartz invariant [51, 52] was slightly simplified by requiring potential (5.1.2) to vanish as $r\to+\infty$ In following [15], superscripts K=2, $\Im$=1 and $\aleph$=1 specify the generic ($a \neq 0$, $b > 0$) radial potentials of the given structure. The shape-invariant h-PT and E/MR potentials ($\Im$=0) associated with two limiting cases a=0 and b=0 are described by the combinations K= $\aleph$=1 and K= $\aleph$=2, respectively, with $\Im$=0 in both cases.

Before considering the general case of the SL problem

$$\left\{\frac{d^2}{dz^2} + {}_1I^{o}_{G}[z;\lambda_o,\mu_o] + \varepsilon\, {}_1\wp_G[z;a,b]\right\}\Phi[z;\varepsilon\,|\,\lambda_o,\mu_o;a,b] = 0\,. \tag{5.1.5}$$

for arbitrary values of parameters a and b, let us first discuss its limiting case a = 0, b = 1 corresponding to the h-PT potential:

$$\left\{\frac{d^2}{dz^2} + {}_1I^{o}_{G}[z;\lambda_o,\mu] + \frac{\varepsilon}{4(1-z)^2}\right\}\Phi[z;\varepsilon\,|\,\lambda_o,\mu] = 0. \tag{5.1.6}$$

Since quantization of (5.1.6) under the boundary conditions

$$\lim_{z\to 0}\left\{z^{-\frac{1}{2}(1+\lambda_o)}\Phi[z;\varepsilon_{c;v}\,|\,\lambda_o,\mu]\right\} = 0 \tag{5.1.7a}$$

and

$$\lim_{z\to 1}\left\{(1-z)^{-\frac{1}{2}(1+\sqrt{-\varepsilon_{c;v}})}\Phi[z;\varepsilon_{c;v}\,|\,\lambda_o,\mu]\right\} = 0 \tag{5.1.7b}$$

is equivalent to the eigenvalue problem

$$(\hat{H}_o - \varepsilon)\Theta[z;\varepsilon\,|\,\lambda_o,\mu;0,1] = 0 \tag{5.1.8}$$

for the self-adjoint operator:

$$\hat{H}_o \equiv -4\frac{d}{dz}(1-z)^2\frac{d}{dz} + V_o[z\,|\,\lambda_o,\mu], \tag{5.1.9}$$

we can immediately apply Kemble's arguments [20] to the function

$$\Theta[z;\varepsilon\,|\,\lambda_o,\mu] = (1-z)^{-\frac{1}{2}}[z]\Phi[z;\varepsilon\,|\,\lambda_o,\mu] \tag{5.1.10}$$

in both LP and LC regions.

Two Fuschian solutions of ordinary differential equation (5.1.4) near the origin,

$${}_1\phi_{a}[z;\varepsilon;\lambda_o,\mu] = z^{\frac{1}{2}(1+\lambda_o)}(1-z)^{\frac{1}{2}(1+\sqrt{-\varepsilon})}F[\alpha(\varepsilon),\beta(\varepsilon);1+\lambda_o;z] \tag{5.1.11}$$

and

$$ {}_1\phi_{\mathbf{d}}[z;\varepsilon;\lambda_o,\mu] = z^{\frac{1}{2}(1-\lambda_o)}(1-z)^{\frac{1}{2}(1+\sqrt{-\varepsilon})} \times F[\alpha(\varepsilon)-\lambda_o,\beta(\varepsilon)-\lambda_o;1-\lambda_o;z], \tag{5.1.11*} $$

where

$$ \beta(\varepsilon)-\alpha(\varepsilon)=\mu>0,\ \alpha(\varepsilon)+\beta(\varepsilon)=\sqrt{-\varepsilon}+\lambda_o+1, \tag{5.1.12} $$

as well as their Wronskian

$$ W\{{}_1\psi_{\mathbf{a}}[z;\varepsilon;\lambda_o,\mu],\,{}_1\psi_{\mathbf{d}}[z;\varepsilon;\lambda_o,\mu]\} = -\lambda_o \tag{5.1.13} $$

have been explicitly introduced by Schafheitlin [53] in his analysis of zeros of the hypergeometric function in the 19th century (under obvious influence of Schwartz's famous work [51]). (Compared with the notation in [53] we set $\lambda=-\lambda_o$ and $\nu=-\sqrt{-\varepsilon}$. In following [52, 54] we [16] also interchanged $\alpha$ and $\beta$ so that $\alpha$ is assumed to be smaller than $\beta$.)

Substituting

$$ c-a-b=\lambda_o>0,\ a=c-\alpha=\beta-\lambda_o,\ b=c-\beta=\alpha-\lambda_o \tag{5.1.14} $$

in (15.3.6) in [55] and changing z for 1–z one can represent the R@O solution

$$ {}_1\phi_{\mathbf{b}}[z;\varepsilon;\lambda_o,\mu] = -\frac{\Gamma(\lambda_o+1)\Gamma(\sqrt{-\varepsilon}+1)}{\Gamma(\lambda_o+\alpha(\varepsilon))\Gamma(\lambda_o+\beta(\varepsilon))} F[\alpha(\varepsilon),\beta(\varepsilon);\sqrt{-\varepsilon}+1;1-z] \tag{5.1.15} $$

as

$$ {}_1\phi_{\mathbf{b}}[z;\varepsilon;\lambda_o,\mu] = -{}_1\phi_{\mathbf{d}}[z;\varepsilon;\lambda_o,\mu]/\lambda_o + {}_1m_0(\varepsilon;\lambda_o,\mu)\,{}_1\phi_{\mathbf{a}}[z;\varepsilon;\lambda_o,\mu], \tag{5.1.16} $$

where

$$ {}_1m_0(\varepsilon;\lambda_o,\mu) = -\frac{\Gamma(-\lambda_o)}{\Gamma(\lambda_o+1)}\,{}_1M_{0,0}(\varepsilon;\lambda_o,\mu). \tag{5.1.17} $$

The eigenvalues

$$ {}_1\varepsilon_{\mathbf{c},v} \equiv -(\mu-\lambda_o-2v+1)^2 \tag{5.1.18} $$

correspond to the poles $\sqrt{-\varepsilon}=\lambda_{1;\mathbf{c},v}$ of the energy-dependent factor

$$ {}_1M_{0,0}(\varepsilon;\lambda_o,\mu) \equiv \frac{\Gamma({}_1\alpha_{++}(\varepsilon;\lambda_o,\mu))\Gamma({}_1\beta_{++}(\varepsilon;\lambda_o,\mu))}{\Gamma({}_1\alpha_{-+}(\varepsilon;\lambda_o,\mu))\Gamma({}_1\beta_{-+}(\varepsilon;\lambda_o,\mu))}, \tag{5.1.19} $$

where

$${}_1\alpha_{\sigma_0\sigma_1}(\varepsilon;\lambda_o,\mu) = \tfrac{1}{2}(\sigma_0\lambda_o + \sigma_1\sqrt{-\varepsilon} + 1 - \mu) \qquad (5.1.20a)$$

and

$${}_1\beta_{\sigma_0\sigma_1}(\varepsilon;\lambda_o,\mu) = \tfrac{1}{2}[\sigma_0\lambda_o + \sigma_1\sqrt{-\varepsilon} + 1 + \mu]\,. \qquad (5.1.20b)$$

Since ${}_1\beta_{++}({}_1\varepsilon;\lambda_o,\mu) > 0$ and none of the gamma-functions has zeros on the real axis, the poles are given by the relation:

$${}_1\alpha_{++}({}_1\varepsilon_{\mathbf{c},v};\lambda_o,\mu) = -v \qquad (5.1.21)$$

which directly leads to (5.1.18). One can immediately verify that the lowest-energy eigenfunction

$${}_1\phi_{\mathbf{c},0}[z;\varepsilon_{\mathbf{c},0};\lambda_o,\mu] = z^{\frac{1}{2}(1+\lambda_o)}(1-z)^{\frac{1}{2}(1+\lambda_{1;\mathbf{c},0})} \qquad (5.1.22)$$

is nodeless, in agreement with Kemble's conclusions [20].

Since both parameters of the hypergeometric functions in the right-hand side of (5.1.11) and (5.1.15), $\alpha(\varepsilon)$ and $\beta(\varepsilon)$, are positive for $\sqrt{-\varepsilon} > \mu - \lambda_o - 1$ the same is true for each term in its Taylor series expansion in z so that the R@O and R@∞ solution in question must be nodeless for any positive value of $\lambda_o$ (including its LC *region*) as far as they lie below the lowest eigenvalue.

We are ready to resume discussion of the general case represented by SL equation (5.1.5). One can easily verify that most of the derived formulas remain valid if we simply change the constant $\mu$ for the following function of ε:

$$\mu(a\varepsilon;\mu_o) \equiv \sqrt{\mu_o^2 - a\varepsilon}\,. \qquad (5.1.23)$$

Below we choose

$$c_1 = a + b = 1 \qquad (5.1.24)$$

and assume that b > 0. The limiting case of the E/MR potential (b=0) will be discussed in Section 7. It is worth mentioning that radial potential (5.1.2) has two branches separated by the h-PT potential. The first (0 < a <1, b = 1–a <1) forms a bridge between the E/MR and h-PT potentials. The second branch (a <0, b = 1–a >1) is exactly solvable only at energies $\varepsilon > -\mu_o^2/|a|$. It is essential that the function

$$\Lambda(t) \equiv t + \lambda_o + 1 - \mu(-at^2;\mu_o) \tag{5.1.25}$$

monotonically increases with t regardless of sign of a. In fact, the assertion is trivial if a <0. As for positive a, one can verify that the function derivative lies between 0 and 1 in the latter case. We thus conclude that the function $\Lambda(\sqrt{-\varepsilon})$ is positive at any energy below the lowest eigenvalue which completes the proof.

By analogy with (5.1.15), the R@∞ superposition of Fuschian solutions

$$_1\phi_{\mathbf{a}}[z;\varepsilon;\lambda_o,\mu_o;1-a] = {_1\phi_{\mathbf{a}}}[z;\varepsilon;\lambda_o,\mu(a\varepsilon;\mu_o)] \tag{5.1.26}$$

and

$$_1\phi_{\mathbf{d}}[z;\varepsilon;\lambda_o,\mu_o;1-a] = {_1\phi_{\mathbf{d}}}[z;\varepsilon;\lambda_o,\mu(a\varepsilon;\mu_o)] \tag{5.1.26*}$$

can be represented as

$$_1\phi_{\mathbf{b}}[z;\varepsilon;\lambda_o,\mu_o;b] = -\frac{1}{\lambda_o}\,{_1\phi_{\mathbf{d}}}[z;\varepsilon;\lambda_o,\mu_o;b] + {_1m_G}(\varepsilon;\lambda_o,\mu_o;b)\,{_1\phi_{\mathbf{a}}}[z;\varepsilon;\lambda_o,\mu_o;b], \tag{5.1.27}$$

where

$$_1m_G(\varepsilon;\lambda_o,\mu_o;1-a) = {_1m_0}\left(\varepsilon;\lambda_o,\mu(a\varepsilon;\mu_o)\right). \tag{5.1.28}$$

The poles $_1\varepsilon_{\mathbf{c},v} \equiv -\lambda^2_{1;\mathbf{c},v}$ of the latter functions coincide with positive roots $\lambda_{1;\mathbf{t}_+,m} = \lambda_{1;\mathbf{c},m}$ of the quadratic equations

$$b\lambda^2_{1;\mathbf{t}_+,m} + 2\lambda_{1;\mathbf{t}_+,m}(2m+1+\lambda_o) + (2m+1+\lambda_o)^2 - \mu_o^2 = 0, \tag{5.1.29}$$

with the common leading coefficient b and the discriminants given by the following second-order polynomials in m:

$$_1\Delta_{+;m}(\lambda_o,\mu_o;b) = 4\left[(1-b)(2m+1+\lambda_o)^2 + b\mu_o^2\right]. \tag{5.1.29†}$$

Negative roots of quadratic equations (5.1.29) describe the primary and secondary sequences of R@O AEH solutions:

$$b\lambda_{1;\mathbf{a},m} = -\lambda_o - 2m - 1 - \tfrac{1}{2}\sqrt{{}_1\Delta_{+;m}(\lambda_o,\mu_o;b)} \qquad \text{for m=0, 1, 2,...} \qquad (5.1.30)$$

and

$$b\lambda_{1;\mathbf{a}',m} = -\lambda_o - 2m - 1 + \tfrac{1}{2}\sqrt{{}_1\Delta_{+;m}(\lambda_o,\mu_o;b)} \qquad \text{for } 2m > \mu_o - \lambda_o - 1, \quad (5.1.30')$$

respectively. Each sequence has an infinite number of members if $0 < b \le 1$. Since discriminant (5.1.29†) monotonically increases with m for $0 < b \le 1$ it immediately follows from (5.1.30) that all the AEH solutions **a**,m lie below the lowest eigenvalue and therefore must be nodeless. The limiting case of the E/MR potential (b=0) has been already studied in great details in [34] (see also [56]). It has some specific features not covered by the general analysis and therefore discussed separately in subsection 8.1 below. The subset of PF beams with b >1 will be briefly discussed in next Section.

To convert m-coefficient (5.1.28) into the TWF function for radial potential (5.1.2) one has to take into account that the variable z(r) is proportional to $r^2$ at small r so that the change of variable from z to r changes the value of the SI. To allow for this effect let us first express SL equation (5.1.5) in terms of an auxiliary variable

$$y(r) \equiv y[z(r)] \equiv \sqrt{z(r)}\,. \qquad (5.1.31)$$

The change of variable converts (5.1.5) into another ST problem

$$\left\{\frac{d^2}{dz^2} + {}_1I^o_H[z;\lambda_o,\mu_o] + \varepsilon\, {}_1\wp_H[z;a,b]\right\} X[z;\varepsilon\,|\,\lambda_o,\mu_o;a,b] = 0\,. \qquad (5.1.32)$$

where (see, i.g., [57])

$${}_1I^o_H[y;\lambda_o,\mu_o] = 4y\, {}_1I^o_G[y^2;\lambda_o,\mu_o] + \tfrac{1}{2}\{z,y\} \qquad (5.1.33)$$

and

$${}_1\wp_H[y;a,b] = 4y^2\, {}_1\wp_G[y^2;a,b] = \frac{ay^2+b}{(1-y^2)^2}\,. \qquad (5.1.33^*)$$

Substituting

$$\{z, y\} = -\tfrac{3}{2} y^{-2}, \tag{5.1.34}$$

into the right-hand side of (5.1.32′) we find that the change of variable y[z] results in the twice larger ExpDiff@O (as expected from the fact that z is square of y). Since

$$\left.\frac{dy}{dr}\right|_{r=0} = 1/\sqrt{b} \tag{5.1.35}$$

we conclude that the parameter $\lambda_o$ coincides with the SI of potential (5.1.2):

$$\lambda_o = \lozenge_o = s_o - \tfrac{1}{2} \qquad (b > 0). \tag{5.1.36}$$

One can verify that the appropriate gauge transformation converts (5.1.32) into the Heun equation [58, 59]

$$\left\{\frac{d^2}{dy^2} + \left[\frac{\gamma_H}{y} + \delta_H \frac{y}{y^2 - 1}\right]\frac{d}{dy} - \frac{4\alpha_H\beta_H}{1 - y^2}\right\} Hf(\alpha_H, \beta_H, \gamma_H, \delta_H, 0; y) = 0, \tag{5.1.37}$$

where

$$\gamma_H + 2\delta_H = \alpha_H + \beta_H + 1. \tag{5.1.38}$$

When solved with the boundary condition

$$Hf(\alpha_H, \beta_H, \gamma_H, \delta_H, 0; y)\Big|_{y=0} = 1 \tag{5.1.39}$$

the corresponding Heun function turns into the hypergeometric series in $y^2$:

$$Hf(2\alpha, 2\beta, 2\gamma - 1, \delta_H, \delta_H; 0; y) \equiv {}_2F_1(\alpha, \beta; \gamma; y^2) \tag{5.1.40}$$

so that the Frobenius solutions near the origin take the form

$$\chi_{\mathbf{a}}[y; \varepsilon; \lozenge_o] = y^{s_o}\, Hf\left(2{}_1\alpha_{+,-}(\varepsilon), 2{}_1\beta_{+,-}(\varepsilon), 2\lambda_o + 1, {}_1\delta_{+,-}(\varepsilon), {}_1\delta_{+,-}(\varepsilon); 0; y\right) \tag{5.1.41}$$

and

$$\chi_{\mathbf{d}}[y; \varepsilon; \lozenge_o] = y^{1-s_o}\, Hf\left(2{}_1\alpha_{-,-}(\varepsilon), 2{}_1\beta_{-,-}(\varepsilon), 2\lambda_o + 1, {}_1\delta_{-,-}(\varepsilon), {}_1\delta_{-,-}(\varepsilon); 0; y\right),$$

(5.1.41*)

where

$$ {}_1\alpha_{\sigma_0,\sigma_1}(\varepsilon) \equiv {}_1\alpha_{\sigma_0,\sigma_1}\left(\varepsilon;\lambda_o,\mu(\mu_o;a\varepsilon)\right), \tag{5.1.42a} $$

$$ {}_1\beta_{\sigma_0,\sigma_1}(\varepsilon) = {}_1\beta_{\sigma_0,\sigma_1}\left(\varepsilon;\lambda_o,\mu(\mu_o;a\varepsilon)\right), \tag{5.1.42b} $$

$$ {}_1\delta_{\sigma_0,\sigma_1}(\varepsilon) \equiv {}_1\delta_{\sigma_0,\sigma_1}(\varepsilon;\lambda_o,\mu(\mu_o;a\varepsilon)) = {}_1\alpha_{\sigma_0,\sigma_1}(\varepsilon) + {}_1\beta_{\sigma_0,\sigma_1}(\varepsilon) - \lambda_o . \tag{5.1.43} $$

(Here and below the notations for the exponent differences $\lambda_o$ and $\Diamond_o = \lambda_o$ associated with the singular points of the Sturm-Liouville equation (5.1.5) and the Schrödinger equation with the radial potential $V[r\,|\,{}_\iota G^{211}]$ are used in a completely interchangeable fashion whichever seems more appropriate in the given context.) Taking into account that the new ST equation has a twice larger SI at the origin, we represent the R@∞ solution as

$$ {}_1\chi_{\mathbf{b}}[z;\varepsilon;\lambda_o,\mu_o;b] = -\frac{1}{2\lambda_o}\,{}_1\chi_{\mathbf{d}}[z;\varepsilon;\lambda_o,\mu_o;b] + {}_1m_{\mathrm{H}}(\varepsilon;\lambda_o,\mu_o;b)\,{}_1\phi_{\mathbf{a}}[z;\varepsilon;\lambda_o,\mu_o;b] $$

(5.1.44)

where

$$ {}_1m_{\mathrm{H}}(\varepsilon;\lambda_o,\mu_o;b) = \tfrac{1}{2}\,{}_1m_{\mathrm{G}}(\varepsilon;\lambda_o,\mu_o;b). \tag{5.1.45} $$

We thus come to the following closed-form expression for the TWF function of the radial potential $V[z;\Diamond_o,\mu_o;1-b,b]$ for $b > 0$:

$$ {}_1m_G(\varepsilon;\Diamond_o,\mu_o;b) = -\frac{\Gamma(-\Diamond_o)}{2b^{\Diamond_o}\,\Gamma(\Diamond_o+1)}\,{}_1M_{0,0}\left(\varepsilon;\Diamond_o,\mu(a\varepsilon;\mu_o)\right). \tag{5.1.46} $$

The partner TWF functions constructed using nodeless Fuschian solutions as FFs for the appropriate DTs will be discussed in section 6.1.

### 5.2 Exactly-solvable radial potential with a Coulomb tail at infinity

Let us now extend the above results to the *c*-GRef radial potential [17, 18] exactly solvable via c-hypergeometric functions by means of the change of variable $\zeta(r)$ such

$$\zeta' = {}_0\wp_G^{-1/2}[\zeta; a, b] \equiv 2\sqrt{\frac{\zeta}{a\zeta + b}}\,, \tag{5.2.1}$$

where prime stands for the first derivative with respect to r. When expressed in terms of $\zeta$, the potential takes the form

$$V[\zeta \mid {}_0\mathcal{G}^{K\Im\aleph}] = -\,{}_0\wp^{-1}[\zeta; a, b]\ \ {}_0I^o[\zeta; \lambda_o, g_o] - 1/2\{\zeta, x\}, \tag{5.2.2}$$

where

$${}_0I^o[z; \lambda_o, g_o] = \frac{1-\lambda_o^2}{4\zeta^2} - \frac{g_o}{4\zeta} - 1/4\,\nu_o^2\,. \tag{5.2.3}$$

Since the isotonic oscillator represented by the potential $V[\zeta \mid {}_0\mathcal{G}^{101}]$ is unbounded from above at large r, it is not covered by the current publication. So we consider only the generic case of a positive leading coefficient a in (5.2.1) which is set to 1 for convenience. An analysis of the asymptotic behavior of potential (5.2.2) at large r shows that it has a Coulomb tail with

$$q_1 = 1/2\, g_o\,. \tag{5.2.4}$$

We can thus make the potential vanish at infinity by choosing $\nu_o = 0$. As a result, the SL equation of our interest, when expressed in terms of the energy-dependent variable

$$\varsigma_\varepsilon \equiv \sqrt{-\varepsilon}\,\zeta\,, \tag{5.2.5}$$

turns into the Whittaker equation [40]

$$\left\{\frac{d^2}{d\varsigma_\varepsilon^2} - 1/4 + \frac{1-\lambda_o^2}{4\varsigma_\varepsilon^2} + \frac{\kappa(\varepsilon)}{\varsigma_\varepsilon}\right\}\Phi[\varsigma_\varepsilon/\sqrt{-\varepsilon}; \varepsilon \mid {}_0\mathcal{G}^{2\Im\aleph}] = 0 \tag{5.2.6}$$

with the energy-dependent index parameter

$$\kappa(\varepsilon) \equiv \kappa(\varepsilon; g_o; b) = \frac{b\varepsilon - g_o}{4\sqrt{-\varepsilon}} \tag{5.2.7}$$

(In this paper we are only interested in the *c*-GRef potentials with $\Im+\aleph=2$, where $\aleph=1$ or 2). By changing the notation $\varsigma_\varepsilon$ for $\varsigma$ in (5.2.6), setting b=0, and making $\varepsilon$ tend to $-\infty$ we come to (2.12). However, in the general case of a positive b the variable $\zeta$ is proportional to $r^2$ so that variable $\varsigma_\varepsilon$ defined via (5.2.5) is not the same as $\varsigma$ appearing in (2.12).

Two Frobenius solutions of the Schrödinger equation at the origin are related to the Frobenius solutions of Whittaker equation (5.2.6) as follows

$$_0\psi_{\mathbf{a}}[\zeta;\varepsilon;\Diamond_o] = r^{s_o}[\zeta]\ {}_0f_{\mathbf{a}}[\zeta;\varepsilon;\Diamond_o] \tag{5.2.8}$$

and

$$_0\psi_{\mathbf{d}}[\zeta;\varepsilon;\Diamond_o] = r^{1-s_o}[\zeta]\ {}_0f_{\mathbf{d}}[\zeta;\varepsilon;\Diamond_o], \tag{5.2.8*}$$

where

$$_0f_{\mathbf{a}}[\zeta;\varepsilon;\Diamond_o] = \sqrt[4]{1/\zeta+1/b}\,|\varepsilon|^{-\frac{1}{4}(\Diamond_o+1)}\left(\sqrt{b}/r[\zeta]\right)^{s_o} M_{\kappa(\varepsilon),\frac{1}{2}\Diamond_o}(\sqrt{-\varepsilon}\,\zeta) \tag{5.2.9}$$

and

$$_0f_{\mathbf{d}}[\zeta;\varepsilon;\Diamond_o] = \sqrt[4]{1/\zeta+1/b}\,|\varepsilon|^{\frac{1}{4}(\Diamond_o-1)}\left(\sqrt{b}/r[\zeta]\right)^{1-s_o}[\zeta]\, M_{\kappa(\varepsilon),-\frac{1}{2}\Diamond_o}(\sqrt{-\varepsilon}\,\zeta]) \tag{5.2.9*}$$

are Taylor series (4.4) expressed in terms of $\zeta$, instead of r. Making use of the conventional relation between Whittaker functions:

$$W_{\kappa,\frac{1}{2}\lambda_o}(\varsigma) = \frac{\Gamma(-\lambda_o)}{\Gamma(\frac{1}{2}(1-\lambda_o)-\kappa)} M_{\kappa,\frac{1}{2}\lambda_o}(\varsigma) + \frac{\Gamma(\lambda_o)}{\Gamma(\frac{1}{2}(1+\lambda_o)-\kappa)} M_{\kappa,-\frac{1}{2}\lambda_o}(\varsigma), \tag{5.2.10}$$

where $|\arg\zeta| < \frac{3}{2}\pi$ and $\lambda_o$ is not an integer (see, i.g., §16.41 in [40]), one can easily verify that the $m$-coefficient in the superposition

$$ {}_0\psi_{\mathbf{b}}[\zeta;\varepsilon;\lozenge_o] = -\frac{1}{2\lambda_o}\, r^{1-s_o}[\varsigma]\, {}_0f_{\mathbf{d}}[\zeta;\varepsilon;\lozenge_o] + r^{s_o}[\zeta]\, {}_0m(\varepsilon;\lozenge_o, g_o; b)\, {}_0f_{\mathbf{a}}[\zeta;\varepsilon;\lozenge_o], \tag{5.2.11} $$

is given by the following expression

$$ {}_0m(\varepsilon;\lozenge_o, g_o; b) = \left(\sqrt{-\varepsilon}\,/\,b\right)^{\lozenge_o} \frac{\Gamma(1-\lozenge_o)}{\lozenge_o\Gamma(\lozenge_o+1)}\, {}_0M_{0,0}\left(\lozenge_o;\kappa(\varepsilon;g_o;b)\right), \tag{5.2.12} $$

where we set

$$ {}_0M_{0,0}(\lambda_o;\kappa) \equiv \frac{\Gamma({}_0\alpha_+(\lambda_o;\kappa))}{\Gamma({}_0\alpha_-(\lambda_o;\kappa))}, \tag{5.2.12′} $$

with

$$ {}_0\alpha_\pm(\lambda_o;\kappa) \equiv \tfrac{1}{2}(1\pm\lambda_o) - \kappa, \tag{5.2.13} $$

As expected [16-18], TWM function (5.2.12) has an infinite number of poles at the roots ${}_0\varepsilon_{\mathbf{c},v}$ of algebraic equations

$$ \kappa({}_0\varepsilon_{\mathbf{c},v}) = \tfrac{1}{2}(1+\lambda_o) + v \tag{5.2.14} $$

describing discrete energy states in potential (5.2.2).

Since parameter (5.2.7) is negative at any energy $\varepsilon<0$ if $g_0 > 0$, TMF function (5.2.12) does not have poles in this case. This implies that any SUSY partner of the potential $V[\zeta\,|\,{}_0G^{211}]$ with a repulsive Coulomb tail has no more than one eigenvalue in the LP region which appears only if the appropriate FF is irregular at both endpoints TMF function (5.2.12) also does not have zeros within the LC range so that anomalous pairs of non-isospectral SUSY partners do not exist either. For this reason we restrict our analysis only to the potentials with an attractive Coulomb tail ($g_0 < 0$).

Converting the algebraic equation

$$\kappa({}_0\varepsilon_{\mathbf{t}_+,m}) = {}^1\!/_2(1+\lambda_o) + m \tag{5.2.14}$$

into the quadratic equations

$$b\,\nu^2_{\mathbf{t}_+,m} + 2\nu_{\mathbf{t}_+,m}(\lambda_o + 2m + 1) + g_o = 0 \tag{5.2.15}$$

with respect to $\nu_{\mathbf{t}_+,m} = \sqrt{-{}_0\varepsilon_{\mathbf{t}_+,m}}$ one can verify that two roots has opposite sign: $\nu_{\mathbf{c},m} > 0$ and $\nu_{\mathbf{a},m} < 0$ for any m iff $g_o < 0$. Since discriminant

$${}_0\Delta_{+,m}(\lambda_o, g_o; b) = (\lambda_o + 2m + 1)^2 - g_o > 0 \tag{5.2.15$^\dagger$}$$

monotonically increases with m the same is true for absolute value of the negative root $\nu_{\mathbf{a},m}$ so that all AEH solutions of type $\mathbf{a}$ for the potential $V[\zeta\,|\,{}_0\mathcal{G}^{211}]$ lie below the ground energy level $\varepsilon_{\mathbf{c},0}$. As a result each such solution can be used to construct a rational SUSY partner $V[z\,|\,{}^1_0\mathcal{G}^{211}_{\mathbf{a},m}]$ quantized via the so-called [15] 'GS confluent Heine' (c-Heine) polynomials. Their detailed analysis will be presented in a separate publication.

By representing (5.2.7) for $g_0 < 0$ as

$$\kappa(\varepsilon; g_o; b) = \frac{|\,g_o\,|}{4\sqrt{-\varepsilon}} - b\sqrt{-\varepsilon} \tag{5.2.16}$$

we find that this parameter monotonically decreases with increase of $\sqrt{-\varepsilon}$ so that

$${}_0\alpha_+\left(\varepsilon; \kappa(\varepsilon; g_o; b)\right) > {}_0\alpha_+\left(\varepsilon_{\mathbf{c};0}; \kappa(\varepsilon_{\mathbf{c};0}; g_o; b)\right) = 0 \;\text{ if }\; \varepsilon < \varepsilon_{\mathbf{c};0}. \tag{5.2.17}$$

Since the Whittaker function

$$M_{\kappa, {}^1\!/_2\lambda_o}(\varsigma_\varepsilon) = \varsigma_\varepsilon^{{}^1\!/_2(\lambda_o+1)}\, e^{-{}^1\!/_2\varsigma_\varepsilon}\, F[{}_0\alpha_+(\varepsilon;\kappa), \lambda_o + 1; \varsigma_\varepsilon] \tag{5.2.18}$$

retains its sign on the positive semi-axis if both parameters ${}_0\alpha_+(\varepsilon;\kappa)$ and $\lambda_o+1$ are positive we conclude that the R@O Frobenius solution below the ground energy level must be nodeless in both LP and LC regions.

Making use of the Milne diagram [60, 52] for positive zeros of the Whittaker function $\Psi(\alpha,\lambda_o+1;\varsigma_\varepsilon)$ for $\alpha$, $\lambda_o+1>0$ one can also directly verify that R@∞ wave function (5.2.11) proportional to the Whittaker function

$$W_{\kappa,\frac{1}{2}\lambda_o}(\varsigma_\varepsilon)=\varsigma_\varepsilon^{\frac{1}{2}(\lambda_o+1)}\,e^{-\frac{1}{2}\varsigma_\varepsilon}\,\Psi(\alpha_+(\varepsilon;\kappa),\lambda_o+1;\varsigma_\varepsilon) \tag{5.2.19}$$

is nodeless at any $\varepsilon<\varepsilon_{\mathbf{c};0}$ regardless of the value of the SI.

## 6. TWF functions for SUSY partners of reflective radial GRef potentials in both LP and LC regions

### 6.1. SUSY pairs of potentials with regular singular points

The DT eliminating the ground energy state results in the radial reduction of the Cooper-Ginocchio-Khare (CGK) potential [61] exactly quantized via the so-called [15] 'GS Heun' polynomials. Substituting (5.1.46) into the right-hand side of (4.20) with $\mathbf{t}_+=\mathbf{c}$ and $\varepsilon_{\mathbf{c}}={}_1\varepsilon_{\mathbf{c};0}$ and taking into account that

$$\begin{aligned}{}_1\alpha_{\pm,+}(\varepsilon;\lambda_o,\mu(a\varepsilon;\mu_o))\,{}_1\beta_{\pm,+}(\varepsilon;\lambda_o,\mu(a\varepsilon;\mu_o))&\\ =\tfrac{1}{4}[(\pm\lambda_o+\sqrt{-\varepsilon}+1)^2-\mu^2(a\varepsilon;\mu_o)]&\end{aligned} \tag{6.1.1}$$

$$=\tfrac{1}{4}\,{}_1\Xi_{\mathbf{t}_\pm,0}(\varepsilon;\lambda_o,\mu_o;1-a)[{}_1\varepsilon_{\mathbf{t}_\pm,0}-\varepsilon], \tag{6.1.1′}$$

where

$${}_1\Xi_{\mathbf{t}_\pm,0}(\varepsilon;\lambda_o,\mu_o;b)\equiv\frac{2(1\pm\lambda_o)}{\sqrt{-\varepsilon}+\sqrt{-{}_1\varepsilon_{\mathbf{t}_\pm,0}}}+b,\ \text{with}\ \mathbf{t}_+=\mathbf{c},\ \mathbf{t}_-=\mathbf{b}, \tag{6.1.2}$$

one can represent the TWF function for the radial CGK potential as

$$\star{}_1m(\varepsilon;\star\Diamond_o,\mu_o;b\,|\,\mathbf{c},{}_1\varepsilon_{\mathbf{c},0}) = \frac{\Gamma(-\star\Diamond_o)}{2b^{\star\Diamond_o-1}\Gamma(\star\Diamond_o+1)}\frac{{}_1M_{0,1}\big(\varepsilon;\star\Diamond_o+1,\mu(a\varepsilon;\mu_o)\big)}{\Xi_{c,0}(\varepsilon;\star\Diamond_o-1,\mu_o;b)}\,, \tag{6.1.3}$$

where the energy-dependent coefficient

$${}_1M_{v,m}(\varepsilon;\lambda_o,\mu) \equiv \frac{\Gamma\big({}_1\alpha_{++}(\varepsilon;\lambda_o,\mu)+v\big)\Gamma\big({}_1\beta_{++}(\varepsilon;\lambda_o,\mu)+v\big)}{\Gamma\big({}_1\alpha_{-+}(\varepsilon;\lambda_o,\mu)+m\big)\Gamma\big({}_1\beta_{-+}(\varepsilon;\lambda_o,\mu)+m\big)} \tag{6.1.4}$$

turns into (5.1.19) for v=0, m=0. [When deriving (6.1.3) we also took into account that ${}_1M_{1,0}(\varepsilon;\lambda_o,\mu) = {}_1M_{0,1}(\varepsilon;\lambda_o+2,\mu)$.] Note that the energy-dependent coefficient ${}_1\Xi_{\mathbf{c},0}(\varepsilon;\lambda_o,\mu_o;b)$ is positive at any negative energy so that TWF function (6.1.3) has poles only if $\alpha_{+,+}\big(\varepsilon;\lambda_o,\mu(\mu_o;a\varepsilon)\big)$ is a negative integer as expected from the fact that the DT in question annihilates the eigenfunction associated with the lowest-energy discrete state in the potential $V[z\,|\,{}_1\mathcal{G}^{211}_{\downarrow\mathbf{c},0}]$.

The eigenfunctions associated with bound energy states in the radial CGK potential $V[z\,|\,{}^1_1\mathcal{G}^{211}_{\mathbf{c},0}]$,

$$\psi_{\mathbf{c},v}[z\,|\,{}^1_1\mathcal{G}^{211}_{\mathbf{c},0}] = \frac{1}{\sqrt[4]{az+b}}\; z^{\frac12\Diamond_o+\frac34}\,|1-z|^{\frac12\lambda_{1;\mathbf{c},v+1}}$$
$$\times\, Hp_{v+1}[z\,|\,{}^1_1\mathcal{G}^{211}_{\mathbf{c},0};\mathbf{c},v] \quad \text{for } 0\le 2v<\mu_o-\lambda_o-3 \tag{6.1.5}$$

are obtained by applying the DT with the FF $\psi_{\mathbf{c},0}[z\,|\,{}_1\mathcal{G}^{211}_{\downarrow\mathbf{c},0}]$ to the eigenfunctions of excited bound energy states in the potential $V[z\,|\,{}_1\mathcal{G}^{211}_{\downarrow\mathbf{c},0}]$. The GS Heun polynomials of order v+1 in the right-hand side of (6.1.5) can be obtained from the Jacobi polynomials using the following differential relations

$$\tfrac12(\lambda_{1;\mathbf{c},0}-\lambda_{1;\mathbf{c},v+1}-v-1)k_{v+1}(\Diamond_o+\lambda_{1;\mathbf{c},v+1})Hp_{v+1}[z\,|\,{}^1_1\mathcal{G}^{211}_{\mathbf{c},0};\mathbf{c},v]$$
$$= {}_1\hat{a}(\lambda_{1;\mathbf{c},v+1}-\lambda_{1;\mathbf{c},0})P^{(\Diamond_o,\lambda_{1;\mathbf{c},v+1})}_{v+1}(2z-1)\,, \tag{6.1.6}$$

where [55]

$$k_m(\nu) = \frac{(\nu+m)_m}{m!\ 2^m}. \tag{6.1.7}$$

and the differential operator

$${}_\iota\hat{a}(\Delta\lambda) = (1-\iota z)\frac{d}{dz} - \tfrac{1}{2}\Delta\lambda\,, \tag{6.1.8}$$

is the particular representative of the so-called ‘abridged Heine-polynomial generators’ ($a$-HPGs) defined via (2.62*) in [15]. (In following [15], the GS Heun polynomials are normalized via the requirement that their leading coefficients are equal to 1.)

The lowest-energy eigenfunction thus has the form:

$$\psi_{c,0}[z\,|\,{}^{1}_{1}G^{211}_{c,0}] = \frac{1}{\sqrt[4]{az+b}}\; z^{\frac{1}{2}\Diamond_o+\frac{3}{4}}\;|1-z|^{\frac{1}{2}\lambda_{1;c,1}}\;(z-{}^{*}z_{c;0,1})$$

$$(\mu_o > 3+\Diamond_o) \tag{6.1.9}$$

where

$${}^{*}z_{c;0,1} = z_{c,1} - \frac{2(1-z_{c,1})}{{}_1\lambda_{1;c,0} - {}_1\lambda_{1;c,1} - 2}\,, \tag{6.1.10}$$

with $z_{c,1}$ standing for the zero of the first-order Jacobi polynomial $P_1^{(\Diamond_o,\lambda_{1;c,1})}(2z-1)$ which is given by generic relation (3.42) in [15]:

$$0 < z_{c,1} = \frac{\Diamond_o+1}{{}_1\lambda_{1;c,1}+\Diamond_o+2} < 1. \tag{6.1.11}$$

It could be explicitly confirmed (as it is done below for the potential $V[z\,|\,{}^{1}_{1}G^{211}_{b,0}]$ with SI within the LC range) that ${}^{*}z_{c;0,1}$ is negative if 0 < b <1 and lies between 1 and +∞ if b >1 As expected, its absolute value becomes infinitely large in the limiting case of the h-PT potential (b=1).

Similarly, one can construct the isospectral rational SUSY partner $V[z\,|\,{}^{1}_{1}G^{211}_{a,0}]$ of $V[z\,|\,{}_1\mathcal{G}^{211}_{\downarrow\mathbf{a},0}]$ by using the basic R@O FF corresponding to the negative root $\lambda_{1;\mathbf{t}_+,0} = \lambda_{1;\mathbf{a},0}$ of quadratic equation (5.1.29) with m set to 0. The eigenfunctions describing bound energy states in the potential $V[z\,|\,{}^1_1\mathcal{G}^{211}_{\mathbf{a},0}]$ thus have the form

$$\psi_{\mathbf{c},v}[z\,|\,{}^1_1\mathcal{G}^{211}_{\mathbf{a},0}] = \frac{1}{\sqrt[4]{az+b}}\, z^{\frac{1}{2}\Diamond_o + \frac{3}{4}}\, |1-z|^{\frac{1}{2}\lambda_{1;\mathbf{a},v}} \times Hp_v[z\,|\,{}^1_1\mathcal{G}^{211}_{\mathbf{a},0};\mathbf{c},v] \quad \text{for } 0 \le 2v < \mu_o - \lambda_o - 3. \tag{6.1.12}$$

Note that the sequence of GS Heun polynomials in the right-hand side of (6.1.12) starts from a constant and that all nonzero-order polynomials in this sequence are related to the Heun polynomials $Hp_v[z\,|\,{}^1_1\mathcal{G}^{211}_{\mathbf{c},0};\mathbf{c},v\text{-}1]$ for the radial CGK potential in a simple fashion:

$$({}_1\lambda_{1;\mathbf{a},0} - {}_1\lambda_{1;\mathbf{c},v} - v)Hp_v[z\,|\,{}^1_1\mathcal{G}^{211}_{\mathbf{a},0};\mathbf{c},v] = -\frac{\lambda_{1;\mathbf{c},0} - \lambda_{1;\mathbf{a},0}}{k_v(\Diamond_o + \lambda_{1;\mathbf{c},v})} P_v^{(\Diamond_o,\lambda_{1;\mathbf{c},v})}(2z-1)$$

$$+ ({}_1\lambda_{1;\mathbf{c},0} - {}_1\lambda_{1;\mathbf{c},v} - v)Hp_v[z\,|\,{}^1_1\mathcal{G}^{211}_{\mathbf{c},0};\mathbf{c},v\text{-}1] \quad \text{for } 0 < 2v < \mu_o - \lambda_o - 3. \tag{6.1.13}$$

In principle one can use solution (5.1.11) at any energy $\varepsilon_{\mathbf{a}} < \varepsilon_{\mathbf{c},0}$ as the FF to generate the isospectral SUSY partner of potential (5.1.1). As already pointed out in previous Section the family of exactly solvable radial IS@O potentials constructed in such a way contains infinitely many rational potentials $V[z\,|\,{}^1_1\mathcal{G}^{211}_{\mathbf{a},m}]$ if $0 < b \le 1$. In the latter case there is also another infinite sequence of rational potentials $V[z\,|\,{}^1_1\mathcal{G}^{211}_{\mathbf{a}',m}]$ associated with the second negative root. One can easily prove this assertion by re-writing quadratic equation (5.1.29) as

$$(|\lambda_{1;\mathbf{t}_+,m}| - \lambda_{1;\mathbf{c},0})[b(|\lambda_{1;\mathbf{t}_+,m}| + \lambda_{1;\mathbf{c},0}) + 2(\lambda_o + 1)] = 4(\lambda_o + m + 1)(|\lambda_{1;\mathbf{t}_+,m}| - m) \quad \text{for } \mathbf{t}_+ = \mathbf{a} \text{ or } \mathbf{a}'. \tag{6.1.14}$$

Any nodeless R@O AEH solutions must thus satisfy the condition

$$m < |\lambda_{1;\mathbf{t}_+,m}| \quad \text{for } \mathbf{t}_+ = \mathbf{a} \text{ or } \mathbf{a}'. \tag{6.1.15}$$

An analysis of (5.1.30) shows that the condition $m < |\lambda_{1;\mathbf{a},m}|$ is trivially fulfilled at least for b <2 whereas a similar analysis of the second root leads to the constraint

$$(\lambda_o + 2m + 1)(1 + \lambda_o) + bm^2 = (\lambda_o + m + 1)^2 + (b - 1)m^2 > \mu_o^2 \tag{6.1.16}$$

so that

$$mb > \sqrt{(1-b)\lambda_o + 1)^2 + b\mu_o^2} - \lambda_o - 1 \quad \text{for } b > 0. \tag{6.1.16′}$$

Two particular cases of constraint (6.1.16): b=0 and b=1 associated with the shape-invariant E/MR and h-PT potentials will be discussed in Section 8. It is worth mentioning in this connection that the derived constraint is formally applicable only to the range of positive b since we had to divide an intermediate inequality by b during the derivation. However, as explicitly demonstrated below it does hold in the limiting case of the E/MR potential.

Each sequence of nodeless R@O AEH solutions turns finite for b larger than 1 and the second sequence $\mathbf{a}', m$ retains for

$$\mu_o < 2m + 1 + \lambda_o < \sqrt{\frac{b}{b-1}}\,\mu_o \quad \text{for } b > 1. \tag{6.1.16*}$$

Condition (6.1.15) is the direct consequence of the following general ‘indexes-of-opposite-sign’ (IOS) rule:

*Any Jacobi polynomial with indexes of opposite sign does not to have zeros between* –1 *and* +1 *iff its order is smaller than absolute value of the negative index.*

In the recent paper [34] Quesne splits this rule into the two depending which index happened to be negative. However, since the indexes can be always interchanged by the reflection of the polynomial argument so that we really deal with a single rule.)

The cited rule directly follows from conditions (C1) and (C4) in Appendix C in [15] -- a convenient compilation of Klein's formulas [62, 63] to select Jacobi polynomials with no zeros between -1 and 1.

Both sequences of the rational potentials $V[z\,|\,{}^{1}_{1}G^{211}_{a,m}]$ and $V[z\,|\,{}^{1}_{1}G^{211}_{a',m}]$ are quantized via the GS Heine polynomials

$$\mathrm{Hi}_{m+v}[z\,|\,{}^{1}_{1}G^{211}_{t_+,m};c,v] = \frac{1}{m - v - \tfrac{1}{2}(\lambda_{1;c,v} - \lambda_{1;t_+,m})} \qquad (6.1.17)$$

$$\times\, {}_1\hat{a}(\lambda_{1;c,v} - \lambda_{1;t_+,m};\bar{z}_{t_+,m})\,\Pi[z;\bar{z}_{c,v}] \quad \text{for } t_+ = a \text{ or } a',$$

where

$$\Pi[z;\bar{z}] \equiv \prod_j (z - z_j) \qquad (6.1.18)$$

and the $a$-HPG is defined via (2.62*) in [15], with $\bar{z}_{t_+,m}$ and $\bar{z}_{c,v}$ standing for complete sets of zeros of the appropriate Jacobi polynomials. (Note that there is a misprint in the order of polynomial (2.64) in [17] – it is equal to m+v regardless of the value of ι.)

It directly follows from (5.1.30) that the primary sequence disappears in the limiting case of the E/MR potential (b = 0). It will be explicitly demonstrated in Section 8 that the secondary sequence contains a finite number of nodeless R@O AEH solutions which have been used by Quesne [34] to construct quantized-by-polynomials SUSY partners of the E/MR potential.

This procedure can be then repeated using the R@O solution

$$^{*}\psi_{a}[z;\varepsilon'_{a}\,|\,a;\varepsilon_{a}) = \frac{z'\,W\{\psi_{a}[z;\varepsilon_{a}],\,\psi_{a}[z;\varepsilon'_{a}]\}}{w(a,\varepsilon'_{a}\,|\,a;\varepsilon_{a})\,\psi_{a}[z;\varepsilon_{a}]}$$

at an energy $\varepsilon'_{a} < \varepsilon_{a}$ as the appropriate FF. Each step increases the singularity strength by 1 and adds a new zero to the TWF function. Again, one can select subsets of rational potentials quantized via multi-index GS Heine polynomials which will be discussed in a separate publication.

As mention in Section 5, R@∞ solution (5.1.44) turns into a Frobenius solution near the origin at any zero of TMF function (5.1.46). One can easily verify that poles of the first gamma-function in the numerator of fraction (5.1.19) coincide with positive roots of the quadratic equations

$$b\lambda^2_{1;\mathbf{t}_-,m} - 2\lambda_{1;\mathbf{t}_-,m}(\lambda_o - 2m - 1) + (2m+1-\lambda_o)^2 - \mu_o^2 = 0 \qquad (6.1.19)$$

whereas the argument of the second gamma-function in the numerator remains finite within the allowed range of $\lambda_o$. As far as the discriminant

$${}_1\Delta_{-;m}(\lambda_o,\mu_o;b) = 4\left[(1-b)(2m+1-\lambda_o)^2 + b\mu_o^2\right] \qquad (6.1.19^\dagger)$$

remains positive the quadratic equation has

i) one positive root iff

$$|2m+1-\lambda_o| < \mu_o \qquad (6.1.20)$$

and

ii) two positive roots iff

$$2m+1 < \lambda_o - \mu_o, \; {}_1\Delta_{-;m}(\lambda_o,\mu_o;b) > 0\,. \qquad (6.1.20^*)$$

To select nodeless R@∞ Frobenius solutions one can again makes use of the IOS rule. We thus conclude that the R@∞ Frobenius solution is nodeless iff the polynomial order m is smaller than the exponent difference $\lambda_o$. In particular this implies that there is no nodeless R@∞ Frobenius solutions other than the basic one if $\lambda_o$ lies within the LC range.

In general one can use any solution of type **b** below the ground energy level $\varepsilon_{\mathbf{c},0}$ to construct a new exactly-solvable radial potential. In this paper we however focus on AEH solutions of type **b** since they represent a unique class of irregular Frobenius solutions near the origin which vanish at the upper end of the quantization interval.

One can independently confirm that the constraint m < $\lambda_o$ is indeed equivalent to the condition for the R@∞ Frobenius solution **b**,m to be nodeless iff $\varepsilon_{\mathbf{b},m} < \varepsilon_{\mathbf{c},0}$. In fact, setting m to 0 in (5.1.29) and subtracting the resultant equation from (6.1.19) shows that

$$\tfrac{1}{4}(\lambda_{1;\mathbf{b},m} - \lambda_{1;\mathbf{c},0})[b(\lambda_{1;\mathbf{b},m} + \lambda_{1;\mathbf{c},0}) + 2(\lambda_o + 1)] = (\lambda_o - m)(\lambda_{1;\mathbf{b},m} + m + 1) \qquad (6.1.21)$$

so that the AEH solution $\mathbf{b},m$ lies below the 'ground' energy $\varepsilon_{\mathbf{c},0}$ iff $m < \lambda_o$. As anticipated, the basic solution $\mathbf{b},0$ always lies below the ground energy level.

We define the primary and secondary sequences of nodeless R@∞ Frobenius solutions within the LP range of $\lambda_o$ as follows

$$b\lambda_{1;\mathbf{b},m} = \lambda_o - 2m - 1 + \tfrac{1}{2}\sqrt{{}_1\Delta_{-;m}(\lambda_o, \mu_o; b)} \tag{6.1.22}$$

$$\text{for } 0 \le m_- \le m < m_+$$

and

$$b\lambda_{1;\mathbf{b}',m} = \lambda_o - 2m - 1 - \tfrac{1}{2}\sqrt{{}_1\Delta_{-;m}(\lambda_o, \mu_o; b)} \tag{6.1.22′}$$

$$\text{for } 0 < m'_- \le m < \tfrac{1}{2}(\lambda_o - \mu_o - 1) < \tfrac{1}{2}(\lambda_o - 1) < \lambda_o,$$

where

$$m_- = \begin{cases} 0 & \text{if } \mu_o > \lambda_o - 1 > 0 \text{ or } 0 < b \le 1, \\ \tfrac{1}{2}(\mu_o / \sqrt{1 - 1/b} - \lambda_o + 1) & \text{otherwise,} \end{cases} \tag{6.1.23}$$

$$m_+ = \begin{cases} \min\{\tfrac{1}{2}(\mu_o + \lambda_o - 1), \lambda_o\} & \text{if } \mu_o > \lambda_o - 1 > 0, \\ \tfrac{1}{2}(\lambda_o - 1) < \lambda_o & \text{otherwise,} \end{cases} \tag{6.1.23$^+$}$$

$$m'_- = \begin{cases} \tfrac{1}{2}(\lambda_o - \mu_o - 1) < \lambda_o & \text{if } \lambda_o > \mu_o + 1 \text{ and } 0 < b \le 1, \\ \tfrac{1}{2}\max\{\mu_o / \sqrt{1 - 1/b} - \lambda_o + 1, \lambda_o - \mu_o - 1\} & \text{otherwise.} \end{cases} \tag{6.1.23′}$$

[The limiting case of the E\MR potential (b=0) requires a separate analysis which will be presented in Section 8 below.] As expected, the upper bound for m is always smaller than $\lambda_o$. Note also that the primary sequence always starts from the basic solution if the potential has the discrete energy spectrum.

Let us reiterate that the secondary sequence exists only in the LP region and only in its part where the potential has no discrete energy states. This implies that the appropriate pair of the rational SUSY partners does not have the discrete energy spectrum either. We have already made a similar observation for the rational SUSY partners of the h-PT potential constructed using DTs with R@∞ FFs labeled as $\tilde{b}'$,m in [15].

By extending linearization (6.1.1′) to an arbitrary AEH solution $t_{\pm,k}$:

$$[{}_1\alpha_{\pm,+}(\varepsilon;\lambda_o,\mu(\mu_o;a\varepsilon))+k][{}_1\beta_{\pm,+}(\varepsilon;\lambda_o,\mu(\mu_o;a\varepsilon))+k]$$
$$= \tfrac{1}{4}\,{}_1\Xi_{t_\pm,k}(\varepsilon;\lambda_o,\mu_o;a)(\varepsilon_{t_\pm,k}-\varepsilon)\,, \qquad (6.1.24)$$

where

$${}_1\Xi_{t_\pm,k}(\varepsilon;\lambda_o,\mu_o;b)\equiv\frac{2(\pm\lambda_o+k)+1}{\sqrt{-\varepsilon}+\lambda_{1,t_\pm,k}}+b \qquad (6.1.25)$$

one can represent the TWF function for the rational SUSY partner $V[z\,|\,{}^1_1G^{211}_{b,m}]$ as

$$*_1m_G(\varepsilon;*\lozenge_o\,|\,b,\varepsilon_{b,m})=-\frac{2^{2m-1}\Gamma(-*\lozenge_o)}{b^{*\lozenge_o+1}\Gamma(*\lozenge_o+1)}\prod_{k=0}^{m-1}[\varepsilon-{}_1\varepsilon_{b,k}(*\lozenge_o-1,\mu_o;b)]$$
$$\times{}_1\Upsilon_{b,m}(\varepsilon;*\lozenge_o-1,\mu_o;b)\,{}_1M_{0,m+1}(\varepsilon;*\lozenge_o-1,\mu(a\varepsilon;\mu_o))$$
$$\text{for } \lozenge_o>1, \qquad (6.1.26)$$

where

$${}_1\Upsilon_{b,m}(\varepsilon;\lambda_o,\mu_o;b)\equiv\prod_{k=0}^{m}\,{}_1\Xi_{b,k}(\varepsilon;\lambda_o,\mu_o;b)\,. \qquad (6.1.27)$$

Note that the DT in question eliminates one zero of the TMF function (5.1.46) while keeping positions of all poles unchanged. Note that zeros of function (6.1.27), if exist, determine the discrete energy spectrum for the rational radial potential obtained from $V[z\,|\,{}_1G^{211}]$ by sequential use of each of $[\lozenge_o]+1$ nodeless wavefunctions $\psi_b[z;\varepsilon_{b;m}]$ (m = $[\lozenge_o]$, $[\lozenge_o]-1,\ldots,0$) as Gauss SSs for the appropriate multi-step DT.

The eigenfunctions of the Schrödinger equation with the potential ${}^1_1V[z\,|\,{}^1_1G^{211}_{b;m}]$ in the LP region are expressible in terms of the following GS Heine polynomials

$$Hi_{m+v+1}[z\,|\,{}^1_1G^{211}_{b,m};c,v]=\frac{1}{m-v-\tfrac{1}{2}(\lambda_{1;c,v}-\lambda_{1;b,m})}$$
$$\times{}_1\hat{g}_{b,m}(2\lambda_o,{}_1\lambda_{c,v}-{}_1\lambda_{b,m})\Pi[z;\bar{z}_{c,v}] \quad \text{for } \lambda_o>1, \qquad (6.1.28)$$

where the generic HPG is defined via (2.57) in [15].

In the LC region the TMF function for the potential $V[z\,|\,{}^{1}_{1}G^{211}_{\mathbf{b},0}]$ takes a different form

$$
{}^{*}{}_{1}m_G(\varepsilon;{}^{*}\Diamond_o\,|\,\mathbf{b},\varepsilon_{\mathbf{b},0}) = b^{{}^{*}\Diamond_o+1}\frac{\Gamma(-{}^{*}\Diamond_o)}{2\Gamma(1-{}^{*}\Diamond_o)} \times \frac{1}{{}_{1}\Xi_{\mathbf{b},0}(\varepsilon;{}^{*}\Diamond_o+1,\mu_o;b)\,{}_{1}M_{0,1}\left(\varepsilon;{}^{*}\Diamond_o+1,\mu(a\varepsilon;\mu_o)\right)} \quad \text{for } 0<{}^{*}\Diamond_o=1-\Diamond_o<1. \tag{6.1.29}
$$

It directly follows from the derived expression that the potential $V[z\,|\,{}^{1}_{1}G^{211}_{\mathbf{b},0}]$ in the LC region has discrete energy states at the energies ${}_{1}\varepsilon_{\mathbf{b},m}$ $(1<2m+1<\lambda_o+\mu_o)$ coincident not with the poles of TWF function (5.1.17), but with its zeros, contrary to the general case. [T he function ${}_{1}\Xi_{\mathbf{b},0}(\varepsilon;\Diamond_o,\mu_o;b)$ is positive by definition if $\Diamond_o<1$.] Note also that, according to (4.26′), the potential $V[z\,|\,{}^{1}_{1}G^{211}_{\mathbf{b},0}]$ does not have the discrete energy state at the energy ${}_{1}\varepsilon_{\mathbf{b},0}$ so that the lowest discrete-energy state in the latter potential has the energy ${}_{1}\varepsilon_{\mathbf{b},1}$. The appropriate eigenfunctions

$$
\psi_{\mathbf{c},v}[z\,|\,{}^{1}_{1}G^{211}_{\mathbf{b},0}] = \frac{1}{\sqrt[4]{az+b}}\, z^{3/4-1/2\Diamond_o}\,|1-z|^{1/2+1/2\lambda_{1;\mathbf{b},v}} \times Hp_{v+1}[z\,|\,{}^{1}_{1}G^{211}_{\mathbf{b},0};\mathbf{c},v] \quad \text{for } \mu_o>2v+1-\lambda_o \tag{6.1.30}
$$

are obtained by applying the DT with the FF $\mathbf{b}$,0 to the Frobenius solutions

$$
\psi_{\mathbf{b},v+1}[z\,|\,{}_{1}G^{211}_{\downarrow\mathbf{b},0}] = \sqrt[4]{az+b}\;{}_{1}C_{\mathbf{b},v+1}\, z^{1/4-1/2\Diamond_o}\,|1-z|^{1/2+1/2\lambda_{1;\mathbf{b},v+1}} \times P_{v+1}^{(-\Diamond_o,\lambda_{1;\mathbf{b},v+1})}(2z-1). \tag{6.1.31}
$$

The GS Heun polynomials of order v+1 in the right-hand side of (6.1.30) can be expressed in terms of the Jacobi polynomials by means of the differential relations

$$Hp_{v+1}[z\,|\,{}^{1}_{1}G^{211}_{\mathbf{b},0};\mathbf{c},v] = -\frac{1}{v+1+\frac{1}{2}(\lambda_{1;\mathbf{b},v+1}-\lambda_{1;\mathbf{b},0})} \times {}_1\hat{a}(\lambda_{1;\mathbf{b},v+1}-\lambda_{1;\mathbf{b},0})\Pi[z;\bar{z}_{\mathbf{b},v+1}] \tag{6.1.32}$$

similar to (6.1.6) above. The eigenfunction associated with the ground-energy state

$$\psi_{\mathbf{c},0}[z\,|\,{}^{1}_{1}G^{211}_{\mathbf{b},0}] = \frac{1}{\sqrt[4]{az+b}}\, z^{3/4-1/2\Diamond_o}\; z^{-1/2\Diamond_o}\, |1-z|^{1/2\lambda_{1;\mathbf{b},1}}\, (z-{}^{*}z_{\mathbf{b},0;1})$$

$$\text{for } \mu_o > 3-\Diamond_o \tag{6.1.33}$$

is obtained by applying the given DT to the Frobenius solution

$$\psi_{\mathbf{b},1}[z\,|\,{}_1G^{211}_{\downarrow\mathbf{b},0}] = {}_1C_{\mathbf{b},1}\sqrt[4]{az+b}\; z^{1/4-1/2\Diamond_o}\, |1-z|^{1/2\lambda_{1;b,1}}\, (z-z_{\mathbf{b},1}), \tag{6.1.34}$$

where the zero $z_{\mathbf{b},1}$ of the first-order polynomial in the right-hand side of (6.1.34) is given by (3.42) in [15]:

$$0 < z_{\mathbf{b},1} = \frac{1-\Diamond_o}{{}_1\lambda_{1;\mathbf{b},1}-\Diamond_o+2} < 1 \quad \text{for } \Diamond_o < 1. \tag{6.1.35}$$

One can verify that

$${}^{*}z_{\mathbf{b},0;1} = z_{\mathbf{b},1} - \frac{2(1-z_{\mathbf{b},1})}{{}_1\lambda_{1;\mathbf{b},0}-{}_1\lambda_{1;\mathbf{b},1}-2}, \tag{6.1.36}$$

by analogy with (6.1.10). To confirm that the zero ${}^{*}z_{\mathbf{b};0,1}$ lies outside the quantization interval (0, 1), let us study more carefully the denominator of the fraction in the right-hand side of (6.1.36) as a function of b at fixed values of $\Diamond_o$ and $\lambda_{1;\mathbf{b},1}$ or, in other words, at a fixed value of $z_{\mathbf{b},1}$. The new parameterization does not result in any complications since the ChExp

$$\lambda_{1;\mathbf{b},1} = [\lambda_o - 3 + \tfrac{1}{2}\,{}_1\Delta_{-;1}(\lambda_o,\mu_o;b)]/b \tag{6.1.37}$$

is a monotonically increasing function of $\mu_o$ at fixed values of $\lambda_o$ and b. An analysis of the difference between quadratic equations (6.1.19), with m set to 0 and 1, shows that

$$\lambda_{1;b,0} - \lambda_{1;b,1} = \frac{4(\lambda_{1;b,1} + 2 - \Diamond_o)}{b(\lambda_{1;b,1} + \lambda_{1;b,0}) + 2(1 - \Diamond_o)} > 0 \qquad (6.1.38)$$

is a positive root

$$\lambda_{1;b,0} - \lambda_{1;b,1} = \frac{4(\lambda_{1;b,1} + 2 - \Diamond_o)}{b\lambda_{1;b,1} + 1 - \Diamond_o + \sqrt{(b\lambda_{1;b,1} + 1 - \Diamond_o)^2 + 4(\lambda_{1;b,1} + 2 - \Diamond_o)b}} \qquad (6.1.39)$$

of the quadratic equation

$$b(\lambda_{1;b,0} - \lambda_{1;b,1})^2 + 2(\lambda_{1;b,0} - \lambda_{1;b,1})(b\lambda_{1;b,1} + 1 - \Diamond_o) - 4(\lambda_{1;b,1} + 2 - \Diamond_o) = 0 \qquad (6.1.40)$$

which monotonically decreases with increase of b at fixed values of $\Diamond_o$ and $\lambda_{1;b,1}$. Therefore this must be also true for $^*z_{b;0,1}$. In the limit b→1 the right-hand of (6.1.39) tends to 2 and a result $^*z_{b;0,1} \to \infty$. This is the direct consequence of the fact that the h-PT potential is form-invariant under basic DTs so that function (6.1.34) must turn into the basic solution at b=1. Similarly, in the limiting case b=0 associated with the shape-invariant E/MR potential

$$\lambda_{1;b,0} - \lambda_{1;b,1} = 2 / z_{b,1} > 2 \qquad (6.1.41)$$

and $^*z_{b,0;1} = 0$ as expected. With increase of b, the positive denominator of the fraction in the right-hand side of (6.1.35) monotonically decreases until it approaches 0 as b→1 in the limiting case of the h-PT potential: We thus conclude that $^*z_{b,0;1}$ monotonically decreases from 0 toward –∞ as b varies from 0 to 1. For b > 1 the denominator becomes negative and as a result $^*z_{b;0,1}$ jumps onto the fragment z > 1 of the positive semi-axis and changes from +∞ to 1 as b continues to grow.

Let us now show that the Schrödinger equation with the *r*-GRef potential in the LC region also has a continuous manifold of nodeless Frobenius solutions of type **d** mentioned in Section 4 and therefore there is a whole family of exactly solvable isospectral potentials with eigenvalues ${}_1\varepsilon_{\mathbf{b},m}$ $(0 \le 2m < \lambda_o + \mu_o - 1)$. Among rational potentials this anomalous family of SUSY partners is represented by the potential $V[z\,|\,{}^1_1G^{211}_{\mathbf{d},0}]$.

To find nodeless Frobenius solutions of type **d**, we need (in following the argumentation presented in the end of Section 4) to select the range of the energies below the lowest eigenvalue $\varepsilon_{\mathbf{c},0}$ where the TWF function is positive. An analysis of (5.1.46) for $0 < \Diamond_o < 1$ reveals that this requirement is satisfied for $\varepsilon < {}_1\varepsilon_{\mathbf{b},0}$ since in the latter case each of the gamma-functions appearing in fraction (5.1.19) has positive arguments. One can directly verify this conclusion by representing Frobenius solution (5.1.11*) in an alternative form:

$$ {}_1\phi_{\mathbf{d}}[z;\varepsilon;\lambda_o,\mu] = z^{\frac{1}{2}(1-\lambda_o)}(1-z)^{\frac{1}{2}(1-\sqrt{-\varepsilon})}\,F[\alpha(\varepsilon),\beta(\varepsilon);1-\lambda_o;z]. \qquad (6.1.38)$$

Taking into account that $\alpha({}_1\varepsilon_{\mathbf{b},0}) = 0$ and therefore $\alpha(\varepsilon) = \sqrt{-\varepsilon} - \sqrt{-{}_1\varepsilon_{\mathbf{b},0}}$, we conclude that all three parameters of the hypergeometric function in the right-hand side of (6.1.38) and therefore all the terms in the appropriate Taylor series are positive if $\varepsilon < {}_1\varepsilon_{\mathbf{b},0}$ and $0 < \Diamond_o < 1$ which completes the proof.

In particular the basic solution of type **d** lies at the energy $\varepsilon = {}_1\varepsilon_{\mathbf{d},0} < {}_1\varepsilon_{\mathbf{b},0}$:

$$ |\lambda_{1;\mathbf{d},0}| = [1-\lambda_o + \tfrac{1}{2}\,{}_1\Delta_{-;0}(\lambda_o,\mu_o;b)]/b > \lambda_{1;\mathbf{b},0}. \qquad (6.1.39)$$

The rational representative $V[z\,|\,{}^1_1G^{211}_{\mathbf{d},0}]$ of the mentioned family of isospectral SUSY partners deserves a special attention since it is exactly quantized via a sequence of GS Heun polynomials

$$\psi_{\mathsf{c},v}[z\,|\,{}^{1}_{1}\mathcal{G}^{211}_{\mathsf{d},0}] = \frac{1}{\sqrt[4]{az+b}}\, z^{3/4-1/2\Diamond_o}\; |1-z|^{1/2\lambda_{1;\mathsf{b},v}} \tag{6.1.40}$$

$$\times\, \mathrm{Hp}_v[z\,|\,{}^{1}_{1}\mathcal{G}^{211}_{\mathsf{d},0};\mathsf{c},v] \quad \text{for } 0 \le 2v < \mu_o + \lambda_o - 3\,.$$

As shown in next Section it can be obtained from $V[z\,|\,{}^{1}_{1}\mathcal{G}^{211}_{\mathsf{b},0}]$ by means of the double-step DT which creates a new lowest-energy eigenstate at the energy ${}_1\varepsilon_{\mathsf{b},0}$ while keeping other eigenvalues unchanged.

Note that the sequence of the GS Heine polynomials in the right-hand side of (6.1.40) starts from a constant and that nonzero-order polynomials are related to the GS Heun polynomials $\mathrm{Hp}_v[z\,|\,{}^{1}_{1}\mathcal{G}^{211}_{\mathsf{b},0};\mathsf{c},v]$ in a simple fashion:

$$[\tfrac{1}{2}({}_1\lambda_{1;\mathsf{b},v} - {}_1\lambda_{1;\mathsf{d},0}) + v]\, \mathrm{Hp}_v[z\,|\,{}^{1}_{1}\mathcal{G}^{211}_{\mathsf{d},0};\mathsf{c},v] = (\lambda_{1;\mathsf{b},0} - \lambda_{1;\mathsf{d},0})\Pi[z;\bar{z}_{\mathsf{b},v}]$$

$$+[\tfrac{1}{2}(\lambda_{1;\mathsf{b},v} - \lambda_{1;\mathsf{b},0}) + v]\mathrm{Hp}_v[z\,|\,{}^{1}_{1}\mathcal{G}^{211}_{\mathsf{d},0};\mathsf{c},v\text{-}1]$$

$$\text{for } v > 0, \tag{6.1.41}$$

An analysis of discriminant (6.1.16$^{\dagger}$) shows that it monotonically increases with m for b<1 and therefore the mentioned family of isospectral SUSY partners contains infinitely many rational potential $V[z\,|\,{}^{1}_{1}\mathcal{G}^{211}_{\mathsf{d},m}]$ quantized via GS Heine polynomials.

**6.2 SUSY pairs of potentials with an attractive Coulomb tail at infinity**

Let us start from the *c*-counterpart $V[\zeta\,|\,{}^{1}_{0}\mathcal{G}^{211}_{\mathsf{c},0}]$ of the radial CGK potential $V[\zeta\,|\,{}^{1}_{0}\mathcal{G}^{211}_{\mathsf{c},0}]$ which is constructed from $V[\zeta\,|\,{}_{0}\mathcal{G}^{211}_{\downarrow\mathsf{c},0}]$ by means of the DT using the lowest-energy eigenfunction

$$\psi_{\mathsf{c},0}[\zeta|{}_0\mathcal{G}^{211}_{\downarrow\mathsf{c},0}] = {}_0\wp_G^{1/4}[\zeta]\; M_{\kappa(\varepsilon),\,1/2\Diamond_o}[\nu_{\mathsf{c},1}\,\zeta] \tag{6.2.1}$$

as its FF. Substituting (5.2.12) into the right-hand side of (4.20) with $\mathsf{t}_+ = \mathsf{c}$ and $\varepsilon_{\mathsf{c}} = {}_0\varepsilon_{\mathsf{c};0}$ and taking into account that

$$\tfrac{1}{2}(1+\lambda_o)-\kappa(\varepsilon;\lambda_o,g_o;b)=\tfrac{1}{4}\,{}_0\Xi_{\mathbf{c},0}(\varepsilon;\lambda_o,g_o;b)({}_0\varepsilon_{\mathbf{c},0}-\varepsilon)\,, \qquad (6.2.2)$$

where

$${}_0\Xi_{\dagger,0}(\varepsilon;;\lambda_o,g_o;b)=-\frac{b+g_o/\sqrt{\varepsilon\,{}_0\varepsilon_{\dagger,0}}}{\sqrt{-\varepsilon}+\nu_{\dagger,0}}\,, \qquad (6.2.3)$$

one can represent the TWF function for the potential $V[\zeta\,|\,{}^{1}_{0}G^{211}_{\mathbf{c},0}]$ as

$${}^{*}{}_0m_G(\varepsilon;{}^*\Diamond_o,g_o;b\,|\,\mathbf{c},0)=\left(\sqrt{-\varepsilon}/b\right)^{{}^*\Diamond_o-1}\frac{\Gamma(-{}^*\Diamond_o)}{\Gamma({}^*\Diamond_o+1)}\times\frac{{}_0M_{0,1}\left({}^*\Diamond_o+1;\kappa(\varepsilon;g_o;b)\right)}{{}_0\Xi_{\mathbf{c},0}(\varepsilon;{}^*\Diamond_o-1,g_o;b)}\,. \qquad (6.2.4)$$

Here

$${}_0M_{v,m}(\lambda_o;\kappa)\equiv\frac{\Gamma\left({}_0\alpha_+(\lambda_o;\kappa)+v\right)}{\Gamma\left({}_0\alpha_-(\lambda_o;\kappa)+m\right)} \qquad (6.2.5)$$

and the energy-dependent parameters ${}_0\alpha_\pm(\varepsilon;\kappa)$ of the gamma-functions are defined via (5.2.13). [Again, we took into account that ${}_0M_{1,0}(\varepsilon;\lambda_o,g_o)={}_0M_{0,1}(\varepsilon;\lambda_o+2,g_o)$.] Substituting (5.2.14) into (5.2.7) with $\varepsilon={}_0\varepsilon_{\mathbf{c},v}$ one can verify that $|g_o|>b\,|\,{}_0\varepsilon_{\mathbf{c},0}\,|$ so that the energy-dependent coefficient $\Xi_{\mathbf{c};0}(\varepsilon;\mu_o;b)$ is negative at any energy $\varepsilon<{}_0\varepsilon_{\mathbf{c},v}$. Therefore TWF function (6.2.4) has a pole at an energy $\varepsilon$ iff ${}_0\alpha_+(\varepsilon;\kappa)$ is a negative integer as expected from the fact that the DT in question annihilates the eigenfunction associated with the lowest-energy discrete state in the potential $V[z\,|\,{}_0G^{211}_{\downarrow\mathbf{c},0}]$.

The eigenfunctions associated with bound energy states in the potential $V[\zeta\,|\,{}^{1}_{0}G^{211}_{\mathbf{c},0}]$,

$$\psi_{\mathbf{c},v}[\zeta\,|\,{}^{1}_{0}G^{211}_{\mathbf{c},0}]=\frac{1}{\sqrt[4]{a\zeta+b}}\,\zeta^{\frac{1}{2}\Diamond_o+\frac{3}{4}}e^{-\frac{1}{2}\nu_{\mathbf{c},v+1}\zeta}\,Hc_{v+1}[\zeta\,|\,{}^{1}_{0}G^{211}_{\mathbf{c},0};\mathbf{c},v]\,, \qquad (6.2.6)$$

are obtained by applying the DT with the FF $\psi_{\mathbf{c},0}[z\,|\,{}_1G^{211}_{\downarrow\mathbf{c},0}]$ to the eigenfunctions of excited bound energy states in the potential $V[z\,|\,{}_0G^{211}_{\downarrow\mathbf{c},0}]$. The GS confluent Heun (*c*-Heun) polynomials of order v+1 in the right-hand side of (6.2.6) are constructed from the generalized Laguerre (GL) polynomials and their derivatives via the following differential relations

$$(\nu_{c,0} - \nu_{c,v+1} - v - 1)\,\mathrm{Hc}_{v+1}[\zeta\,|\,{}^{1}_{0}G^{211}_{c,0};c,v]$$

$$= \frac{2(v+1)!}{(-\nu_{c,v+1})^{v+1}}\,{}_0\hat{a}(\nu_{c,v+1} - \nu_{c,0})\,L^{(\lozenge_o)}_{v+1}(\nu_{c,v+1}\zeta)\,, \qquad (6.2.7)$$

where we took into account (4.3.23) in [15] to compensate for the leading coefficient of the GL polynomial. (In following [15], GS c-Heun polynomials are normalized via the requirement that their leading coefficients are equal to 1.)

The lowest-energy eigenfunction can be thus represented as

$$\psi_{c,0}[\zeta\,|\,{}^{1}_{0}G^{211}_{c,0}] = \frac{1}{\sqrt[4]{a\zeta + b}}\,\zeta^{\frac{1}{2}\lozenge_o + \frac{3}{4}} e^{-\frac{1}{2}\nu_{c,1}\zeta}(\zeta - {}^{*}\zeta_{c;0,1})\,, \qquad (6.2.8)$$

where

$${}^{*}\zeta_{t,0;1} \equiv \zeta_{t,1} + \frac{2}{\nu_{t,1} - \nu_{t,0}}\,. \qquad (6.2.9)$$

with $t = c$. The zero of the first-order GL polynomial,

$$\zeta_{c,1} = (\lozenge_o + 1)/\nu_{c,1} > 0\,, \qquad (6.2.10)$$

is given by generic relation (3.42) in [15] for ι=0. An analysis of the difference between quadratic equations (5.2.15) with m set to 0 and 1 shows that

$$\frac{2}{\nu_{c,0} - \nu_{c,1}} = b(\nu_{c,0}/\nu_{c,1} + 1) + 2\zeta_{c,1} \qquad (6.2.11)$$

so that zero (6.2.9) lies at the negative semi-axis, as expected.

Similarly, one can construct the isospectral rational SUSY partner $V[\zeta\,|\,{}^{1}_{0}G^{211}_{a,0}]$ of $V[\zeta\,|\,{}_{1}G^{211}_{\downarrow a,0}]$ by using the basic R@O FF associated with the negative root $\nu_{t_+,0} = \nu_{a,0}$ of quadratic equation (5.2.15) with m set to 0. The eigenfunctions describing bound energy states in the potential $V[\zeta\,|\,{}^{1}_{0}G^{211}_{a,0}]$ thus have the form

$$\psi_{c,v}[\zeta\,|\,{}^{1}_{0}G^{211}_{a,0}] = \frac{1}{\sqrt[4]{a\zeta + b}}\,\zeta^{\frac{1}{2}\lozenge_o + \frac{3}{4}}\,e^{\frac{1}{2}\nu_{c,v}\zeta}\,\mathrm{Hc}_v[\zeta\,|\,{}^{1}_{1}G^{211}_{a,0};c,v] \qquad (6.2.12)$$

By analogy with the regular case discussed in subsection 6.1, the sequence of GS *c*-Heun polynomials in the right-hand side of (6.2.12) starts from a constant and that polynomials describing excited bound energy states are related to those for the *c*-counterpart $V[\zeta \mid {}^{1}_{0}\mathcal{G}^{211}_{\mathbf{c},0}]$ of the radial CGK potential $V[\zeta \mid {}^{1}_{1}\mathcal{G}^{211}_{\mathbf{c},0}]$ in a simple fashion:

$$(\nu_{\mathbf{a},0} - \nu_{\mathbf{c},v} - v)\mathrm{Hc}_v[\zeta \mid {}^{1}_{0}\mathcal{G}^{211}_{\mathbf{a},0};\mathbf{c},v] = -\frac{v!}{(-\nu_{\mathbf{c},v})^{v}}\frac{\nu_{\mathbf{c},0} - \nu_{\mathbf{a},0}}{\nu_{\mathbf{a},0} - \nu_{\mathbf{c},v} - v}L_v^{(\lozenge_o)}(\nu_{\mathbf{c},v}\zeta)$$
$$+(\nu_{\mathbf{c},0} - \nu_{\mathbf{c},v} - v)\mathrm{Hc}_v[\zeta \mid {}^{1}_{0}\mathcal{G}^{211}_{\mathbf{c},0};\mathbf{c},v\text{-}1] \qquad (6.2.13)$$

for v > 0.

Again, one can use any solution (5.2.8) below $\varepsilon_{\mathbf{c};0}$ as the FF for the DT to construct a new radial potential allowing one to write the closed-form expression for the appropriate TWF function. In particular this family of exactly solvable IS@O radial potentials contains infinitely many rational potentials $V[z \mid {}^{1}_{0}\mathcal{G}^{211}_{\mathbf{a},m}]$ mentioned in previous Section.

By analogy with the analysis presented for the regular case, one can also write the closed-form expression for the TWF function of the *rational* radial potentials $V[z \mid {}^{1}_{0}\mathcal{G}^{211}_{\mathbf{b},m}]$ constructed using an infinite number of nodeless R@ ∞ Frobenius solutions near the origin. Similarly to the discussion presented above for SUSY partners of the *r*-GRef radial potential, Frobenus solutions (5.2.8*) change their type at zeros of TWM function (5.2.12) which coincide with positive roots $\nu_{\mathbf{b},m}$ of the quadratic equations

$$b\,\nu^2_{\mathbf{b},m} + 2\nu_{\mathbf{b},m}(2m+1-\lambda_o) + g_o = 0\,. \qquad (6.2.14)$$

Each equation has a single positive root if $g_o < 0$.

Subtracting (6.2.14) from quadratic equation (5.2.15) with m=0 gives

$$(\nu_{\mathbf{c},0} - \nu_{\mathbf{b},m})[b(\nu_{\mathbf{b},m} + \nu_{\mathbf{c},0}) + 2(1+\lambda_o)] = 4(\lambda_o - m)\nu_{\mathbf{b},m} \qquad (6.2.15)$$

we conclude that the FF $\psi_{\mathbf{b},m}[\zeta \mid {}_{0}\mathcal{G}^{211}_{\downarrow\mathbf{b},m}]$ is nodeless iff $m < \lambda_o$. In particular, this implies that only the basic solution $\mathbf{b}$,0 can be used as the FF if the SI lies within the LC range.

By substituting (5.2.12) into the right-hand side of (4.23) one can represent the TWF function for the rational SUSY partner $V[z\,|\,{}_0^1G^{211}_{\mathbf{b},m}]$ as

$$*_0m_G(\varepsilon;*\Diamond_o,g_o,b\,|\,\mathbf{b},m)=\left(\sqrt{-\varepsilon}/b\right)^{\Diamond_o}\frac{\Gamma(-*\Diamond_o)}{4^m\Gamma(*\Diamond_o+1)}\,{}_0\Upsilon_{\mathbf{b},m}(\varepsilon;*\Diamond_o+1,g_o,b)$$

$$\times\prod_{k=0}^{m-1}[\varepsilon-{}_0\varepsilon_{\mathbf{b},k}(*\Diamond_o+1,g_o,b)]\;{}_0M_{v,m}\left(\varepsilon;*\Diamond_o+1,\kappa(\varepsilon;g_o,b)\right)$$

$$\text{for } *\Diamond_o>2, \qquad (6.2.16)$$

where the products

$${}_0\Upsilon_{\mathbf{t}_\pm,m}(\varepsilon;\Diamond_o,g_o;b)\equiv\prod_{k=0}^{m}{}_0\Xi_{\mathbf{t}_\pm,k}(\varepsilon;\Diamond_o,g_o;b) \qquad (6.2.17)$$

are formed by the energy-dependent coefficients of proportionality

$${}_0\Xi_{\mathbf{t}_\pm,k}(\varepsilon;;\lambda_o,g_o;b)=-\frac{b+g_o/\sqrt{\varepsilon\;{}_0\varepsilon_{\mathbf{t}_\pm,k}}}{\sqrt{-\varepsilon}+\nu_{\mathbf{t}_\pm,k}} \qquad (6.2.18)$$

in the linearized expressions

$$\tfrac{1}{2}(1+\lambda_o)-\kappa(\varepsilon;\lambda_o,g_o;b)+k=\tfrac{1}{4}\,{}_0\Xi_{\mathbf{t}_\pm,k}(\varepsilon;\lambda_o,g_o;b)({}_0\varepsilon_{\mathbf{t}_\pm,k}-\varepsilon),. \qquad (6.2.19)$$

By analogy with the regular case ι=1, the TMF function for the potential $V[z\,|\,{}_0^1G^{211}_{\mathbf{b},0}]$ in the LC region,

$$*_0m_G(\varepsilon;1-\Diamond_o\,|\,\mathbf{b},\varepsilon_{\mathbf{b},0})=\frac{4^{m-1}\Gamma(\Diamond_o)}{(\Diamond_o-1)\Gamma(2-\Diamond_o)}$$

$$\times\frac{\left(b/\sqrt{-\varepsilon}\right)^{\Diamond_o}}{{}_0\Xi_{\mathbf{b},0}(\varepsilon;\Diamond_o,g_o;b)\;{}_0M_{0,1}\left(\Diamond_o;\kappa(\varepsilon;g_o;b)\right)} \qquad (6.2.20)$$

$$\text{for } 0<\Diamond_o<1,$$

has infinitely many poles at the energies $\varepsilon = {}_0\varepsilon_{\mathbf{b},m}$ with m > 1. [The function $_0\Xi_{\mathbf{b},0}(\varepsilon;\Diamond_o,g_o;b)$ is positive by definition if $\Diamond_o$<1.] The appropriate eigenfunctions

$$\psi_{\mathbf{c},v}[\zeta\,|\,{}_0^1\mathcal{G}^{211}_{\mathbf{b},0}] = \frac{1}{\sqrt[4]{a\zeta+b}}\,\zeta^{3/4-1/2\Diamond_o}\,e^{-1/2\nu_{\mathbf{b},v+1}\zeta}\,Hc_{v+1}[\zeta\,|\,{}_0^1\mathcal{G}^{211}_{\mathbf{b},0};\mathbf{c},v]$$

$$\text{for } 0<\Diamond_o<1 \qquad (6.2.21)$$

are obtained by applying the DT with the FF $\mathbf{b}$,0 to the Frobenius solutions

$$\psi_{\mathbf{b},v+1}[\zeta\,|\,{}_0\mathcal{G}^{211}_{\downarrow\mathbf{b},0}] = \sqrt[4]{a\zeta+b}\;\,{}_0C_{\mathbf{b},v+1}\zeta^{1/4-1/2\Diamond_o}\,e^{-1/2\nu_{\mathbf{b},v+1}\zeta}\,L^{(-\Diamond_o)}_{v+1}(\nu_{\mathbf{b},v+1}\zeta)$$

$$\text{for } 0<\Diamond_o<1. \qquad (6.2.22)$$

The GS c-Heun polynomials of order v+1 in the right-hand side of (6.2.21) can be obtained from GL polynomials using the differential relations

$$(\nu_{\mathbf{b},0}-\nu_{\mathbf{b},v+1}-v)Hc_{v+1}[\zeta\,|\,{}_0^1\mathcal{G}^{211}_{\mathbf{b},0};\mathbf{c},v]$$

$$= \frac{2(v+1)!}{(-\nu_{\mathbf{b},v+1})^{v+1}}\,{}_1\hat{a}(\nu_{\mathbf{b},v+1}-\nu_{\mathbf{b},0})L^{(-\Diamond_o)}_{v+1}(\nu_{\mathbf{b},v+1}\zeta)$$

$$\text{for } 0<\Diamond_o<1. \qquad (6.2.23)$$

The lowest-energy eigenfunction

$$\psi_{\mathbf{c},0}[\zeta\,|\,{}_0^1\mathcal{G}^{211}_{b,0}] = \frac{1}{\sqrt[4]{a\zeta+b}}\,\zeta^{3/4-1/2\Diamond_o}\,e^{-1/2\nu_{\mathbf{b},v+1}\zeta}(\zeta-{}^*\zeta_{\mathbf{b};0,1}) \qquad (6.2.24)$$

$$\text{for } 0<\Diamond_o<1$$

is again obtained by applying the DT in question to the Frobenius solution

$$\psi_{\mathbf{b},1}[\zeta\,|\,{}_0\mathcal{G}^{211}_{\downarrow\mathbf{b},1}] = \sqrt[4]{a\zeta+b}\;\,{}_0C_{\mathbf{b},1}\zeta^{1/4-1/2\Diamond_o}\,e^{-1/2\nu_{\mathbf{b},1}\zeta}(\zeta-\zeta_{\mathbf{b},1}) \qquad (6.2.25)$$

$$\text{for } 0<\Diamond_o<1,$$

where the zero $\zeta_{\mathbf{b},1}$ of the first-order polynomial in the right-hand side of (6.2.25) is again obtained using generic formula (3.42) in [15] for ι=0:

$$0 < \zeta_{\mathbf{b},1} = \frac{1-\Diamond_o}{\nu_{\mathbf{b},1}} < 1 \quad \text{for } \Diamond_o < 1. \tag{6.2.26}$$

By analyzing the difference between quadratic equations (6.2.14) with m set to 0 and 1 one can verify that, similarly to (6.2.11) ,

$$\frac{2}{\nu_{\mathbf{b},0} - \nu_{\mathbf{b},1}} = b(\nu_{\mathbf{b},0} / \nu_{\mathbf{b},1} + 1) + 2\zeta_{\mathbf{b},1}. \tag{6.2.27}$$

Since the zero $^{*}\zeta_{\mathbf{b};0,1}$ of the first-order GS c-Heun polynomial is again given by (6.2.9) with **t** = **c**, it must lie on the negative semi-axis as predicted.

Keeping in mind that

$$_0\alpha_-(\lambda_o; \kappa(_0\varepsilon_{\mathbf{b},0}; g_o; b) = 0 \tag{6.2.28}$$

one finds that

$$_0\alpha_-(\lambda_o; \kappa(\varepsilon_{\mathbf{d}}; g_o; b) = b(\sqrt{-\varepsilon_{\mathbf{d}}} + \sqrt{-\varepsilon_{\mathbf{b};0}})](1 - \sqrt{\varepsilon_{\mathbf{b};0} / \varepsilon_{\mathbf{d}}}) > 0 \tag{6.2.29}$$

so that the Frobenius solution of type **d** below the energy $_0\varepsilon_{\mathbf{b},0}$

$$\psi_{\mathbf{d}}[\zeta \,|\, _0G^{211}] = (\sqrt{b}\zeta)^{\frac{1}{2}(1-\lambda_o)} e^{\frac{1}{2}\zeta} \times F[\alpha_-(\lambda_o; k(\varepsilon_{\mathbf{d}}; g_o; b), 1-\lambda_o; \zeta] \tag{6.2.30}$$

is necessarily nodeless. By examining sign of (5.2.12′) under constraint (6.2.29) one can directly verify $m$-coefficient (5.2.12) is positive, in agreement with the conclusion made in the end of Section 4. We thus conclude that in the LC region there is a continuous family of isospectral SUSY partners with the infinite discrete energy spectrum $_0\varepsilon_{\mathbf{b},m}$ (m=0, 1, ...).

An analysis of quadratic equation (6.2.14) show that its negative root monotonically decreases with m so that the mentioned family of isospectral SUSY partners contains infinitely many rational potentials $V[\zeta \,|\, {}_{0}^{1}G^{211}_{\mathbf{d},m}]$ quantized via GS *c*-Heine polynomials. Similarly to the regular case, the first potential $V[\zeta \,|\, {}_{0}^{1}G^{211}_{\mathbf{d},0}]$ in this sequence of isospectral rational SUSY partners is exactly quantized via a sequence of GS *c*-Heun polynomials

$$\psi_{\mathbf{c},v}[\zeta \,|\, {}_{0}^{1}G^{211}_{\mathbf{d},0}] = \frac{1}{\sqrt[4]{a\zeta + b}}\, \zeta^{3/4 - 1/2\,\Diamond_o}\, e^{-1/2\,\nu_{\mathbf{b},v}\,\zeta}\, Hc_v[\zeta \,|\, {}_{0}^{1}G^{211}_{\mathbf{d},0}; \mathbf{c}, v]. \qquad (6.2.31)$$

As shown in next Section it can be obtained from $V[\zeta \,|\, {}_{0}^{1}G^{211}_{\mathbf{b},0}]$ by means of the double-step DT which creates a new lowest-energy eigenstate at the energy ${}_{0}\varepsilon_{\mathbf{b},0}$ while keeping other eigenvalues unchanged.

## 7. Form-invariance of radial GRef potentials under double-step Darboux transformations with basic seed solutions

As demonstrated in previous Section, the radial potential $V[\xi \,|\, {}_{\iota}G^{211}_{\downarrow\mathbf{c},0}]$ with at least one bound energy state has four basic AEH solutions, each of its own type $\mathbf{t} = \mathbf{a}, \mathbf{b}, \mathbf{c}$, and $\mathbf{d}$. Since any basic AEH solution ${}_{\iota}\psi_{\mathbf{t}}[\xi(r); {}_{\iota}\varepsilon_{\mathbf{t},0}]$ is nodeless by definition each of them can be used to construct the exactly-solvable SUSY partner $V[\xi \,|\, {}_{\iota}^{1}G^{211}_{\mathbf{t},0}]$. In a separate publication [35] we have proven existence of such a quartet of exactly-solvable SUSY partners for the generic *r*-GRef potential $V[z \,|\, {}_{1}G^{220}_{\downarrow\mathbf{c},0}]$ on the line provided that the density function ${}_{\iota}\wp[\xi; a, b]$ in the appropriate SL equation has two distinct real zeros lying outside the quantization interval [0, 1]. (In particular, the CGK potential [61] represents the SUSY partner $V[z \,|\, {}_{1}^{1}G^{211}_{\mathbf{c},0}]$ obtained from $V[z \,|\, {}_{1}G^{220}_{\downarrow\mathbf{c},0}]$ by eliminating its ground energy state.) A similar result (though with some additional restrictions) can be proven for the generic *c*-GRef potential $V[\zeta \,|\, {}_{0}G^{220}_{\downarrow\mathbf{c},0}]$. The family of the radial potentials of our current interest arises as the limiting case when one of the outer roots tends to 0. The remarkable feature of this transition illuminated below is that the pairs of potentials $V[\xi(r)\,|\,{}^{1}_{\iota}G^{220}_{t_+,0}]$ with $t_+ = a$ or $c$ and $V[\xi(r)\,|\,{}^{1}_{\iota}G^{220}_{t_-,0}]$ with $t_- = b$ or $d$ collapse into the single potential

$$V[\xi\,|\,{}_{\iota}H^{21}_{+}] \equiv V[\xi\,|\,{}^{1}_{\iota}G^{211}_{a,0}] = V[\xi\,|\,{}^{1}_{\iota}G^{211}_{d,0}] \tag{7.1a}$$

and

$$V[\xi\,|\,{}_{\iota}H^{21}_{-}] \equiv V[\xi\,|\,{}^{1}_{\iota}G^{211}_{b,0}] = V[\xi\,|\,{}^{1}_{\iota}G^{211}_{c,0}], \tag{7.1b}$$

respectively. [It should be stressed that this analogy between potentials on the real axis $(-\infty, +\infty)$ and those on the semi-axis $(0, \infty)$ only works for the LP range of the resultant radial potential and, as a general rule, any such transition has to be performed with a great caution.] As shown in [35], a similar collapse takes place in case of the basic SUSY partners of the so-called ‘linear-tangent-polynomial’ (LTP) potential $V[\xi\,|\,{}_{\iota}G^{110}_{\downarrow c,0}]$ in the limiting case when one of the real roots of the density function tends to infinity. In both limiting cases the Schrödinger equation is exactly quantized via Heun polynomials so that we refer to these SUSY partners as radial Heun-reference (HRef) potentials and use superscripts K and $\aleph$ in ${}_{\iota}H^{K\aleph}_{\pm}$ to distinguish between them.

The radial HRef potentials constructed in such a way are represented by PFs

$$V[\xi\,|\,{}^{1}_{\iota}G^{211}_{t_\pm,0}] = -\,{}_{\iota}\wp^{-1}[\xi; a, b]\;\; {}_{\iota}I^{o}[\xi\,|\,{}^{1}_{\iota}G^{211}_{t_\pm,0}] - \tfrac{1}{2}\{\xi, x\}, \tag{7.2}$$

where

$$I^{o}[\xi\,|\,{}^{1}_{\iota}G^{211}_{t_\pm,0}] = \frac{1-\lambda^2_{o;t_\pm}}{4\xi^2} + \frac{1}{4(1-\iota\xi)^2} - \frac{3}{4(\xi-\xi_T)^2} + (-1)^{\iota}\,\frac{O^{o}_{1}[\xi\,|\,{}^{1}_{\iota}G^{211}_{t_\pm,0}]}{4\xi(1-\iota\xi)(\xi-\xi_T)} \tag{7.3}$$

and

$$\xi_T \equiv -b/a = -b/(1-b)\,. \tag{7.4}$$

As pointed out above the coefficients of the first-order polynomials

$$O^{o}_{1}[\xi\,|\,{}^{1}_{\iota}G^{211}_{t_\pm,0}] = O^{o}_{1,1}\,|\,{}^{1}_{\iota}G^{211}_{t_\pm,0})\,(\xi-\xi_T) + o_o\,|\,{}^{1}_{\iota}G^{211}_{t_\pm,0}) \tag{7.5}$$

are automatically chosen in such a way that the radial potential in question is quantized via Heun polynomials. (We will present an accurate proof of this result in one of the following publications).

By analyzing the Wronskian of two basic functions ${}_\iota\psi_{\dagger}[\xi(r);\,{}_\iota\varepsilon_{\dagger,0}]$ and ${}_\iota\psi_{\dagger'}[\xi(r);\,{}_\iota\varepsilon_{\dagger',0}]$, one can prove that the SL equation

$$\left\{\frac{d^2}{d\xi^2} + {}_\iota I^o[\xi\,|\,{}^1_\iota G^{211}_{\dagger,0}] + \varepsilon\frac{(1-b)\xi+b}{4\xi(1-\iota\xi)^2}\right\}\Phi[\,|\,{}^1_\iota G^{211}_{\dagger,0}] = 0 \qquad (7.6)$$

has the solution

$$ {}_1\phi_{\dagger\dagger'}[z;\lambda_{o;\dagger},\mu_o;\xi_T] \equiv \sqrt{\frac{z(1-z)}{|\,z-z_T\,|}}\, z^{\frac{1}{2}(\sigma_{0;\dagger'}\lambda_{o;\dagger}-1)}\,(1-z)^{\frac{1}{2}\lambda_{1;\dagger'}} $$

$$ \times[(\lambda_{1;\dagger}-\lambda_{1;\dagger'})z+(\sigma_{0;\dagger}-\sigma_{0;\dagger'})\lambda_o(z-1)] \qquad (7.7)$$

($z_T = \xi_T$) or

$$ {}_0\phi_{\dagger\dagger'}[\zeta;\lambda_{o;\dagger},g_o;\zeta_T] \equiv \sqrt{\frac{\zeta}{|\,\zeta-\zeta_T\,|}}\,\zeta^{\frac{1}{2}(\sigma_{0;\dagger'}\lambda_o-1)}\,e^{-\frac{1}{2}\lambda_{1;\dagger'}\zeta} \times[(\lambda_{1;\dagger}-\lambda_{1;\dagger'})\zeta-(\sigma_{0;\dagger}-\sigma_{0;\dagger'})\lambda_o] \qquad (7.7')$$

($\zeta_T = \xi_T$) for $\iota$ = 1 or 0, respectively. If $\sigma_{0;\dagger} = -\sigma_{0;\dagger'}$, then

$$\sigma_{0;\dagger'}\lambda_o - 1 = \begin{cases} \sigma_{0;\dagger'}(\lambda_o+\sigma_{0;\dagger}) & \text{for } \lambda_o > 1 \text{ or } \dagger = \dagger_+, \\ \lambda_o - 1 & \text{for } \lambda_o < 1 \text{ and } \dagger = \dagger_-. \end{cases} \qquad (7.8)$$

On other hand, if $\sigma_{0;\dagger} = \sigma_{0;\dagger'}$, then

$$\sigma_{0;\dagger'}\lambda_o + 1 = \begin{cases} \sigma_{0;\dagger'}(\lambda_o+\sigma_{0;\dagger}) & \text{for } \lambda_o > 1 \text{ or } \dagger = \dagger_+, \\ 1-\lambda_o & \text{for } \lambda_o < 1 \text{ and } \dagger = \dagger_-. \end{cases} \qquad (7.8^*)$$

In both cases the exponents of z or $\zeta$ in (7.7) and (7.7′), respectively, agree with the known value of the exponent difference at the origin,

$$\lambda_{o;\dagger} = \begin{cases} \lambda_o + \sigma_{0;\dagger} & \text{for } \lambda_o > 1 \text{ or } \dagger = \dagger_+, \\ 1 - \lambda_o & \text{for } \lambda_o < 1 \text{ and } \dagger = \dagger_- \end{cases} \tag{7.9}$$

for the HRef PF beam ${}^{1}_{\iota}\mathcal{G}^{211}_{\dagger,0}$. Making use of Suzko's reciprocal formula [67, 68]

$$^{*}{}_{\iota}\phi_{\dagger\dagger'}[\xi;\lambda_{o;\dagger},{}_{\iota}\mu_o;\xi_T] = \frac{{}_{\iota}\wp^{-1/2}[\xi;1-b,b]}{{}_{\iota}\phi_{\dagger\dagger'}[\xi;\lambda_{o;\dagger},{}_{\iota}\mu_o;\xi_T]}, \tag{7.10}$$

one can easily verify that the latter solution do not have a singularity at $\xi = \xi_R$ if $\sigma_{0;\dagger} = \sigma_{0;\dagger'}$. Here we treat $g_o$ as a counter part of $\mu_o$:

$${}_{1}\mu_o = \mu_o, \; {}_{0}\mu_o = g_o, \tag{7.11}$$

because both of them, as explained below, are unaffected by double-step DTs in question. (For the *c*-GRef radial potential this assertion directly follows from the fact that $g_o$ determines the Coulomb tail which, as pointed out in Section 3, remains unchanged under any DT [23, 24].) We thus conclude that double-step DTs with basic SSs of types $\mathsf{a}$ and $\mathsf{c}$ or $\mathsf{b}$ and $\mathsf{d}$ convert radial GRef potentials (5.1.2) and (5.2.2) onto themselves. (In the region with no discrete energy levels this is also true for double-step DTs with basic SSs s of types $\mathsf{a}$ and $\mathsf{a}'$ or types $\dagger_-$ and $\dagger'_-$ if $\sigma_{1;\dagger_-} = \sigma_{1;\dagger'_-}$.)

At this point it seems convenient to split again our analysis into two parts treating SUSY partners of each radial potential (5.1.2) or (5.2.2) separately.

### 7.1. SUSY triplets of radial potentials solvable via hypergeometric and Heun functions

Let us start our analysis from the proof that radial *r*-GRef potential (5.1.2) is form-invariant under the double-step DT with basic SSs of type $\mathsf{c}$ and $\mathsf{a}$ (and therefore under its reverse – the double-step DT with basic SSs of type $\mathsf{b}$ and $\mathsf{d}$). To do it first note that the DT may convert SL equation (7.6) into the differential equation with three regular singular points for ι=1 only if the appropriate FF has the 'basic' form:

$$ {}^{\mathrm{H}}_{1}\phi_{*\mathbf{t},0}[z;{}^{*}\lambda_o,\nu_{\mathbf{t}}] = \sqrt{\frac{z(1-z)}{|z-z_T|}}\, z^{-\sigma_{0;\mathbf{t}}\,{}^{1}\!/_{2}\,{}^{*}\lambda_o}\,(1-z)^{-{}^{1}\!/_{2}\nu_{\mathbf{t}}}, \qquad (7.1.1) $$

where the type $*\mathbf{t}$ 'opposite' of $\mathbf{t}$ is determined via the relations:

$$ \sigma_{0;*\mathbf{t}} = -\sigma_{0;\mathbf{t}},\ \sigma_{1;*\mathbf{t}} = -\sigma_{1;\mathbf{t}}. \qquad (7.1.2) $$

SL equation (7.5) does have function (7.1.1) as its solution iff

$$ \begin{aligned} O_1[z\,|\,{}^{1}_{1}G^{211}_{\mathbf{t},0}] - b\nu^2_{\mathbf{t}}(z-z_T) &= 2(z-z_T)(1-\sigma_{0;\mathbf{t}}\,{}^{*}\lambda_o)(1-\nu_{\mathbf{t}}) \\ &\quad - 2[(1-\sigma_{0;\mathbf{t}}\,{}^{*}\lambda_o)(z-1)+(1-\nu_{\mathbf{t}})z]. \end{aligned} \qquad (7.1.3) $$

which gives

$$ O_1[z\,|\,{}^{1}_{1}G^{211}_{\mathbf{t},0}] = O_{1;1}\,|\,{}^{1}_{1}G^{211}_{\mathbf{t},0})(z-z_T) + o\,|\,{}^{1}_{1}G^{211}_{\mathbf{t},0}), \qquad (7.1.3') $$

where

$$ O_{1;1}\,|\,{}^{1}_{1}G^{211}_{\mathbf{t},0}) = b\nu^2_{\mathbf{t}} + 2\sigma_{0;\mathbf{t}}\,{}^{*}\lambda_o\,\nu_{\mathbf{t}} - 2 \qquad (7.1.4a) $$

and

$$ {}^{1}\!/_{2}(1-b)\ o\,|\,{}^{1}_{1}G^{211}_{\mathbf{t},0}) = b(1-\nu_{\mathbf{t}}) + 1 - \sigma_{0;\mathbf{t}}\,{}^{*}\lambda_o. \qquad (7.1.4b) $$

On other hand, consolidation of (5.1.29) and (6.1.19) for m=0 leads to two quadratic equations

$$ b\lambda^2_{1;\mathbf{t}_\pm,0}(\lambda_o,\mu_o;b) \pm 2\lambda_{1;\mathbf{t}_\pm,0}(\lambda_o,\mu_o;b)(\lambda_o \pm 1) = \mu_o^2 - (\lambda_o \pm 1)^2. \qquad (7.1.5) $$

In the LP region of the potential $V[z\,|\,{}^{1}_{1}G^{211}_{\mathbf{t},0}]$, i.e. for ${}^{*}\lambda_o > 1$, they can be represented as a single quadratic equation

$$ b\nu^2_{\pm}({}^{*}\lambda_o,\mu_o;b) + 2\,{}^{*}\lambda_o\nu_{\pm}({}^{*}\lambda_o,\mu_o;b) + {}^{*}\lambda_o^2 - \mu_o^2 = 0, \qquad (7.1.6) $$

where

$$ \begin{aligned} {}_1\nu_{-}({}^{*}\lambda_o,\mu_o;b) &= {}_1\lambda_{1;\mathbf{a},0}({}^{*}\lambda_o-1,\mu_o;b) &&\text{for}\quad {}^{*}\lambda_o > 1, &&(7.1.7a)\\ &= {}_1\lambda_{1;\mathbf{d},0}({}^{*}\lambda_o+1,\mu_o;b) &&\text{for}\quad {}^{*}\lambda_o \neq 1 &&(7.1.7d) \end{aligned} $$

and

$$ {}_1\nu_+(^*\lambda_o,\mu_o;b) = {}_1\lambda_{1;\mathbf{c},0}(^*\lambda_o - 1,\mu_o;b) \quad \text{for} \quad ^*\lambda_o > 1, \qquad (7.1.7c)$$

$$ = {}_1\lambda_{1;\mathbf{b},0}(^*\lambda_o + 1,\mu_o;b) \;\; (^*\lambda_o > 1) \quad \text{for} \quad ^*\lambda_o \neq 1 \qquad (7.1.7b)$$

are its lower and upper real roots. Comparing quadratic polynomial (7.1.4a) with the right-hand side of (7.1.6) gives

$$\nu_{\mathbf{t}} = \sigma_{0;\mathbf{t}}\; {}_1\nu_{\pm}(^*\lambda_o > 1,\mu_o;b) \qquad (7.1.8)$$

or, to be more precise,

$$\nu_{\mathbf{a}} = -\nu_{\mathbf{b}} = {}_1\nu_{-}(^*\lambda_o > 1,\mu_o;b) \qquad (7.1.8a)$$

and

$$\nu_{\mathbf{c}} = -\nu_{\mathbf{d}} = {}_1\nu_{+}(^*\lambda_o > 1,\mu_o;b). \qquad (7.1.8b)$$

Keeping in mind that

$$b\nu_+(^*\lambda_o,\mu_o;b) + b\nu_-(^*\lambda_o,\mu_o;b) = 2\,^*\lambda_o \qquad (7.1.9)$$

we conclude that there are only two distinctive polynomials (7.1.3): ${}_1O_1[z\,|\,{}_1\mathbf{H}^{21}_{-}]$ for $\mathbf{t} = \mathbf{a}$ or $\mathbf{d}$ and ${}_1O_1[z\,|\,{}_1\mathbf{H}^{21}_{+}]$ for $\mathbf{t} = \mathbf{b}$ or $\mathbf{c}$, with the coefficients

$${}_1O_{1;1}\,|\,{}_1\mathbf{H}^{21}_{\pm}) = \mu_o^2 - {}^*\lambda_o^2 - 2 \qquad (7.1.10a)$$

and

$$\tfrac{1}{2}(1-b)\;o\,|\,{}^1_1\mathcal{G}^{211}_{\mathbf{t},0}) = b[1 + {}_1\nu_{\pm}(^*\lambda_o,\mu_o;b)] + 1 - {}^*\lambda_o. \qquad (7.1.10b)$$

For $\mu_o > {}^*\lambda_o$ the subscript '+' or '–' in the notation ${}_1\nu_{\pm}(^*\lambda_o,\mu_o)$ coincides with the root sign. This is the region where the isospectral SUSY partner $V[z\,|\,{}_1\mathbf{H}^{21}_{-}]$ of the potential $V[z\,|\,{}_1\mathcal{G}^{211}]$ necessarily has the discrete energy spectrum while the number of energy levels in the potential $V[z\,|\,{}_1\mathbf{H}^{21}_{+}]$ is less by one.

It directly follows from (7.1.7a) and (7.1.7b) that the double-step DT using pairs of basic SSs $\mathbf{c}$,0 and $\mathbf{a}$,0 (or $\mathbf{b}$,0 and $\mathbf{d}$,0) does not affect the value of the parameter $\mu_o$, assuming that $\mu_o > \lambda_o + 1$ so that the potential $V[z\,|\,{}^1_1\mathcal{G}^{211}_{\mathbf{c},0}]$ has at least one bound energy state. Another interesting observation followed from (7.1.7a) and (7.1.7b) is that the energies of two basic solutions are the same

$$_1\varepsilon_{0,-} \equiv -_1\nu_-^2(^*\lambda_o,\mu_o;b) = {_1\varepsilon_{\mathbf{a},0}}(^*\lambda_o-1,\mu_o;b) = {_1\varepsilon_{\mathbf{d},0}}(^*\lambda_o+1,\mu_o;b) \quad (7.1.11a)$$

and

$$_1\varepsilon_{0,+} \equiv -_1\nu_+^2(^*\lambda_o,\mu_o;b) = {_1\varepsilon_{\mathbf{c},0}}(^*\lambda_o-1,\mu_o;b) = {_1\varepsilon_{\mathbf{b},0}}(^*\lambda_o+1,\mu_o;b) \quad (7.1.11b)$$

for both potentials $V[z\,|\,{_1}\mathbf{H}_+^{21}]$ and $V[z\,|\,{_1}\mathbf{H}_-^{21}]$.

The LC region of the potentials $V[z\,|\,{_1^1}G_{\mathbf{t},0}^{211}]$ can be reached by a Darboux deformation of the radial GRef potential only if the appropriate FF is irregular at the origin ($\mathbf{t}$ = $\mathbf{t}_-$= $\mathbf{b}$ or $\mathbf{d}$). Let us define the potentials $V[z\,|\,{_1}\mathbf{H}_\pm^{21}]$ using the DT which reduce the SI $2-\lambda_o$ by 1 in the same way as it has been done above for the LP region. We thus need to prove that the DTs keeping the potential in the LC region lead to the same polynomials $O_1[z\,|\,{_1^1}G_{\mathbf{t}_-,0}^{211}]$ if $0<{^*\lambda_o}=1-\lambda_o<1$. Representing (7.1.5) for $\mathbf{t}_-$= $\mathbf{b}$ or $\mathbf{d}$ as

$$b\lambda_{1;\mathbf{t}_-,0}^2(\lambda_o,\mu_o;b)+2\lambda_{1;\mathbf{t}_-,0}(\lambda_o,\mu_o;b)\,(1-\lambda_o)=\mu_o^2-(1-\lambda_o)^2 \quad (7.1.12)$$

and comparing the latter with (7.1.6) shows that

$$\lambda_{1;\mathbf{d},0}(1-{^*\lambda_o},\mu_o;b)=\nu_-({^*\lambda_o}<1,\mu_o;b) \quad (7.1.13d)$$

and

$$\lambda_{1;\mathbf{b},0}(1-{^*\lambda_o},\mu_o;b)=\nu_+({^*\lambda_o}<1,\mu_o;b)\,. \quad (7.1.13b)$$

If the SI of the GRef potential lies within the LC range then (as shown in previous Section) the potential $V[z\,|\,{_1^1}G_{\mathbf{d},0}^{211}]$ has an extra energy level ${_1\varepsilon_{\mathbf{b},0}}$, compared with the discrete energy spectrum of the potential. According to (7.1.13b), this is the energy level inserted by the DT of type $\mathbf{d}$ for the transition LP→LC.

This completes the proof that the radial potential $V[z\,|\,{_1}G^{211}]$ has no more than two SUSY partners. The SUSY pairs $V[z\,|\,{_1}\mathbf{H}_\pm^{21}]$ always exist if the potential $V[z\,|\,{_1}G^{211}]$ has the discrete energy spectrum. Since quadratic equation (7.1.4) has real roots iff

$$\mu_o^2 > (1/b-1)^*\lambda_o^2 \equiv -^*\lambda_o^2/z_T \quad (7.1.14)$$

a pair of the HRef potentials exists for any positive value of $\mu_o$ if $z_T > 1$.

Substituting (7.1.7a), (7.1.7d), and(7.1.7b) into (4.20), (4.20*), and (4.23) with $\varepsilon_{\mathbf{a},0}$, $\varepsilon_{\mathbf{d},0}$, and $\varepsilon_{\mathbf{b},0}$ standing for $\varepsilon_{\mathbf{a}}$, $\varepsilon_{\mathbf{d}}$, and $\varepsilon_{\mathbf{b},m}$, respectively, shows that the TWF functions of HRef potentials $V[z\,|\,{}_1\mathbf{H}^{21}_{+}]$ and $V[z\,|\,{}_1\mathbf{H}^{21}_{-}]$ can be represented in the following dual form

$$ {}^{*}{}_1m_{\pm}(\varepsilon;{}^{*}\lozenge_o,\mu_o;b) = -\frac{{}_1\nu^2_{\pm}({}^{*}\lozenge_o,\mu_o;b)+\varepsilon}{4\,{}^{*}\lozenge_o^2}\;{}_1m_G(\varepsilon;{}^{*}\lozenge_o-1,\mu_o;b) \quad \text{for} \quad {}^{*}\lozenge_o > 1 \tag{7.1.15} $$

$$ = -\frac{4({}^{*}\lozenge_o+1)^2}{{}_1\nu^2_{\mp}({}^{*}\lozenge_o,\mu_o;b)+\varepsilon}\;{}_1m_G(\varepsilon;{}^{*}\lozenge_o+1,\mu_o;b) \quad \text{for} \quad {}^{*}\lozenge_o \neq 1. \tag{7.1.15*} $$

Note that the first expression is not applicable to the LC region. In the latter case one has to use (4.22*) and (4.23), coupled with (7.1.13d) and (7.1.13b), which gives

$$ {}^{*}{}_1m_{\pm}(\varepsilon;{}^{*}\lozenge_o,\mu_o;b) = \frac{\varepsilon-{}_1\nu^2_{\mp}({}^{*}\lozenge_o,\mu_o;b)}{16\,{}^{*}\lozenge_o^2(1-{}^{*}\lozenge_o)^2}\;{}_1m_G^{-1}(\varepsilon;1-{}^{*}\lozenge_o,\mu_o;b) \tag{7.1.15′} $$

$$ \text{for} \quad {}^{*}\lozenge_o < 1. $$

The argument presented above can be also applied to the Jost function [39]

$$ {}_1f(k;\lozenge_o,\mu_o;b) = \frac{2b^{\frac{1}{2}\lozenge_o}\,e^{-ikr_o}\,\Gamma(1-ik)\Gamma(\lozenge_o+1)}{\Gamma\big({}_1\alpha_{++}(k^2;\lozenge_o,\mu(\mu_o;1-b))\big)\Gamma\big({}_1\beta_{++}(k^2;\lozenge_o,\mu(\mu_o;1-b))\big)}, \tag{7.1.16} $$

where the parameter $r_o$ ($x^+$ in [17]) is determined by the asymptotic behavior of the variable z(r) at infinity

$$ ln[1-z(r)] \approx -2(r+r_o) \quad \text{as} \quad r\to\infty. \tag{7.1.17} $$

Note that we included the additional factor $2s_o+1 = 2\lozenge_o = 2\lambda_o$ which was erroneously omitted in the definition of the Jost function in [17]. (As clarified in next Section, two definitions coincide in the particular case of the Hulthén potential [66] which was used as an illustrative example.) Substituting (7.1.8a) and (7.1.8b) into (3.5a) - (3.5d) thus gives

$$ {}_1f(k;{}^*\Diamond_o \mid {}_1\mathbf{H}^{21}_{\pm}) = -\frac{2({}^*\Diamond_o - 1)}{ik + {}_1\nu_{\pm}({}^*\Diamond_o, \mu_o; b)}\, {}_1f(k;{}^*\Diamond_o - 1, \mu_o; b) \quad \text{for } {}^*\Diamond_o > 1 \tag{7.1.18} $$

$$ = -\frac{ik - {}_1\nu_{\mp}({}^*\Diamond_o, \mu_o; b)}{2\,{}^*\Diamond_o}\, {}_1f(k;{}^*\Diamond_o + 1) \quad \text{for } {}^*\Diamond_o \neq 1. \tag{7.1.18*} $$

Our next step is to explicitly demonstrate that TWF function (5.1.46) is indeed form-invariant under the double-step DTs of our interest for two scenarios: i) ${}^*\Diamond_o > 1$ and ii) $0 < {}^*\Diamond_o = \Diamond_o - 1 < 1$.

**i) Double-step DTs of types LP→LP→LP or LC→LP→LP**

**(**$1 < {}^*\lambda_o = {}^*\Diamond_o = \Diamond_o + 1,\ \ 2 < {}^{**}\lambda_o = {}^{**}\Diamond_o = \Diamond_o + 2$)

Due to the form-invariance of the potential $V[z \mid {}_1G^{211}]$ under the double-step DT with basic SSs **c**,0 and **a**,0 TWF function (5.1.46) transformed according to (4.20) at each step must satisfy the following ladder relation:

$$ \frac{{}_1m_G(\varepsilon;\Diamond_o + 2, \mu_o; b)}{{}_1m_G(\varepsilon;\Diamond_o, \mu_o; b)} = \frac{[{}_1\varepsilon_{\mathbf{a},0}(\Diamond_o, \mu_o; b) - \varepsilon][{}_1\varepsilon_{\mathbf{c},0}(\Diamond_o, \mu_o; b) - \varepsilon]}{16(\Diamond_o + 1)^2(\Diamond_o + 2)^2}. \tag{7.1.19} $$

The derived relation is closely related to a similar ladder relation for function (5.1.19):

$$ \frac{{}_1M_{0,0}(\varepsilon;\nu + 1, \mu)}{{}_1M_{0,0}(\varepsilon;\nu - 1, \mu)} = \tfrac{1}{16}\, {}_1G(\varepsilon;\nu, \mu), \tag{7.1.20} $$

where

$$ {}_1G(\varepsilon;\nu, \mu) \equiv (\nu^2 - \varepsilon - \mu^2)^2 + \varepsilon\nu^2. \tag{7.1.21} $$

$$ = [(\sqrt{-\varepsilon} + \nu)^2 - \mu^2] \times [(\sqrt{-\varepsilon} - \nu)^2 - \mu^2]. \tag{7.1.21'} $$

To verify (7.1.20) one simply needs to substitute the ladder relations

$$ {}_1\alpha_{\pm,\sigma_1}\left(\varepsilon;\lambda_o+2,\mu\right) = {}_1\alpha_{\pm,\sigma_1}\left(\varepsilon;\lambda_o,\mu\right)\pm 1 \qquad (7.1.22a) $$

and

$$ {}_1\beta_{\pm,\sigma_1}\left(\varepsilon;\lambda_o+2,\mu\right) = {}_1\beta_{\pm,\sigma_1}\left(\varepsilon;\lambda_o,\mu\right)\pm 1 \qquad (7.1.22b) $$

into (5.1.19) with $\lambda_o$ changed for $\lambda_o + 2$ and then re-arrange the products of $\alpha_{\pm,+}$- and $\beta_{\pm,+}$-coefficients according to (6.1.1).

Defining a new quadratic polynomial in ε via the relation

$$ {}_1G_2^0(\varepsilon;\nu,\mu_o;b) \equiv {}_1G\left(\varepsilon;\nu,\mu(\mu_o;1-b)\right) \qquad (7.1.23) $$

and taking into account the identity

$$ \frac{\Gamma(-\lambda_o-2)}{\Gamma(\lambda_o+3)} \equiv \frac{1}{(\lambda_o+1)^2(\lambda_o+2)^2}\times\frac{\Gamma(-\lambda_o)}{\Gamma(\lambda_o+1)} \qquad (7.1.24) $$

we find that (7.1.19) is equivalent to the following recurrence relation for the TWF function

$$ \frac{{}_1m_G(\varepsilon;\lozenge_o+2,\mu_o;b)}{{}_1m_G(\varepsilon;\lozenge_o,\mu_o;b)} = \frac{{}_1G_2^0(\varepsilon;\lozenge_o+1,\mu_o;b)}{16b^2(\lozenge_o+1)^2(\lozenge_o+2)^2}, \qquad (7.1.25) $$

By representing quadratic equations (7.1.5) as

$$ {}_1G_2^0\left({}_1\varepsilon_{\dagger\pm,0}(\lozenge_o,\mu_o;b);\lambda_o\pm 1,\mu;b\right) = 0\,. \qquad (7.1.26) $$

one can then directly verify that (7.1.19) and (7.1.25) are indeed equivalent. Taking into account (7.1.11a) and (7.1.11b) we can just decompose quadratic polynomial (7.1.23) as follows

$$ {}_1G_2^0(\varepsilon;{}^{*}\nu,\mu_o;b) = b^2[{}_1\nu_{+}^2({}^{*}\lozenge_o,\mu_o;b)+\varepsilon]\times[{}_1\nu_{-}^2({}^{*}\lozenge_o,\mu_o;b)+\varepsilon]\,. \qquad (7.1.27) $$

Re-arranging recurrent relation (7.1.25) as

$$ \frac{{}_1\nu_{\mp}^2(\lozenge_o+1,\mu_o;b)+\varepsilon}{4(\lozenge_o+1)^2}\,{}_1m_G(\varepsilon;\lozenge_o,\mu_o;b) = \frac{4(\lozenge_o+2)^2\,{}_1m(\varepsilon;\lozenge_o+2,\mu_o;b)}{{}_1\nu_{\pm}^2(\lozenge_o+1,\mu_o;b)+\varepsilon]} \qquad (7.1.28) $$

and substituting $\lozenge_o$ for ${}^*\lozenge_o - 1 > 0$ one can also confirm that the TWF functions of HRef potentials $V[z\,|\,{}_1\mathbf{H}^{21}_{+}]$ and $V[z\,|\,{}_1\mathbf{H}^{21}_{-}]$ allow dual representation (7.1.15) - (7.1.15*).

By analogy, substitution of both (7.1.15) and a similar relation for ${}^*{}_1m_{\pm}(\varepsilon; {}^*\lozenge_o + 2, \mu_o; b)$ into (7.1.25), with $\lozenge_o = {}^*\lozenge_o - 1$, leads to the following double-step recurrence relations for the TWF functions of $V[z\,|\,{}_1\mathbf{H}^{21}_{+}]$ and $V[z\,|\,{}_1\mathbf{H}^{21}_{-}]$:

$$\frac{{}^*{}_1m_{\pm}(\varepsilon; {}^*\lozenge_o + 2, \mu_o; b)}{{}^*{}_1m_{\pm}(\varepsilon; {}^*\lozenge_o, \mu_o; b)} = \frac{[{}_1\nu^2_{\pm}({}^*\lozenge_o + 2, \mu_o; b) + \varepsilon] \times [{}_1\nu^2_{\mp}({}^*\lozenge_o, \mu_o; b) + \varepsilon]}{16 b^2 ({}^*\lozenge_o + 1)^2 ({}^*\lozenge_o + 2)^2}$$
$$\text{for } {}^*\lozenge_o > 1. \qquad (7.1.29)$$

Note that subscripts '+' and '–' in the second bracket in the numerator are interchanged compared with the first. This implies $V[z\,|\,{}_1\mathbf{H}^{21}_{+}]$ and $V[z\,|\,{}_1\mathbf{H}^{21}_{-}]$: are related via double-step DTs so that each HRef potential is form-invariant under the four-step DTs.

It should be stressed that we has to use two single-step DTs (instead of a single second-order DT [67]) to prove that the resultant potentials do not have singularities inside the quantization interval. After that we can consider two families of GRef and HRef potentials in an independent fashion and use second-order DTs to completely decouple their analysis.

Taking into account that

$${}_1\alpha_{+,+}\left(k^2; \lambda_o, \mu(\mu_o; a)\right) {}_1\beta_{+,+}\left(k^2; \lambda_o, \mu(\mu_o; a)\right)$$
$$= \tfrac{1}{4}[-bk^2 - 2(\lambda_o + 1)ik + (\lambda_o + 1)^2 - \mu_o^2] \qquad (7.1.30)$$

and comparing the quadratic polynomial in the right-hand side with (7.1.4) gives

$${}_1f(k; \lozenge_o + \tfrac{5}{2}, \mu_o; b\,|\,\mathsf{a}, \varepsilon_{\mathsf{a},0}; \mathsf{c}, \varepsilon_{\mathsf{c},0}) = \frac{4(\lozenge_o + 1)(\lozenge_o + 2)}{(-ik - \lambda_{1;\mathsf{c},0})(-ik - \lambda_{1;\mathsf{a},0})}$$
$$\times {}_1f(k; \lozenge_o + \tfrac{1}{2}, \mu_o; b), \qquad (7.1.31)$$

as predicted by sequential use of DTs (3.5a) and (3.5c).

ii) **Double-step DTs of type LP→LC→LC**

**(**$1 < \Diamond_o < 2,\ {}^{*}\Diamond_o = \Diamond_o - 1 < 1,\ {}^{**}\Diamond_o = 2 - \Diamond_o < 1$)

Let us now apply the double-step DTs with basic SSs $\mathbf{b}$,0 and $\mathbf{d}$,0 to the GRef potential $_1V[z;\Diamond_o,\mu_o;1-b,b]$ with 1 < $\Diamond_o$ <2. The first step leads to HRef potentials $V[z\,|\,_1\mathbf{H}^{21}_{-}]$ or $V[z\,|\,_1\mathbf{H}^{21}_{+}]$ with the TWF functions given by (7.1.15*). It is essential that the SI for both potentials lies within the LC range: 0 < ${}^{*}\Diamond_o$ < $\Diamond_o$-1. The second DT converts the resultant HRef potential back into the GRef potential $_1V[z;2-\Diamond_o,\mu_o;1-b,b]$ while keeping the SI within the LC range. Making use of (4.26) and (4.26′) with $\varepsilon_{\mathbf{d},0}$ and $\varepsilon_{\mathbf{b},0}$ standing for $\varepsilon_{\mathbf{d}}$ and $\varepsilon_{\mathbf{b},m}$ the TMF function for the latter potential can be thus represented as

$$_1m_G(\varepsilon;1-{}^{*}\Diamond_o,\mu_o;b) = \frac{\varepsilon - {}_1\varepsilon_{\mathbf{b},0}(1-{}^{*}\Diamond_o,\mu_o;b)}{16\,{}^{*}\Diamond_o^2(1-{}^{*}\Diamond_o)^2}\ {}^{*}\ {}_1m_{+}^{-1}(\varepsilon;{}^{*}\Diamond_o,\mu_o;b) \qquad (7.1.32)$$

for ${}^{*}\Diamond_o < 1$

$$_1m_G(\varepsilon;1-{}^{*}\Diamond_o,\mu_o;b) = \frac{\varepsilon - {}_1\varepsilon_{\mathbf{d},0}(1-{}^{*}\Diamond_o,\mu_o;1-a)}{16\,{}^{*}\Diamond_o^2(1-{}^{*}\Diamond_o)^2}\ {}^{*}\ {}_1m_{-}^{-1}(\varepsilon;{}^{*}\Diamond_o,\mu_o;b) \qquad (7.1.32^{*})$$

for ${}^{*}\Diamond_o < 1$.

Taking into account that the DTs with the FFs $\mathbf{d}$,0 and $\mathbf{b}$,0 lead to the potentials with the with the lowest eigenvalues $_1\varepsilon_{\mathbf{b},0}$ and $_1\varepsilon_{\mathbf{b},1} > {}_1\varepsilon_{\mathbf{b},0}$, i.e., to $V[z\,|\,_1\mathbf{H}^{21}_{-}]$ and $V[z\,|\,_1\mathbf{H}^{21}_{+}]$, respectively, we find

$$_1\varepsilon_{\mathbf{b},0}(1-{}^{*}\Diamond_o,\mu_o;b) = -{}_1\nu_{+}^2({}^{*}\Diamond_o,\mu_o;b) \qquad (7.1.33)$$

and

$$_1\varepsilon_{\mathbf{d},0}(1-{}^{*}\Diamond_o,\mu_o;b) = -{}_1\nu_{-}^2({}^{*}\Diamond_o,\mu_o;b) \qquad (7.1.33^{*})$$

which brings us back to (7.1.15′).

Combining (7.1.15) and (7.1.15′).with (7.1.27) shows that the double-step DT of our interest should convert the TWF function of the potential ${}_1V[z;\lozenge_o,\mu_o;a,b]$ for $1 < \lozenge_o < 2$ in an atypical fashion:

$$ {}_1m_G(\varepsilon;\lozenge_o,\mu_o;b)\,{}_1m_G(\varepsilon;2-\lozenge_o,\mu_o;b) = -\frac{{}_1G_2^0(\varepsilon;1-\lozenge_o,\mu;b)}{64\lozenge_o^2(\lozenge_o-1)^2\;(\lozenge_o-2)^2} \qquad (7.1.34) $$

for $1 < \lozenge_o < 2$.

To derive this anomalous transformation rule from explicit expression (5.1.46) for the TWF functions in the left side of (7.1.34), let us first represent the latter as

$$ {}_1M_{0.0}(\varepsilon;1+\nu,\mu)\,{}_1M_{0.0}(\varepsilon;1-\nu,\mu) = \tfrac{1}{16}\,{}_1G(\varepsilon;\nu,\mu)\,, \qquad (7.1.35) $$

taking advantage of the identity

$$ \frac{\Gamma(-1-\lozenge_o)\Gamma(\lozenge_o-1)}{\Gamma(\lozenge_o+2)\Gamma(2-\lozenge_o)} = -\frac{1}{\lozenge_o^2(\lozenge_o^2-1)^2}\,. \qquad (7.1.36) $$

Substituting (7.1.22a) and (7.1.22b) into the right-hand side of (7.1.35) and re-arranging products (6.1.1) of $\alpha_{\pm,+}$- and $\beta_{\pm,+}$-coefficients into the quadratic polynomial ${}_1G(\varepsilon;\nu,\mu)$ decomposed according to (7.1.21′) we come to the symmetry relation:

$$ {}_1M_{0.0}(\varepsilon;\nu,\mu)\,{}_1M_{0.0}(\varepsilon;-\nu,\mu) = 1 \qquad (7.1.37) $$

directly followed from the interrelations:

$$ {}_1\alpha_{++}(\varepsilon;\nu,\mu) = {}_1\alpha_{-+}(\varepsilon;-\nu,\mu) \qquad (7.1.38a) $$

and

$$ {}_1\beta_{++}(\varepsilon;\nu,\mu) = {}_1\beta_{-+}(\varepsilon;-\nu,\mu)\,. \qquad (7.1.38b) $$

The author was unable to write the general relations between partner Jost functions for DTs keeping the SI within the LC range. However, since the Jost function for the resultant potential ${}_1V[z;2-\lozenge_o,\mu_o;a,b]$ is known one can explicitly verify that zeros of the function ${}_1f(k;2-\lozenge_o,\mu_o;b)$ defined via (7.1.16) coincide not with poles of TWF function (5.1.19) but with its zeros -- contrary to the conventional rule for the LP region.

In fact, by combining (6.1.1) with (7.1.36a) we find

$$\Gamma({}_1\alpha_{++}(k^2;2-\Diamond_o,\mu)=\Gamma({}_1\alpha_{-+}(k^2;\Diamond_o,\mu))/{}_1\alpha_{-+}(k^2;\Diamond_o-2,\mu)\,, \qquad (7.1.39)$$

as expected from the fact that discrete energy levels in question must coincide with the energies of AEH solutions b,m of the SL equation (5.1.5).

**7.2. SUSY doublets of radial potentials solvable via *c*-Heun functions**

Form-invariance of the radial c-GRef potential under double-step DTs with basic SSs of type c and a (or d and b) is performed in a very similar way. Namely, SL equation (7.6), with ι set to 0, has the basic solution

$${}^{H}_{0}\phi_{*\dagger,0}[z;{}^*\lambda_o,\nu_\dagger]=\sqrt{\frac{\zeta}{|\zeta-\zeta_T|}}\,\zeta^{-\frac{1}{2}\sigma_{0;\dagger}{}^*\lambda_o}\,e^{\frac{1}{2}\nu_\dagger\zeta} \qquad (7.2.1)$$

iff

$$O_1[\zeta\,|\,{}^{1}_{0}\mathcal{G}^{211}_{\dagger,0}]-b\nu^2_\dagger(\zeta-\zeta_T)=-2[(\zeta-\zeta_T)(1-\sigma_{0;\dagger}{}^*\lambda_o)\nu_\dagger+(1-\sigma_{0;\dagger}{}^*\lambda_o)\zeta-\nu_\dagger] \qquad (7.2.2)$$

so that the coefficients of the first-order polynomials

$$O_1[\zeta\,|\,{}^{1}_{0}\mathcal{G}^{211}_{\dagger,0}]=O_{1;1}\,|\,{}^{1}_{0}\mathcal{G}^{211}_{\dagger,0})(\zeta-\zeta_T)+o\,|\,{}^{1}_{0}\mathcal{G}^{211}_{\dagger,0}) \qquad (7.2.2')$$

are given by the relations:

$$O_{1;1}\,|\,{}^{1}_{0}\mathcal{G}^{211}_{\dagger,0})=b\nu^2_\dagger-2(1-\sigma_{0;\dagger}{}^*\lambda_o)\nu_\dagger+2(1-\sigma_{0;\dagger}{}^*\lambda_o) \qquad (7.2.3a)$$

and

$$o^{\pm}\,|\,{}^{1}_{0}\mathcal{G}^{211}_{\dagger,0})=2[\nu_\dagger+(1-\sigma_{0;\dagger}{}^*\lambda_o)\zeta_T] \qquad (7.2.3b)$$

Again, it is convenient to represent quadratic equations (5.2.15) and (6.2.14) for basic (m=0) solutions of type $\dagger_-$ or $\dagger_+$ in a uniform fashion

$$b\,{}_0\nu^2_{\dagger_\pm,0}(\lambda_o,g_o;b)+2\,{}_0\nu_{\dagger_\pm,0}(\lambda_o,g_o;b)(1\pm\lambda_o)+g_o=0\,, \qquad (7.2.4)$$

where

$$ {}_0\nu_-({}^*\lambda_o, g_o; b) = \nu_{\mathbf{a},0}({}^*\lambda_o - 1, g_o; b) \quad \text{for} \quad {}^*\lambda_o > 1, \tag{7.2.5a} $$

$$ = \nu_{\mathbf{d},0}({}^*\lambda_o + 1, g_o; b) \quad \text{for} \quad {}^*\lambda_o \neq 1 \tag{7.2.5d} $$

and

$$ {}_0\nu_+({}^*\lambda_o, g_o; b) = {}_0\nu_{\mathbf{c},0}({}^*\lambda_o - 1, g_o; b) \quad \text{for} \quad {}^*\lambda_o > 1, \tag{7.2.5c} $$

$$ = {}_0\nu_{\mathbf{b},0}({}^*\lambda_o + 1, g_o; b) \ ({}^*\lambda_o > 1) \quad \text{for} \quad {}^*\lambda_o \neq 1 \tag{7.2.5b} $$

are its lower and upper real roots. Comparing (7.2.4) with (7.2.3a) one can directly confirm that polynomials (7.2.2) has the common leading coefficient:

$$ O_{1;1} \mid {}_0\mathbf{H}^{21}_{\pm}) = g_o + 1 - {}^*\lambda_o^2 . \tag{7.2.6} $$

As for their free terms (7.2.3b), they can be represented as

$$ o \mid {}_0\mathbf{H}^{21}_{\pm}) = (1 \pm {}^*\lambda_o)\frac{b}{b-1} - {}_0\nu_{\pm}({}^*\lambda_o, g_o), \tag{7.2.7} $$

where ${}_0\nu_+({}^*\lambda_o, g_o)$ and ${}_0\nu_-({}^*\lambda_o, g_o)$ are upper and lower real roots of the quadratic equation

$$ b * \nu^2 - 2\nu * \lambda_o + {}^*\lambda_o^2 + g_o = 0 . \tag{7.2.8} $$

[For $g_o < -{}^*\lambda_o^2$ the subscript '+' or '–' in the notation ${}^*\nu_{\pm}({}^*\lambda_o, \mu_o)$ coincides with the root sign.] We thus proved that the radial potential $V[\zeta \mid {}_0G^{211}]$ has no more than two SUSY partners as asserted at the beginning of this Section. Since quadratic equation (7.2.8) has real roots iff

$$ \lambda_o^2 > b g_o \tag{7.2.9} $$

a pair of the HRef potentials exists for any negative value of $g_o$ .

Substituting (7.2.5a), (7.2.5d), and (7.2.5b) into (4.20), (4.20*), and (4.23) with $\varepsilon_{\mathbf{a},0}$, $\varepsilon_{\mathbf{d},0}$, and $\varepsilon_{\mathbf{b},0}$ standing for $\varepsilon_{\mathbf{a}}$, $\varepsilon_{\mathbf{d}}$, and $\varepsilon_{\mathbf{b},m}$, respectively, shows that dual representations (7.1.15) and (7.1.15*) for the TWF functions of the HRef potentials $V[z\,|\,{}_1\mathbf{H}_{+}^{21}]$ and $V[z\,|\,{}_1\mathbf{H}_{-}^{21}]$ remains valid for their *c*-counterparts:

$$ {}^{*}{}_0 m_{\pm}(\varepsilon;{}^{*}\Diamond_o,g_o;b) = -\frac{{}_0\nu_{\pm}^2({}^{*}\Diamond_o,g_o;b)+\varepsilon}{4\,{}^{*}\Diamond_o^2}\ {}_0 m_G(\varepsilon;{}^{*}\Diamond_o-1,g_o;b) \quad \text{for } {}^{*}\Diamond_o>1 \qquad (7.2.10)$$

$$ = -\frac{4({}^{*}\Diamond_o+1)^2}{{}_0\nu_{\mp}^2({}^{*}\Diamond_o,\mu_o;b)+\varepsilon}\ {}_0 m_G(\varepsilon;{}^{*}\Diamond_o+1,g_o;b) \quad \text{for } {}^{*}\Diamond_o\neq 1. \qquad (7.2.10^*)$$

Again, the first expression is inapplicable to the LC region and one has to use (4.22*) and (4.23) instead. This gives

$$ {}^{*}{}_0 m_{\pm}(\varepsilon;{}^{*}\Diamond_o,g_o;b) = \frac{\varepsilon-{}_0\nu_{\mp}^2({}^{*}\Diamond_o,g_o;b)}{16\,{}^{*}\Diamond_o^2(1-{}^{*}\Diamond_o)^2}\ {}_0 m_G^{-1}(\varepsilon;1-{}^{*}\Diamond_o,g_o;b) \quad \text{for } {}^{*}\Diamond_o<1 \qquad (7.2.10')$$

taking into account that

$$ \nu_{\mathbf{d},0}(1-{}^{*}\lambda_o,g_o;b)=\nu_{-}({}^{*}\lambda_o<1,g_o;b) \qquad (7.2.11d)$$

and

$$ \nu_{\mathbf{b},0}(1-{}^{*}\lambda_o,g_o;b)=\nu_{+}({}^{*}\lambda_o<1,g_o;b)\,. \qquad (7.2.11b)$$

The formulas for the Jost functions of the radial c-HRef potentials are derived in exactly the same way as (7.1.18a) and (7.1.18a) above which gives

$$ f(k;{}^{*}\Diamond_o,g_o;b\,|\,{}_0\mathbf{H}_{\pm}^{21}) = -\frac{2({}^{*}\Diamond_o-1)}{ik+{}_1\nu_{\pm}({}^{*}\Diamond_o,\mu_o;b)}\ {}_0 f(k;{}^{*}\Diamond_o-1,g_o;b) \quad \text{for } {}^{*}\Diamond_o>1 \qquad (7.2.12)$$

$$ = -\frac{ik-{}_1\nu_{\mp}({}^{*}\Diamond_o,g_o;b)}{2\,{}^{*}\Diamond_o}\ {}_0 f(k;{}^{*}\Diamond_o+1,g_o;b) \quad \text{for } {}^{*}\Diamond_o\neq 1, \qquad (7.2.12^*)$$

where the Jost function of the c-GRef potential is obtained via (A.11) in Appendix A:

$$ {}_0 f(k;\Diamond_o,g_o;b) = 2\sqrt[4]{k}\ e^{-ikr_o(b)-\frac{1}{2}i\pi\,\mu(-ik;g_o;b)}\,(2k)^{\frac{1}{4}ikb} \times \left(e^{\frac{1}{2}i\pi}b/k\right)^{\frac{1}{2}(\Diamond_o-1)}\ \frac{\Gamma(\Diamond_o+1)}{\Gamma\left(\alpha_{+}(\Diamond_o;\mu(-ik;g_o;b))\right)}. \qquad (7.2.13)$$

**i) Double-step DTs of types LP→LP→LP or LC→LP→LP**

$(1 < {}^{*}\lambda_o = {}^{*}\Diamond_o = \Diamond_o + 1, \;\; 2 < {}^{**}\lambda_o = {}^{**}\Diamond_o = \Diamond_o + 2)$

An analysis of the transformation property of TWF function (5.2.12) under action of the double-step DTs our interest is performed along the same line as done above for the regular case. First, making use of the ladder relations

$$ {}_0\alpha_{\pm}(\lambda_o + 2, \kappa) = {}_0\alpha_{\pm}(\lambda_o, \kappa) \pm 1, \tag{7.2.14}$$

similar to (7.1.22a) in the regular case, and taking into account that

$$ {}_0\alpha_{+}(\lambda_o;\kappa)[{}_0\alpha_{-}(\lambda_o;\kappa) - 1] = \kappa^2 - \tfrac{1}{4}(1+\lambda_o)^2 \tag{7.2.15}$$

one can verify that

$$ \frac{{}_0M_{0,0}\left(\lambda_o + 2; \kappa(g_0;b)\right)}{{}_0M_{0,0}\left(\lambda_o; \kappa(g_0;b)\right)} = -\frac{{}_0G_2^0(\varepsilon;\lambda_o + 1, g_0;b)}{16\varepsilon}, \tag{7.2.16}$$

where

$$ {}_0G_2^0(\varepsilon;\nu, g_0;b) \equiv (b\varepsilon - g_o)^2 + 4\nu^2\varepsilon. \tag{7.2.17}$$

Taking into account the identity

$$ \frac{\Diamond_o \Gamma(-\Diamond_o - 1)\Gamma(\Diamond_o + 1)}{(\Diamond_o + 2)\Gamma(\Diamond_o + 3)\Gamma(1 - \Diamond_o)} \equiv \frac{1}{(\Diamond_o + 2)^2 (\Diamond_o + 1)^2} \tag{7.2.18}$$

we find that the appropriate recurrence relation for TWF function (5.2.12) has the form very similar to (7.1.19)

$$ \frac{{}_0m_G(\varepsilon;\Diamond_o + 2, g_o;b)}{{}_0m_G(\varepsilon;\Diamond_o, g_o;b)} = -\frac{{}_0G_2^0(\varepsilon;\Diamond_o + 1, g_0;b)}{16b^2(\Diamond_o + 1)^2(\Diamond_o + 2)^2}. \tag{7.2.19}$$

By representing (7.2.4) as quadratic equations with respect to energies:

$$ {}_0G_2^0\left({}_0\varepsilon_{\dagger\pm,0}(\Diamond_o, \mu_o;b);\lambda_o \pm 1, g_0;b\right) = 0 \tag{7.2.20}$$

one can verify that the quadratic polynomial in the numerator of the fraction in the right-hand side of (7.2.16) has energies ${}_0\varepsilon_{\dagger_+,0}(\lambda_o, g_o; b)$ as its roots. (Again, since both leading coefficient and free term of this quadratic polynomial are positive it may have only two negative roots if any.) We thus conclude that (7.2.16) is indeed nothing but the condition for the radial *c*-GRef potential $V[\zeta\,|\,{}_0\mathcal{G}^{211}]$ to be form-invariant under the double-step DT with basic SSs of type $\mathsf{c}$ and $\mathsf{a}$.

Making use of (7.2.19) one can verify that dual representation (7.1.15) and (7.1.15*) for the TWF functions of the HRef potentials $V[z\,|\,{}_1\mathbf{H}_+^{21}]$ and $V[z\,|\,{}_1\mathbf{H}_-^{21}]$ remains valid for their *c*-counterparts:

$$\star{}_0m_\pm(\varepsilon;\star\lozenge_o, g_o; b) = -\frac{{}_0\nu_\pm^2(\star\lozenge_o, g_o; b)+\varepsilon}{4\star\lozenge_o^2}\,{}_0m_G(\varepsilon;\star\lozenge_o-1, g_o; b)$$

$$\text{for } \star\lozenge_o > 1 \tag{7.2.21}$$

$$= -\frac{4(\star\lozenge_o+1)^2}{{}_0\nu_\mp^2(\star\lozenge_o,\mu_o;b)+\varepsilon}\,{}_0m_G(\varepsilon;\star\lozenge_o+1, g_o; b) \quad \text{for } \star\lozenge_o \neq 1. \tag{7.2.21*}$$

Again, substitution of both (7.2.21) and a similar relation for $\star{}_0m_\pm(\varepsilon;\star\lozenge_o+2, g_o; b)$ with with $\lozenge_o = \star\lozenge_o - 1$ into (7.2.19), coupled with decomposition of quadratic polynomial (7.2.17) as

$${}_0G_2^0(\varepsilon;\star\nu, g_o; b) = b^2[{}_0\nu_+^2(\star\lozenge_o, g_o; b)+\varepsilon]\times[{}_0\nu_-^2(\star\lozenge_o, g_o; b)+\varepsilon], \tag{7.2.22}$$

leads to the following double-step recurrence relations for the TWF functions of $V[z\,|\,{}_0\mathbf{H}_+^{21}]$ and $V[z\,|\,{}_0\mathbf{H}_-^{21}]$:

$$\frac{\star{}_0m_\pm(\varepsilon;\star\lozenge_o+2, g_o; b)}{\star{}_0m_\pm(\varepsilon;\star\lozenge_o, g_o; b)} = \frac{[{}_0\nu_\pm^2(\star\lozenge_o+2, g_o; b)+\varepsilon]\times[{}_0\nu_\mp^2(\star\lozenge_o, g_o; b)+\varepsilon]}{16b^2(\star\lozenge_o+1)^2(\star\lozenge_o+2)^2}$$

$$\text{for } \star\lozenge_o > 1. \tag{7.2.23}$$

Similarly to the analysis presented in subsection 7.1, one can directly verify that the recurrence relation

$$\frac{{}_0f(k;\Diamond_o+2,g_o;b)}{{}_0f(k;\Diamond_o,g_o;b)} = -\frac{b(\Diamond_o+1)(\Diamond_o+2)}{i\,k[\,{}^1\!/_2(\Diamond_o+1)-\mu(-ik,g_o;b)]} \qquad (7.2.24)$$

matches the double-step transformation of the Jost function under Darboux deformations of the potential $V[\zeta\,|\,{}_0G^{211}]$ using basic SSs of type **c** and **a**. Indeed, representing quadratic equation (7.2.4) with respect to $\nu_{\mathbf{t}_+,0}$ as

$$\mu\Big(\nu_{\mathbf{t}_+,0}(\lambda_o,g_o;b)\Big) = {}^1\!/_2(\Diamond_o+1) \qquad (7.2.25)$$

shows that the denominator of the fraction in the right-hand side of (7.2.24) can be decomposed as

$$-ik[\,{}^1\!/_2(1+\lambda_o)-\mu(-ik;\lambda_o,g_o;b)] = b[-ik-\nu_{\mathbf{c},0}(\lambda_o,g_o;b)]\times[-ik-\nu_{\mathbf{a},0}(\lambda_o,g_o;b)] \qquad (7.2.26)$$

which completes the proof.

**ii) Double-step DTs of type LP→LC→LC**

**(**$1<\Diamond_o<2,\ {}^{\star}\Diamond_o=\Diamond_o-1<1,\ {}^{\star\star}\Diamond_o=2-\Diamond_o<1$)

Again, while (7.2.10*) are applicable to the whole range of ${}^{\star}\lambda_o$, except the boundary point ${}^{\star}\lambda_o = 1$ between the LC and LP regions, one has to re-write (7.2.10) in the LC region should be re-written as

$${}^{\star}{}_0m_+(\varepsilon;{}^{\star}\Diamond_o,g_o;b) = \frac{16\,{}^{\star}\Diamond_o^2(1-{}^{\star}\Diamond_o)^2}{\varepsilon-0\varepsilon_{\mathbf{b},0}(1-{}^{\star}\Diamond_o,g_o;b)}\ {}_0m_G^{-1}(\varepsilon;1-{}^{\star}\Diamond_o,g_o;b) \qquad (7.2.27a)$$

for ${}^{\star}\Diamond_o<1$

and

$${}^{\star}{}_0m_-(\varepsilon;{}^{\star}\Diamond_o,g_o;b) = \frac{16\,{}^{\star}\Diamond_o^2(1-{}^{\star}\Diamond_o)^2}{\varepsilon-{}_0\varepsilon_{\mathbf{d},0}(1-{}^{\star}\Diamond_o,g_o;b)}\ {}_1m_G^{-1}(\varepsilon;1-{}^{\star}\Diamond_o,g_o;b) \qquad (7.2.27c)$$

for ${}^{\star}\Diamond_o<1$,

making use of general expressions (4.26) and (4.26′).

Using (7.2.10) and (7.2.10*) one can also show that the double-step DT of our interest converts the TWF function of the potential ${}_0V[z;\lozenge_o,g_o;a,b]$ for $1<\lozenge_o<2$ in an atypical fashion:

$$ {}_0m_G(\varepsilon;\lozenge_o,g_o;b)\,{}_0m_G(\varepsilon;2-\lozenge_o,g_o;b)=-\frac{{}_0G_2^0(\varepsilon;1-\lozenge_o,g_o;b)}{64\lozenge_o^2(\lozenge_o-1)^2\,(\lozenge_o-2)^2} \tag{7.2.28}$$

for $1<\lozenge_o<2$.

To prove (7.2.28) based on explicit expression (5.2.12) for the TWF functions in the left side of (7.2.28), we again first represent it as

$$ {}_0M_{0.0}\left(\varepsilon;1+\nu,\kappa\right){}_0M_{0.0}\left(\varepsilon;1-\nu,\kappa\right)=-\frac{{}_0G(\varepsilon;\nu,\kappa)}{16\varepsilon}\ . \tag{7.2.29}$$

taking advantage of the identity

$$\frac{\Gamma(1-\lozenge_o)\Gamma(\lozenge_o-1)}{\Gamma(\lozenge_o+1)\Gamma(3-\lozenge_o)}=-\frac{1}{\lozenge_o(\lozenge_o-2)(\lozenge_o-1)^2}\,. \tag{7.2.30}$$

Making use of (7.2.14) followed by (7.2.15) and then decomposing the quadratic polynomial ${}_0G(\varepsilon;\nu,\kappa)$ according to (7.2.22), we find that symmetry relation (7.1.37) remains valid in the *c*-limit:

$$ {}_0M_{0.0}\left(\varepsilon;\nu,\kappa\right){}_0M_{0.0}\left(\varepsilon;-\nu,\kappa\right)=1\,. \tag{7.2.31}$$

as the straightforward consequence of the interrelation between $\alpha_+$- and $\alpha_-$-coefficients

$$ {}_1\alpha_+(\varepsilon;\nu,o)={}_1\alpha_-(\varepsilon;-\nu,\kappa)\,. \tag{7.2.32}$$

similar to (7.1.38a) for their regular counter-parts.

As mentioned above, the author was unable to write the general relations between partner Jost functions for DTs keeping the SI within the LC range. But again, since the Jost function for the resultant potential ${}_0V[z;2-\lozenge_o,g_o;a,b]$ is known can explicitly verify that zeros of the function ${}_0f(k;2-\lozenge_o,g_o;b)$ defined via (7.2.13) coincide not with poles of TWF function (5.2.12) but with its zeros -- contrary to the conventional rule for the LP region. In fact, by combining (7.2.14) with (7.2.32) we find

$$\Gamma\big(\alpha_+(2-\lozenge_o;\mu(-ik;g_o;b)\big)=\Gamma\big(\alpha_-(\lozenge_o;\mu(-ik;g_o;b)+1\big)\ . \tag{7.2.33}$$

Making use of (A.2) in Appendix A this confirms that discrete energy levels in the potential $_0V[z;2-\Diamond_o,g_o;a,b]$ coincide with the energies of AEH solutions b,m of Whittaker equation (5.2.6) in the general case b>0 (ℵ=1). [Some specific features of the E/MR potential (b=0) will be discussed in next Section.]

## 8. Transformation properties of TWF functions under basic Darboux deformations of shape-invariant reflective potentials on the half-line

The purpose of this Section is to illustrate the general analysis presented in Section 3 and 4 using three ‘explicit’ reflective radial potentials as illustrative examples. Namely, two (h-PT and E/MR) potentials solvable via hypergeometric functions as well as the centrifugal KC potential – the trivial representative of the c-GRef potentials which does not require any additional change of variable. Since all three potentials preserve their form under DTs with basic FFs the appropriate TWF functions must satisfy the following ladder relations.

$$_\iota m(\varepsilon;\Diamond_o+1,\nu_{o,\iota};b)=\frac{\varepsilon-{}_\iota\varepsilon_{\dagger_+,0}}{4\Diamond_o(\Diamond_o+1)}\,{}_\iota m(\varepsilon;\Diamond_o,\nu_{o,\iota};b) \quad \text{for any } \Diamond_o>0, \tag{8.1a}$$

$$_\iota m(\varepsilon;\Diamond_o-1,\nu_{o,\iota};b)=\frac{4\Diamond_o(\Diamond_o-1)}{\varepsilon-{}_\iota\varepsilon_{\dagger_-,0}}\,{}_\iota m(\varepsilon;\Diamond_o,\nu_{o,\iota};b) \quad \text{for } \Diamond_o>1, \tag{8.1b}$$

and

$$^*{}_\iota m(\varepsilon;\Diamond_o,\nu_{o,\iota};b)=\frac{\varepsilon-{}_\iota\varepsilon_{\dagger_-,0}}{4\Diamond_o(\Diamond_o-1)\,{}_\iota m(\varepsilon;\Diamond_o,\nu_{o,\iota};b)} \quad \text{for } 0<\Diamond<1 \tag{8.1b*}$$

where $\nu_{o,\iota}=\mu_o$ or $g_o$ for ι=1 (b = 0 or 1) or ι=0 (b = 0), respectively (b = 0). For completeness we also present similar ladder relations for the appropriate Jost functions in cases when either initial or final value of the SI lies within the LP range:

$$_\iota f(k;\Diamond_o+1,\nu_{o,\iota};b)=\frac{2\Diamond_o}{-\sigma_{1;\dagger_+}\sqrt{-\varepsilon_{\dagger_+}}-ik}\,{}_\iota f(k;\Diamond_o,\nu_{o,\iota};b), \tag{8.2a}$$

$$_\iota f(k;\Diamond_o-1,\nu_{o,\iota};b)=\frac{\sigma_{1;\dagger_-}\sqrt{-\varepsilon_{\dagger_-,0}}-ik}{2(\Diamond_o-1)}\,{}_\iota f(k;\Diamond_o,\nu_{o,\iota};b) \quad \text{for } \Diamond_o>1. \tag{8.2b}$$

## 8.1 Reflective radial potentials with exponential tails

### I) The h-PT potential

As initially found in our earlier works [21, 49], the Schrödinger equation with the h-PT potential

$$V_{s,t}(r) = \frac{s(s-1)}{sh^2 r} - \frac{(s+t)(s+t-1)}{ch^2 r} \qquad (8.1.1)$$

$$= \frac{\lambda_o^2 - 1}{sh^2 r} - \frac{\mu_o^2 - 1}{ch^2 r}, \qquad (8.1.1')$$

where $s \equiv s_o = \lambda_o + 1/2 > 1/2$, $s+t \equiv s_o + t_o = \mu_o + 1/2 > 1/2$ in our current terms, has 4 basic solutions

$$\psi_{a}(r;s,t) = sh^{s} r\ ch^{s+t} r, \qquad (8.1.2a)$$

$$\psi_{t_-}(r;s,t) = sh^{1-s} r\ ch^{1-s-t} r \ \ (t_- = b \text{ for } t > 2-2s \text{ and } t_- = d' \text{ otherwise}) \qquad (8.1.2b)$$

$$\psi_{t_+}(r;s,t) = sh^{s} r\ ch^{1-s-t} r \ \ (t_+ = c \text{ for } t > 1 \text{ and } t_+ = a' \text{ otherwise}), \qquad (8.1.2c)$$

$$\psi_{d}(r;s,t) = sh^{1-s} r\ ch^{s+t} r \ \ (t_- = d \text{ for } t > -1 \text{ and } t_- = b' \text{ otherwise}) \qquad (8.1.2d)$$

such that the DT using one of these solutions as the FF converts the h-PT potential onto itself. It was years before Gendenshtein [30] put forward the suggestion that the form-invariance of the generic potential under the DT using the ground-state eigenfunctions as the FF is the intrinsic feature of the exact solvability. This claim was quickly refuted by Cooper, Ginocchio, and Khare [61] who explicitly demonstrated that the potential $V[z|_1G^{220}]$ in our current terms does not retain its form under action of the DT in question -- the straightforward consequence of the fact [35] that the density function $\wp[z]$ in the appropriate rational SL equation analytically extended to the finite complex plane has zeros outside the quantization interval [0 , 1], except a few specific cases [16] referred to below as '*shape-invariant*' GRef potentials. The common feature of the latter potentials is that the appropriate density functions $\wp_G[z;a,b,c_0]$ remain finite when continued into the complex plane. As a result each *shape-invariant* GRef potential retains its form under action of any of four DTs with basic FFs.

In case of radial *r*-GRef potentials ($c_0 = 0$) the outer singularity in (5.1.1) disappears if either a=0 or b=0, leading to two (h-PT and EMR, respectively) shape-invariant potentials of our current interest. An analysis of quadratic equations (7.1.5) with b = 1 reveals the following possible cases for the roots $\lambda_{1;\mathbf{t}_{\pm},0}$:

$$_1\lambda_{1;\mathbf{a},0} = -\lambda_o - 1 - \mu_o < 0 \leftrightarrow {}_1\beta_{+,-}({}_1\varepsilon_{\mathbf{a},0};\lambda_o,\mu_o) = 0, \quad (8.1.3a)$$

$$_1\lambda_{1;\mathbf{a}',0} = -\lambda_o - 1 + \mu_o < 0 \leftrightarrow {}_1\alpha_{+,-}({}_1\varepsilon_{\mathbf{a}',0};\lambda_o,\mu_o) = 0, \quad (8.1.3a')$$

$$_1\lambda_{1;\mathbf{b},0} = \lambda_o - 1 + \mu_o > 0 \leftrightarrow {}_1\alpha_{-,+}({}_1\varepsilon_{\mathbf{b},0};\lambda_o,\mu_o) = 0, \quad (8.1.3b)$$

$$_1\lambda_{1;\mathbf{b}',0} = \lambda_o - 1 - \mu_o > 0 \text{ (LP)} \leftrightarrow {}_1\beta_{-,+}({}_1\varepsilon_{\mathbf{b}',0};\lambda_o,\mu_o) = 0, \quad (8.1.3b')$$

$$_1\lambda_{1;\mathbf{c},0} = -\lambda_o - 1 + \mu_o > 0 \leftrightarrow {}_1\alpha_{+,+}({}_1\varepsilon_{\mathbf{c},0};\lambda_o,\mu_o) = 0, \quad (8.1.3c)$$

$$_1\lambda_{1;\mathbf{d},0} = \lambda_o - 1 - \mu_o < 0 \leftrightarrow {}_1\beta_{-,-}({}_1\varepsilon_{\mathbf{d},0};\lambda_o,\mu_o) = 0, \quad (8.1.3d)$$

$$_1\lambda_{1;\mathbf{d}',0} = \lambda_o - 1 + \mu_o < 0 \leftrightarrow {}_1\alpha_{-,-}({}_1\varepsilon_{\mathbf{d}',0};\lambda_o,\mu_o) = 0, \quad (8.1.3d')$$

where the α- and β-coefficients are defined via (5.1.20a) and (5.1.20b), accordingly.

The second basic solution $\mathbf{a}',0$ obviously exists only in the region $\mu_o < \lambda_o + 1$, where the h-PT potential does not have the discrete energy spectrum. The pair of the basic solutions $\mathbf{b},0$ and $\mathbf{d},0$ co-exist iff $\mu_o > |\lambda_o - 1|$.

Keeping in mind that the FF for the reverse DT has the form

$$^{\star}\psi_{\mathbf{t}}(r;s,t) = \psi_{\mathbf{t}}^{-1}(r;s,t) \quad (8.1.4)$$

one finds

$$\lambda_{o;\mathbf{a}} = \lambda_o + 1, \; \mu_{o;\mathbf{a}} = \mu_o + 1; \quad (8.1.5a)$$

$$\lambda_{o;\mathbf{a}'} = \lambda_o + 1, \; \mu_{o;\mathbf{a}'} = \begin{cases} \mu_o - 1 \text{ for } \mu_o > 1 \\ \text{or} \\ 1 - \mu_o \text{ for } \mu_o < 1; \end{cases} \quad (8.1.5a')$$

$$\lambda_{o;\mathbf{b}} = \begin{cases} \lambda_o - 1 \text{ for } \lambda_o > 1 \\ \text{or} \\ 1 - \lambda_o \text{ for } \lambda_o < 1, \end{cases} \quad \mu_{o;\mathbf{b}} = \begin{cases} \mu_o - 1 \text{ for } \mu_o > 1 \\ \text{or} \\ 1 - \mu_o \text{ for } \mu_o < 1; \end{cases} \tag{8.1.5b}$$

$$\lambda_{o;\mathbf{b}'} = \lambda_o - 1 > 0, \quad \mu_{o;\mathbf{b}'} = \mu_o + 1; \tag{8.1.5b$'$}$$

$$\lambda_{o;\mathbf{c}} = \lambda_o + 1, \quad \mu_{o;\mathbf{c}} = \mu_o - 1 > 0; \tag{8.1.5c}$$

$$\lambda_{o;\mathbf{d}} = \begin{cases} \lambda_o - 1 \text{ for } \lambda_o > 1 \\ \text{or} \\ 1 - \lambda_o \text{ for } \lambda_o < 1, \end{cases} \quad \mu_{o;\mathbf{d}} = \mu_o + 1; \tag{8.1.5d}$$

$$\lambda_{o;\mathbf{d}'} = 1 - \lambda_o > 0, \quad \mu_{o;\mathbf{d}'} = 1 - \mu_o > 0 \tag{8.1.5d$'$}$$

(There is a misprint in (4.2.6b) in [15]: the sum $\lambda_o + \mu_o$ is equal to $\lambda_{o;\tilde{\mathbf{t}}_-} + \mu_{o;\tilde{\mathbf{t}}_-} + 2$.)

Now we are ready to discuss the crucial gap in the arguments presented in the pioneering work of Gangopadhyaya, Panigrahi, and Sukhatme [1] (see also Chapter 12 in [10]) which prevented them from discovering the anomaly of our interest.. An analysis of superpotential (9) used to generate the appropriate DT in [1] shows that it is nothing but the logarithmic derivative of lowest-energy eigenfunction (8.1.2c) taken with opposite sign ($B = s$, $B = s + t - 1$). The inverse DT has basic solution (8.1.2d) as its FF. The important element overlooked both in [1] and later in {10] is that the larger ChExp, B+1, for the potential $V_+(r)$ in the pair of SUSY partners (10) in [1] is restricted to the LP range by the very way in which this potential was constructed. On other hand, basic solution (8.1.2d) also exist in the LC region. When applied to the h-PT potential in the LC region, the DT using the latter solution as its FF keeps the SI $0 < \Diamond_o = B - \frac{1}{2} < 1$ within the LC range: $\Diamond_o \to 1 - \Diamond_o$ $(B \to 2 - B)$. The inverse of the DT with the basic FF of type $\mathbf{d}$ has the FF of type $\mathbf{b}$. when applied to the h-PT potential in the LC region. This anomalous pair of SUSY partners with completely different energy spectra was simply omitted from their analysis.

It directly follows from the expressions for the transformed energy-dependent coefficients ${}_1\alpha_\sigma(\varepsilon;\lambda_{o;\mathbf{t}},\mu_{o;\mathbf{t}})$ and ${}_1\beta_\sigma(\varepsilon;\lambda_{o;\mathbf{t}},\mu_{o;\mathbf{t}})$ listed in Appendix B that

$$M(\varepsilon;\lambda_{o;\mathbf{b}},\mu_{o;\mathbf{b}}) = \frac{1}{{}_1\alpha_{--}(\varepsilon;\lambda_o,\mu_o)\,{}_1\alpha_{-+}(\varepsilon;\lambda_o,\mu_o)} M(\varepsilon;\lambda_o,\mu_o) \quad (8.1.6)$$

for $\mu_o > 1$

and

$$M(\varepsilon;\lambda_{o;\mathbf{t}_},\mu_{o;\mathbf{t}_}) \equiv \frac{1}{{}_1\beta_{--}(\varepsilon;\lambda_o,\mu_o)\,{}_1\beta_{-+}(\varepsilon;\lambda_o,\mu_o)} M(\varepsilon;\lambda_o,\mu_o) \quad (8.1.6')$$

for $\mathbf{t}_ = \mathbf{b}$, $\mu_o < 1$, $\mathbf{t}_ = \mathbf{b}'$ or $\mathbf{d}$

within the LP range ($\lambda_o > 1$). On other hand,

$$M(\varepsilon;\lambda_{o;\mathbf{t}_},\mu_{o;\mathbf{t}_}) \equiv \frac{{}_1\beta_{--}(\varepsilon;\lambda_o,\mu_o)\,{}_1\beta_{-+}(\varepsilon;\lambda_o,\mu_o)}{M(\varepsilon;\lambda_o,\mu_o)} \quad (8.1.7)$$

for $\mathbf{t}_ = \mathbf{b}$, $\mu_o > 1$ or $\mathbf{t}_ = \mathbf{d}$

and

$$M(\varepsilon;\lambda_{o;\mathbf{t}_},\mu_{o;\mathbf{t}_}) \equiv \frac{{}_1\alpha_{--}(\varepsilon;\lambda_o,\mu_o)\,{}_1\alpha_{-+}(\varepsilon;\lambda_o,\mu_o)}{M(\varepsilon;\lambda_o,\mu_o)} \quad (8.1.7')$$

for $\mathbf{t}_ = \mathbf{b}$, $\mu_o < 1$ or $\mathbf{t}_ = \mathbf{d}'$.

An analysis of (8.1.3b), (8.1.3b′), (8.1.3d), and (8.1.3d′) shows that

$${}_1\alpha_{--}(\varepsilon;\lambda_o,\mu_o)\,{}_1\alpha_{-+}(\varepsilon;\lambda_o,\mu_o) = \begin{cases} \varepsilon - {}_1\varepsilon_{\mathbf{b};0} & \text{for } \lambda_o + \mu_o > 1, \\ \varepsilon - {}_1\varepsilon_{\mathbf{d}';0} & \text{otherwise} \end{cases} \quad (8.1.8b)$$

and

$${}_1\beta_{--}(\varepsilon;\lambda_o,\mu_o)\,{}_1\beta_{-+}(\varepsilon;\lambda_o,\mu_o) = \begin{cases} \varepsilon - {}_1\varepsilon_{\mathbf{d};0} & \text{for } \lambda_o < \mu_o + 1, \\ \varepsilon - {}_1\varepsilon_{\mathbf{b}';0} & \text{otherwise.} \end{cases} \quad (8.1.8d)$$

(Similar relations can be written for R@O basic solutions but they are of no need for our current discussion.) Combining (8.1.8b) and (8.1.8d) with the energy-independent relations

$$\frac{\Gamma(-\lambda_{o;\mathbf{t}_})}{\Gamma(\lambda_{o;\mathbf{t}_})} = \frac{\Gamma(1-\lambda_o)}{\Gamma(\lambda_o - 1)} = -\lambda_o(\lambda_o - 1)\frac{\Gamma(-\lambda_o)}{\Gamma(\lambda_o)} \quad \text{for } \lambda_o > 1, \quad (8.1.9)$$

$$\frac{\Gamma(-\lambda_{o;\dagger_-})}{\Gamma(\lambda_{o;\dagger_-})} = \frac{\Gamma(\lambda_o - 1)}{\Gamma(1-\lambda_o)} = -\frac{1}{\lambda_o(\lambda_o - 1)}\frac{\Gamma(\lambda_o)}{\Gamma(-\lambda_o)} \quad \text{for } \lambda_o < 1 \tag{8.1.9*}$$

one can directly verify that the TWF functions in question satisfy symmetry relations (8.1b) and (8.1b*), as expected. As for symmetry relation (8.1a), it is simply the reverse of (8.1b) and therefore does not require a separate proof.

A similar analysis can be performed for Jost function (7.1.16) in the limiting case a=0, b=1:

$$f(k;\Diamond_o,\mu_o;1) = \frac{2e^{-ik\,r_o}\;\Gamma(1-ik)\Gamma(\Diamond_o+1)}{\Gamma\big({}_1\alpha_{+,+}(k^2;\Diamond_o,\mu_o)\big)\Gamma\big({}_1\beta_{+,+}(k^2;\Diamond_o,\mu_o)\big)}. \tag{8.1.10}$$

Making use of (B.3b) and (B*.3b) in Appendix B shows that

$$f(k;\Diamond_o+1,\mu_o-\sigma_{1;\dagger_+};1\,|\,\dagger_+;\varepsilon_{\dagger_+,0}) = -\frac{2s_o+1}{i\,k+\sigma_{1;\dagger_+}\sqrt{-\varepsilon_{\dagger_+,0}}}f(k;\Diamond_o,\mu_o;1) \tag{8.1.11}$$

in agreement with our original results [21] for the transformation properties of Jost functions under Darboux deformations of IS@O potential (3.2) in the LP region.

On other hand,

$$f(k;1-\Diamond_o;\mu_o-\sigma_{1;\dagger_-};1)$$

$$= \frac{2e^{-ik\,r_o}\;\Gamma(1-ik)\Gamma(2-\Diamond_o)}{\Gamma\big({}_1\alpha_{-+}(k^2;1-\Diamond_o,\mu_o-\sigma_{1;\dagger_-})\big)\Gamma\big({}_1\beta_{-+}(k^2;1-\Diamond_o,\mu_o-\sigma_{1;\dagger_-})+1\big)}$$

$$\text{for } \dagger_- = \mathsf{b},\ \mu_o > 1 \ \text{ or } \ \dagger_- = \mathsf{d} \tag{8.1.12}$$

and

$$f(k;1-\Diamond_o;1-\mu_o;1) = \frac{2e^{-ik\,r_o}\;\Gamma(1-ik)\Gamma(2-\Diamond_o)}{\Gamma\big({}_1\beta_{-+}(k^2;1-\Diamond_o,1-\mu_o)\big)\Gamma\big({}_1\alpha_{-+}(k^2;1-\Diamond_o,1-\mu_o)+1\big)}$$

$$\text{for } \dagger_- = \mathsf{b},\ \mu_o < 1 \ \text{ or } \ \dagger_- = \mathsf{d}'. \tag{8.1.12′}$$

within the LC range of the SI, $0 < \Diamond_o < 1$. One can easily verify that the function ${}_1\beta_{-+}(k^2;1-\Diamond_o,\mu)$ for $\mu > 0$ is positive on the upper part of the imaginary axis so that discrete energy states in transformed potential (3.2), with $V(r;\Diamond_o)$ standing for h-PT potential (8.1.1′), are determined by the conditions:

$$
{}_1\alpha_{-+}({}^{*}\varepsilon_{c;v};\Diamond_o,\mu_o-\sigma_{1;t_-}) \equiv \tfrac{1}{2}(-\lambda_o+\sigma_{1;t_-}\sqrt{-{}^{*}\varepsilon_{c;v}}+1-\mu_o+\sigma_{1;t_-}) = -v
$$

$$
\text{for } t_- = b,\ \mu_o > 1 \text{ or } t_- = d \qquad (8.1.13)
$$

Unless $\Diamond_o = \tfrac{1}{2}$ the energy levels calculated via (8.1.13) have no correlation with the discrete energy spectrum given by the conventional formula:

$$
{}_1\alpha_{++}(\varepsilon_{c;v};\Diamond_o,\mu_o) \equiv \tfrac{1}{2}(\sqrt{-\varepsilon_{c;v}}-\lambda_o-\mu_o+\sigma_{1;t_-}) = -v \qquad (8.1.13^{*})
$$

for potential (8.1.1′),

Before repeating similar arguments for the E/MR potential let us point to a very specific feature of the h-PT potential which distinguishes the latter from any other radial *r*-GRef potential. Namely, linear-fractional transformation

$$
\breve{z}(r) = \frac{z(r)}{z(r)-1} \qquad (8.1.14)
$$

makes energy-independent the characteristic exponents near the singular points of the quantization interval. As a result [15] the potential is quantized via a finite set of orthogonal ‘Romanovski-Jacobi’ polynomials [68, 69]. The remarkable consequence of this observation first noticed by Odake and Sasaki [70, 71] is that rational SUSY partners of the h-PT potential are also quantized via finite sets of orthogonal polynomials referred to below as ‘orthogonal GS Heine polynomials’.

To relate the general analysis performed in previous Section to our previous results [15] using variable (8.1.14), first note that quadratic equations (5.1.29) and (6.1.19) share the same discriminant at the limit b=1:

$$
{}_1\Delta_{\pm;m}(\lambda_o,\mu_o;1) = 4\mu_o^2 \qquad (8.1.15)
$$

so that the pairs of positive roots defined via (5.1.30), (5.1.30′) and (6.1.22), (6.1.22′) take the form:

$$\lambda_{1;\mathbf{a},m} = -2m-1-\lambda_o-\mu_o \quad \text{for m=0, 1, 2,... ,} \tag{8.1.16a}$$

$$\lambda_{1;\mathbf{a}',m} = \mu_o - 2m-1-\lambda_o \quad \text{for } 2m > \mu_o - \lambda_o - 1, \tag{8.1.16a$'$}$$

and

$$\lambda_{1;\mathbf{b},m} = \mu_o + \lambda_o - 2m - 1 > 0 \qquad \text{for } m < \lambda_o \tag{8.1.16b}$$

$$\lambda_{1;\mathbf{b}',m} = \lambda_o - \mu_o - 2m - 1 > 0, \tag{8.1.16b$'$}$$

respectively. As already mentioned in subsection 6.1, the sequence of R@ ∞ Frobenius solutions $\mathbf{b}'$,m ($\tilde{\mathbf{b}}'$,m in [15]) exists only in the part of the LP region with no discrete energy spectrum and therefore cannot be used to construct the orthogonal GS Heine polynomials in question.

The condition $m < |\lambda_{1;\mathbf{a},m}|$ for the Jacobi polynomials with indexes $\lambda_o > 0$ and $\lambda_{1;\mathbf{a},m} < 0$ not to have zeros between -1 and 1 is automatically fulfilled for the sequence (8.1.16a). Similarly, the constraint

$$m < |\lambda_{1;\mathbf{a}',m}| = 2m + 1 + \lambda_o - \mu_o \tag{8.1.17}$$

directly followed from (6.1.16) for b=1 leads to condition (4.2.15a′) in [15] which determines another infinite (secondary) sequence of nodeless R@O AEH solutions.

The corresponding sequences of orthogonal GS Heine polynomials are obtained via the relations

$$\mathrm{Hi}_{m+v}[\tilde{z}\,|\,{}^{1}_{1}\breve{\mathcal{G}}^{201}_{\mathbf{t}_+,m};\mathbf{c},v] = c_{\mathbf{t}_+,m;\mathbf{c},v}(\breve{z}-1)^{m+v} \times \mathrm{Hi}_{m+v}[\breve{z}/(\breve{z}-1)\,|\,{}^{1}_{1}\mathcal{G}^{101}_{\mathbf{t}_+,m};\mathbf{c},v] \quad \text{for } \mathbf{t}_+ = \mathbf{a} \text{ or } \mathbf{a}', \tag{8.1.18a}$$

where we use symbol $\breve{\mathcal{G}}$ in the right-hand side of (8.1.18), instead of $\tilde{\mathcal{G}}$, to indicate that we deal with the PF beam defined on the negative semi-axis $\breve{z} < 0$. The GS Heine polynomials in the right-hand side are obtained via (6.1.17) and the scale factor is computed from the requirement that the leading coefficient of polynomial of (8.1.18a) is equal to 1. The sequence of orthogonal GS Heine polynomials $\mathrm{Hi}[\tilde{z}\,|\,{}^{1}_{1}\breve{\mathcal{G}}^{201}_{\mathbf{a}',m};\mathbf{c},v]$

has been recently constructed by Grandati [56]. A more detailed analysis between our and his results is postponed for a separate publication.

Finally, (8.1.16b) defines a *finite* sequence of nodeless AEH solutions which, according to (4.2.13b) in [15], all lie below the lowest discrete energy level, as expected. Within the LP range of the SI $\Diamond_o$ the corresponding sequences of orthogonal GS Heine polynomials can be constructed via the relations

$$\mathrm{Hi}_{m+v+1}[\breve{z}\,|\,{}^{1}_{1}\breve{G}^{201}_{\mathbf{b},m};\mathbf{c},v] = c_{\mathbf{b},m;\mathbf{c},v}(\breve{z}-1)^{m+v+1} \times \mathrm{Hi}_{m+v+1}[\breve{z}/(\breve{z}-1)\,|\,{}^{1}_{1}G^{101}_{\mathbf{b},m};\mathbf{c},v] \text{ for } \lambda_o > 1, \tag{8.1.18b}$$

where the GS Heine polynomials in the right-hand side are obtained via (6.1.28).

$$\mathrm{Hi}_{m+m'}[\breve{z}\,|\,{}^{1}_{1}\breve{G}^{201}_{\mathbf{b},m};\mathbf{c},m'+\theta_{m'-m}] = c_{\mathbf{b},m;\mathbf{c},v}(\breve{z}-1)^{m+m'} \times \mathrm{Hi}_{m+m'}[\breve{z}/(\breve{z}-1)\,|\,{}^{1}_{1}G^{101}_{\mathbf{b},m};\mathbf{c},m'+\theta_{m'-m}] \text{ for } 0<\lambda_o<1. \tag{8.1.18b*}$$

An analysis of the indexes of Jacobi polynomials (42) in [**70**] shows that Odake and Sasaki deal with the sequence $\mathbf{b}$,m, namely,

$$g_\ell = \lambda_o - \ell - 1/2,\; h_\ell = \mu_o + \ell - 3/2,$$

where we added subscript $\ell$ to their parameters g and h to stress that both depend on the order of the Jacobi polynomial used to construct appropriate AEH solution. Contrary to the statement made in [70] and repeated later in [72], the number of the potentials in the sequence is equal not to the number of bound states, but to $[\lambda_o]$. As already mentioned [15] in connection with rational SUSY partners of the trigonometric version of the PT potential, Odake and Sasaki in their breakthrough study [70] simply assumed without a proof that the appropriate Jacobi polynomials do not have zeros within the quantization interval. In the following paper [71] they tried to justify this assumption by pointing to the fact that all the terms in the appropriate hypergeometric polynomial are all positive if its argument is restricted to the negative semi-axis. However, the analysis of the parameters of the hypergeometric polynomial: $\alpha = -\ell$, $\beta = -\mu_o + \lambda_o + \ell - 1$, and $\gamma = 1 - \lambda_o$ shows that both $\beta + k$ and $\gamma + k$ retain negative for $0 \le k < \ell$ iff $\ell < \lambda_o$ and $0 \le 2\ell < \mu_o - \lambda_o + 1)$ so that $\ell$ must indeed be smaller than $\lambda_o$, in agreement with our results.

There may be also sub-sequences of orthogonal GS Heine polynomials associated with nodeless AEH solutions of types **d** and **d′** ; however existence of the latter solutions requires a more specific analysis beyond the scope of this paper.

**II) The E/MR potential**

So far we completely stayed away from any discussion of the E/MR potential (b=0, a=1)

$$V[z(r) \mid {}_1\mathcal{G}^{202}] \equiv V_{E/MR}[z(r); s_o, \mu_o] \equiv \tfrac{1}{2}\mu_o^2 [1 - cth\, r] + \frac{s_o(s_o - 1)}{sh^2 r} \qquad (8.1.19)$$

(using Infeld and Hull's representation[x)] [73]) . The main reason for such a discriminatory treatment of this potential is that the variable z(r) in the limit: b=0, a=1:

$$z(r) = 1 - e^{-2r} \qquad (8.1.20)$$

becomes proportional to r (instead of $\sqrt{r}$ ) for r <<1 and as a result one has to distinguish between the exponent difference $\lambda_o = 2s_o - 1 > 0$ [17, 18] at the singular point z = 0 of SL equation (5.1.5) and the potential SI, $\Diamond_o = s_o - \tfrac{1}{2} > 0$, namely, $\lambda_o = 2\Diamond_o$ if b=0. *This explains why we kept interchanging the parameters* $\Diamond_o$ *and* $\lambda_o$ *in different formulas for* b > 0, *except simply using one of them.*

Since z $\approx$ 2 r near the origin, two Fuschian solutions of the Schrödinger equation with potential (8.1.19) are expressed in terms of Fuschian solutions (5.1.11) and (5.1.11*) of

---

x) In Infeld and Hull's notation $\beta = s_o$ and $\nu = \tfrac{1}{4}\mu_o^2$. Note also that the centrifugal term in our expression (15) for this potential in [16] has wrong sign ($\lambda_o^2 = h_0 + 1$, $h_1 = -1$, $\mu_o^2 = f + 1$ in our current terms).

SL equation (5.1.5) in the straightforward way:

$$_1\psi_{\mathbf{a}}(r;\varepsilon;\Diamond_o,\mu_o;0) = 2^{\Diamond_o}\ {}_1\phi_{\mathbf{a}}[z(r);\varepsilon;2\Diamond_o,\mu_o;0]/\sqrt{1-z(r)} \tag{8.1.21}$$

and

$$_1\psi_{\mathbf{d}}(r;\varepsilon;\Diamond_o,\mu_o;0) = 2^{-\Diamond_o}\ {}_1\phi_{\mathbf{d}}[z(r);\varepsilon;2\Diamond_o,\mu_o;0]/\sqrt{1-z(r)}\,. \tag{8.1.21*}$$

Substituting (8.1.21) and (8.1.21*) into the right-hand side of (4.1) and comparing the resultant expression with (5.1.27) one finds

$$_1\psi_{\mathbf{b}}(r;\varepsilon;\Diamond_o,\mu_o;0) = 2^{-\frac{1}{2}\Diamond_o}\ {}_1\phi_{\mathbf{b}}[z(r);\varepsilon;2\Diamond_o,\mu_o;0]/\sqrt{1-z(r)} \tag{8.1.22}$$

or, combined with (5.1.28),

$$_1m(\varepsilon;\Diamond_o,\mu_o;0) = 2^{2\Diamond_o}\ {}_1m_0\left(\varepsilon;2\Diamond_o,\sqrt{\mu_o^2-\varepsilon}\right), \tag{8.1.23}$$

where the auxiliary function $_1m_0(\varepsilon;\lambda_o,\mu)$ is defined via (5.1.17). It needs to be stressed that the derived expression for the TWF function of the E/MR potential cannot be obtained from (5.1.45) by setting b equal to 0.

Since the common leading coefficient of quadratic equations (5.1.29) and (6.1.19) vanishes for b=0 the energies of AEH solutions are described by the elementary formulas

$$_1\varepsilon_{\mathbf{t}_\pm,m}(\lambda_o,\mu_o) = -\lambda^2_{1;\mathbf{t}_\pm,m}, \tag{8.1.24}$$

where

$$\lambda_{1;\mathbf{t}_\pm,m} = \frac{\mu_o^2}{2(2m+1\pm\lambda_o)} - \tfrac{1}{2}(2m+1\pm\lambda_o)\,. \tag{8.1.25}$$

These AEH solutions were recently used by Quesne [34] and Grandati [56] to construct rational SUSY partners of the E/MR potential quantized via 'Gauss-seed Heine polynomials' [15]. An analysis of (3.1) in [34] shows that $s_o = A$, $\lambda_o + 1 = 2A$, and $\mu_o = 2\sqrt{B}$ (A = *a* +1, B = b in [56]) so that the condition for the E/MR potential to have the discrete energy spectrum is B > $A^2$ in Quesne's notation. However a direct correlation between her Types I, II and III and our labels **a**, **b**, and **d** is slightly complicated by the fact that (in following [74, 12, 13]) both she and Grandati [56] express Jacobi polynomials in terms of another variable

$$\tilde{\eta}(r) = \coth r = \frac{1+e^{-2r}}{1-e^{-2r}} = \frac{2-z(r)}{z(r)}, \tag{8.1.26}$$

so that the appropriate hypergeometric functions have $\tilde{z}(r) = z^{-1}(r)$ as their argument. The quantization interval is then defined as (1, +∞). (Instead of using the hypergeometric equation in $\tilde{z}$, Quesne [34] considers the second- order equation with irregular singular points both at 0 and 1 which brings an additional complication to the comparison.) One can directly confirm that solutions given by (3.3) and (3.4) behave near the origin as $\coth^{-A} r$ and $\coth^{A-1} r$, respectively, so that Quesne's Type I stands for solutions regular at the origin (type $\mathbf{a}$ in our terms) whereas Types II and III correspond to the labels $\mathbf{b}$ and $\mathbf{d}$ here.

It can be also easily verified that the two conditions

$$2m < 2|\lambda_{1;\mathbf{a}',m}| = 2m+1+\lambda_o - \frac{\mu_o^2}{2m+1+\lambda_o} \tag{8.1.27}$$

and

$$|\lambda_{1;\mathbf{a}',m}| - \lambda_{1;\mathbf{c},0} = (\lambda_o + m + 1)\left[1 - \frac{\mu_o^2}{(2m+1+\lambda_o)(\lambda_o+1)}\right] > 0 \tag{8.1.27*}$$

are indeed equivalent and coincide with (6.1.16) in the limiting case b=0. In Quesne's notation [34] the lower bound for m is equal to $B/A - A$.

It is crucial that derivative (5.1.1) does not vanish at the origin in the limiting case of the E/MR potential (b=0), contrary to the general case b > 0. Since

$$\lambda_o = 2s_o - 1 = 2s_{0;\mathbf{a}'} - 3 \tag{8.1.28}$$

for b=0, instead of the generic expression $\lambda_o = s_{0;\mathbf{a}'} - 3/2$ for b > 0, the appropriate AEH Frobenius solution of the Schrödinger equation has the ChExp@)O $s_{0;\mathbf{a}'}$ iff GS Heine polynomial (6.1.17) has the single zero root:

$$\lim_{b\to 0} \mathrm{Hi}_{m+v}[z\,|\,{}^{1}_{1}\boldsymbol{G}^{211}_{\mathbf{t}_{+},m};\mathbf{c},v] = z\mathrm{Hi}_{m+v-1}[z\,|\,{}^{1}_{1}\boldsymbol{G}^{202}_{\mathbf{t}_{+},m};\mathbf{c},v]. \tag{8.1.29}$$

The direct proof of this expression is presented in Appendix C. The remarkable feature of the GS Heine polynomials in the right-hand side of (8.1.29) is that each sequence starts from the polynomial of order m -1. It was Quesne [34] who first made this observation by analyzing similar polynomial solutions $y^{(A,B)}_{m+v-1}(\tilde{z})$ in the reciprocal argument $\tilde{z}(r)$. We refer the reader to Appendix C below for some additional details between these two sets of polynomials in mutually reciprocal arguments.

Here we are mostly interested in nodeless R@∞ AEH solutions ($\mathbf{t}_{-} = \mathbf{b}$ or $\mathbf{b}'$) such that

$$\lambda_{1;\mathbf{t}_{-},m}(\lambda_o,\mu_o;0) = \frac{\mu_o^2}{2(2m+1-\lambda_o)} - \tfrac{1}{2}(2m+1-\lambda_o) \equiv \frac{B-(m+1-A)^2}{m+1-A} > 0 \tag{8.1.30}$$

since they form the unique class of Frobenius solutions near the singular point r = 0 which, contrary to eigenfunctions, are regular only at infinity, not at the origin. Inequality (8.1.30) holds if either

$$0 < m+1-A < \sqrt{B} \tag{8.1.31}$$

or

$$m+1 < A-\sqrt{B} \equiv \tfrac{1}{2}(\lambda_o+1-\mu_o) < \tfrac{1}{2}(\lambda_o+1) < \lambda_o+1. \tag{8.1.31′}$$

The second condition is satisfied only in the region where the E/MR potential does not have the discrete energy spectrum. This is nothing but secondary sequence (6.1.22′) in the limit b→0. As expected, the order m of Jacobi polynomials in this sequence satisfy the constraint $m < \lambda_o$.

The solutions from the primary sequence described by exponents (6.1.22) are nodeless if either there is no discrete energy spectrum [i.e., if $\lambda_o - 1 \le \mu_o < \lambda_o + 1$ and $m < \tfrac{1}{2}(\lambda_o + \mu_o - 1) < \lambda_o$] or if they lie below the ground energy level:

$$ {}_1\varepsilon_{\mathbf{c},0}(\lambda_o,\mu_o) = -\tfrac{1}{4}\frac{[\mu_o^2-(1+\lambda_o)^2]^2}{(1+\lambda_o)^2} = \frac{(B-A^2)^2}{4A^2}, \tag{8.1.32} $$

i. e. if

$$ B-(m+1-A)^2 > \frac{(B-A^2)(m+1-A)}{A} > 0\,. \tag{8.1.33} $$

The latter condition holds iff m+1–A lies between two roots of the quadratic equation:

$$ A(m+1-A)^2 + (B-A^2)(m+1-A) - AB = 0, \tag{8.1.34} $$

i.e., iff

$$ -\tfrac{1}{2}(B-A^2)/A - \sqrt{\Delta_{\mathbf{b}}} < m+1-A < -\tfrac{1}{2}(B-A^2)/A + \sqrt{\Delta_{\mathbf{b}}} = A\,, \tag{8.1.35} $$

where

$$ \Delta_{\mathbf{b}} = \tfrac{1}{4}(B+A^2)^2/A^2. \tag{8.1.36} $$

We thus find that the polynomial order m in the primary sequence of nodeless AEH solutions **b**,m is restricted to the range [34]

$$ A-1 < m < 2A-1 = \lambda_o < A+\sqrt{B}-1. \tag{8.1.37} $$

Surprisingly the primary sequence **b**,m for the E/MR potential $V[z\,|\,{}_1\mathcal{G}^{202}]$ does not start from the basic solution m=0, in contrast with the generic radial *r*-GRef potential $V[z\,|\,{}_1\mathcal{G}^{211}]$. An analysis of behavior of root (6.1.22) shows that it infinitely grows as $b\to 0$ if the linear coefficient of quadratic equation (6.1.19) is negative so that only solutions with $m > \tfrac{1}{2}(\lambda_o - 1) = A-1$ retain in the limiting case b=0.

The exponent differences for singularities at the origin in the SL equation for the potential $V[z\,|\,{}_1^1\mathcal{G}^{202}_{\mathbf{t}_,m}]$ are equal to

$$ \lambda_{o;\mathbf{t}_} = |\,2-\lambda_o\,|. \tag{8.1.38} $$

Substituting (8.1.36b) into (8.1.28) thus gives

$$ {}_1\lambda_{1;\mathbf{t}_-,0} = \frac{\mu_o^2 - (1+\lambda_{o;\mathbf{t}_-})^2}{2(\lambda_{o;\mathbf{t}_-}+1)} \quad \text{for} \quad \lambda_o > 1 \tag{8.1.39} $$

and

$$ {}_1\lambda_{1;\mathbf{t}_-,0} = \frac{\mu_o^2 - (1-\lambda_{o;\mathbf{t}_-})^2}{2(1-\lambda_{o;\mathbf{t}_-})} \quad \text{within the LC range.} \tag{8.1.39*} $$

We thus conclude that DTs with the basic FFs do not change the parameter $\mu_o$ so that the energy-dependent arguments of gamma-functions in (5.1.19) are simply shifted by 1:

$$ {}_1\alpha_{+,+}(\varepsilon;\lambda_{o;\mathbf{t}_\pm},\mu_o;1) = {}_1\alpha_{+,+}(\varepsilon;\lambda_o,\mu_o;1) \pm 1, \tag{8.1.40a} $$

$$ {}_1\beta_{+,+}(\varepsilon;2-\lambda_o,\mu_o;1) = {}_1\beta_{+,+}(\varepsilon;\lambda_o,\mu_o;1) \pm 1 \tag{8.1.40b} $$

and

$$ {}_1\alpha_{-,+}(\varepsilon;\lambda_{o;\mathbf{t}_\pm},\mu_o;1) = {}_1\alpha_{-,+}(\varepsilon;\lambda_o,\mu_o;1) \mp 1, \tag{8.1.40a′} $$

$$ {}_1\beta_{-,+}(\varepsilon;\lambda_{o;\mathbf{t}_\pm},\mu_o;1) = {}_1\beta_{-,+}(\varepsilon;\lambda_o,\mu_o;1) \mp 1 \tag{8.1.40b′} $$

as far as the exponent differences $\lambda_{o;\mathbf{t}_-}$ lie within the LP range. Combining the latter relations with the identities

$$ \begin{aligned} [{}_1\alpha_{+,+}(\varepsilon;\lambda_o,\mu_o;1)-1]\times[{}_1\beta_{+,+}(\varepsilon;\lambda_o,\mu_o;1)-1] &\equiv \tfrac{1}{4}[(\lambda_o+\sqrt{-\varepsilon}-1)^2-\mu^2(\varepsilon;\mu_o)] \\ &\equiv \tfrac{1}{4}[(\lambda_o-1)^2-\mu_o^2+2\sqrt{-\varepsilon}(\lambda_o-1)] \end{aligned} \tag{8.1.41} $$

and

$$ \begin{aligned} {}_1\alpha_{-,+}(\varepsilon;\lambda_o,\mu_o;1)\,{}_1\beta_{-,+}(\varepsilon;\lambda_o,\mu_o;1) &\equiv \tfrac{1}{4}[(1-\lambda_o+\sqrt{-\varepsilon})^2-\mu^2(\varepsilon;\mu_o)] \\ &\equiv \tfrac{1}{4}[(\lambda_o-1)^2-\mu_o^2-2\sqrt{-\varepsilon}(\lambda_o-1)] \end{aligned} \tag{8.1.41′} $$

thus gives

$$ \begin{aligned} &\frac{\Gamma({}_1\alpha_{+,+}(\varepsilon;\lambda_{o;\mathbf{t}_-},\mu_o;1))\,\Gamma({}_1\beta_{+,+}(\varepsilon;\lambda_{o;\mathbf{t}_-},\mu_o;1))}{\Gamma({}_1\alpha_{-,+}(\varepsilon;\lambda_{o;\mathbf{t}_-},\mu_o;1))\,\Gamma({}_1\beta_{-,+}(\varepsilon;\lambda_{o;\mathbf{t}_-},\mu_o;1))} \\ &\qquad = \frac{4(\lambda_o-1)^2}{\varepsilon-\varepsilon_{\mathbf{t}_-,0}} \times \frac{\Gamma({}_1\alpha_{+,+}(\varepsilon;\lambda_o,\mu_o;1))\,\Gamma({}_1\beta_{+,+}(\varepsilon;\lambda_o,\mu_o;1))}{\Gamma({}_1\alpha_{-,+}(\varepsilon;\lambda_o,\mu_o;1))\,\Gamma({}_1\beta_{-,+}(\varepsilon;\lambda_o,\mu_o;1))}. \end{aligned} \tag{8.1.42} $$

Substituting (8.1.42) into (5.1.19), with $\lambda_o$ changed for $\lambda_{o;\dagger_}$ and taking into account that

$$\frac{\Gamma(-\lambda_{o;\dagger_})}{\Gamma(\lambda_{o;\dagger_})} = \frac{\Gamma(2-\lambda_o)}{\Gamma(\lambda_o-2)} = (\lambda_o-1)^2[4(s_o-1)^2-1]\frac{\Gamma(-\lambda_o)}{\Gamma(\lambda_o)} \quad \text{for } \lambda_o > 1 \qquad (8.1.43)$$

we then directly come to (8.1b).

Within the LC range

$$_1\alpha_{\pm,+}(\varepsilon;\lambda_{o;\dagger_},\mu_o;1) = {}_1\alpha_{\mp,+}(\varepsilon;\lambda_o,\mu_o;1) \pm 1 \qquad (8.1.44a)$$

and

$$_1\beta_{\pm,+}(\varepsilon;\lambda_{o;\dagger_},\mu_o;1) = {}_1\beta_{\mp,+}(\varepsilon;\lambda_o,\mu_o;1) \pm 1. \qquad (8.1.44b)$$

Thereby, changing $\lambda_o$ in (5.1.19) for $2-\lambda_o$ and making use of (8.1.41) and (8.1.41′), coupled with the energy-independent relation

$$\frac{\Gamma(-\lambda_{o;\dagger_})}{\Gamma(\lambda_{o;\dagger_})} = \frac{\Gamma(\lambda_o-2)}{\Gamma(2-\lambda_o)} = \frac{1}{(\lambda_o-1)^2[4(s_o-1)^2-1]}\frac{\Gamma(\lambda_o)}{\Gamma(-\lambda_o)} \quad \text{for } \lambda_o < 1, \qquad (8.1.45)$$

leads to (8.1b*), as expected.

As mentioned in previous Section, right-sides of formulas (13) and (14) for the Jost function in [13] must be multiplied by the additional factor $2s_o-1$, i. e., by $\lambda_o$ for the E/MR potential which gives

$$f(k;\tfrac{1}{2}\lambda_o+\tfrac{1}{2}) = \frac{e^{-ik\,r_o}\;\Gamma(1-ik)\Gamma(\lambda_o+1)}{2^{\frac{1}{2}(\lambda_o-1)}\Gamma\big({}_1\alpha_{++}(k^2;\lambda_o,\mu_o;1)\big)\Gamma\big({}_1\beta_{++}(k^2;\lambda_o,\mu_o;1)\big)}. \qquad (8.1.46)$$

Representing (8.1.41) as

$$[{}_1\alpha_{+,+}(k^2;\lambda_o,\mu_o;1)-1]\times[{}_1\beta_{+,+}(k^2;\lambda_o,\mu_o;1)-1] = \tfrac{1}{2}(\lambda_o-1)(-ik-{}_1\lambda_{1;\dagger_,0}) \qquad (8.1.47)$$

we confirm that the Jost function for the E/MR potential is transformed according to (3.4b) and (3.4d), as expected [21].

In the particular case of the Hulthén potential $V_{E/MR}[z(r);1,\mu_o]$:

$$ {}_{1}f\,(k;\Diamond_{o},;\mu_{o}1) = \frac{e^{-ik\,r_{o}}\;\Gamma(1-ik)\Gamma(1)}{\Gamma(\,{}_{1}\alpha_{++}(k^{2};1,\mu_{o};1)\,)\,\Gamma(\,{}_{1}\beta_{++}(k^{2};1,\mu_{o};1))}, \tag{8.1.48} $$

in agreement with (14.16) in [39]. (Since the Hulthén potential is simply the particular case of the E/MR potential there is obviously no sense in adding the two [75-78].[x)])
As explicitly demonstrated by Laha et al [79] the DT with the lowest-energy normalizable FF does not convert the Hulthén potential onto itself. Its SUSY partner constructed in such a way is nothing but Haeringen's reduction [80] of the Eckart potential $V_{E/MR}[z(r);s_{o},0]$. By applying the state-erasing DT to the latter potential one finally comes to the generic E/MR potential.

Among the exactly-solvable IS@O potentials with an exponential asymptotics at infinity, the Hulthén potential is the only one which represents Fulton's Case II [27]: $q_0 = 0$, $q_1 \neq 0$. When applied to Hulthén potential multi-step DTs with R@O SSs generate Kratzer potentials which all represent Case II A in terms of [48]. We thus see a lot of similarity between Hulthén and Coulomb potentials first noticed by Ma [81] in the fifties. Apparently this similarity deserves a more serious attention in connection with the very specific features of their TWF functions.

---

[x)] Meyur and Debnath [77] as well as Deta et al [78] erroneously refer to the E/MR potential as the Scarf potential which covers the fact that they first represent the same potential in two different forms and then simply add up both expressions.

## 8.2 The centrifugal Kepler-Coulomb potential

Setting a = 1 and b =0 in (5.2.1) gives $\zeta = 2r$ so that the exponent differences for the common singular point $\zeta = r = 0$ in the Whittaker equation (5.2.6) and in the Schrödinger equation with the KC potential coincide. As a result $\lambda_o = 2\Diamond_o$, by analogy with E/MR potential. Since $\zeta$ is no more proportional to $r^2$ at small r one first has to re-write (5.2.9) and (5.2.9*) as

$$ {}_0f_{\mathbf{a}}(r;\varepsilon;\Diamond_o) = |\varepsilon|^{-\frac{1}{4}(2\Diamond_o+1)}\, r^{-s_o}\, M_{\kappa(\varepsilon),\Diamond_o}(\sqrt{-\varepsilon}\, r) \tag{8.2.1} $$

and

$$ {}_0f_{\mathbf{d}}(r;\varepsilon;\Diamond_o) = |\varepsilon|^{\frac{1}{4}(2\Diamond_o-1)}\, r^{s_o-1}[\zeta]\, M_{\kappa(\varepsilon),-\Diamond_o}(\sqrt{-\varepsilon}\, r]). \tag{8.2.1*} $$

Consequently the TWM function for the KC potential also has a slightly different form

$$ {}_0m(\varepsilon;\Diamond_o,g_o;0) = |2\varepsilon|^{\Diamond_o}\frac{\Gamma(1-2\Diamond_o)}{2\Diamond_o\Gamma(2\Diamond_o+1)}\,{}_0M_{0,0}\big(2\Diamond_o;\kappa(\varepsilon;g_o;0)\big) \tag{8.2.2} $$

compared with the general case. Note that the derived expression differs by sign from (5.11) in [48]. The discrepancy apparently comes due to a misprint in sign of the coefficient of $M_{\beta,\mu}(\varsigma)$ in (5.7) in [48] which was verified by directly comparing the latter formula with its original version [82].

If the potential has the discrete energy spectrum ($g_o < 0$) then TWF function (8.2.2) has a finite number of zeros at the energies

$$ {}_0\varepsilon_{\mathbf{b},k} = -\frac{g_1^2}{4(\lambda_o - 2k - 1)^2} \tag{8.2.3} $$

such that

$$ \frac{g_o}{2\sqrt{-\varepsilon_{\mathbf{b},k}}} = \lambda_o - 2k - 1 < 0, \tag{8.2.4} $$

i. e., for $k > s_o - 1$. The appropriate AEH solutions lie below the ground energy level if

$k < \lambda_o$. The Schrödinger equation with the attractive effective Coulomb potential thus has nodeless AEH solutions at energies (8.2.3) for $s_o - 1 < k < 2s_o - 1$, in agreement with Grandati's results [83] who came to the same conclusion by analyzing zeros of Laguerre polynomials with negative index $-\lambda_o$.

Since the common leading coefficient of the quadratic equations (7.2.4) specifying energies of basic solutions vanishes each equation has a single root

$$\nu_{\mathsf{t}_\pm,0}(\lambda_o, g_o; 0) = -\frac{g_o}{2(1 \pm \lambda_o)} \tag{8.2.5}$$

so that the Schrödinger equation with the centrifugal Kepler-Coulomb potential has only two basic solutions at the energies

$$_0\varepsilon_{\mathsf{t}_\pm,0}(\lambda_o, g_1) = -\frac{g_1^2}{4(1 \pm \lambda_o)^2}, \tag{8.2.6}$$

where

$$\mathsf{t}_+ = \begin{cases} \mathsf{c} \text{ if } g_1 < 0, \\ \mathsf{a} \text{ if } g_1 > 0; \end{cases} \quad \mathsf{t}_- = \begin{cases} \mathsf{b} \text{ if } g_1(\lambda_o - 1) > 0, \\ \mathsf{d} \text{ if } g_1(\lambda_o - 1) < 0. \end{cases} \tag{8.2.7}$$

The SUSY partner of TWM function (8.2.2) generated using the basic FF $\mathsf{t}_-, 0$ thus takes the form

$$_0m(\varepsilon; \Diamond_{o;\mathsf{t}_-}, g_o; 0) = -\frac{(2\sqrt{-\varepsilon})^{\lambda_o - 2}\Gamma(2-\lambda_o)}{(2-\lambda_o)\Gamma(\lambda_o - 2)} \times \frac{\Gamma(\tfrac{1}{2}(\lambda_o - 1) - \kappa(\varepsilon))}{\Gamma(\tfrac{1}{2}(3-\lambda_o) - \kappa(\varepsilon))} \quad (\lambda_o > 2) \tag{8.2.8}$$

and

$$_0m(\varepsilon; \lambda_{o;\mathsf{t}_-}, g_o; 0) = -\frac{(2\sqrt{-\varepsilon})^{2-\lambda_o}\Gamma(\lambda_o - 2)}{(2-\lambda)\Gamma(2-\lambda_o)} \times \frac{\Gamma(\tfrac{1}{2}(2-\lambda_o) - \kappa(\varepsilon))}{\Gamma(\tfrac{1}{2}(\lambda_o - 1) - \kappa(\varepsilon))} \quad (\lambda_o < 2). \tag{8.2.8*}$$

Substituting

$$k^2(\varepsilon; g_1; 0) - \tfrac{1}{4}(1-\lambda_o)^2 = \tfrac{1}{4}\frac{(1-\lambda_o)^2}{\varepsilon}[\,_0\varepsilon_{\mathsf{t}_-,0}(\lambda_o, g_1) - \varepsilon] \tag{8.2.9}$$

into the identities:

$$\frac{\Gamma(\frac{1}{2}(\lambda_o - 1) - \kappa(\varepsilon))}{\Gamma(\frac{1}{2}(3-\lambda_o) - \kappa(\varepsilon))} = \frac{1}{k^2(\varepsilon) - \frac{1}{4}(1-\lambda_o)^2} \frac{\Gamma(\frac{1}{2}(1+\lambda_o) - k(\varepsilon))}{\Gamma(\frac{1}{2}(1-\lambda_o) - k(\varepsilon))} \qquad (8.2.10)$$

and

$$\frac{\Gamma(\frac{1}{2}(2-\lambda_o) - \kappa(\varepsilon))}{\Gamma(\frac{1}{2}(\lambda_o - 1) - \kappa(\varepsilon))} = [k^2(\varepsilon) - \frac{1}{4}(1-\lambda_o)^2] \frac{\Gamma(\frac{1}{2}(1-\lambda_o) - k(\varepsilon))}{\Gamma(\frac{1}{2}(1+\lambda_o) - k(\varepsilon))}, \qquad (8.2.10^*)$$

coupled with a similar expression

$$\frac{\Gamma(2-\lambda_o)}{\Gamma(\lambda_o - 2)} = (\lambda_o - 1)^2 [(\lambda_o - 1)^2 - 1] \frac{\Gamma(-\lambda_o)}{\Gamma(\lambda_o)} \qquad (8.2.11)$$

for gamma-functions with energy-independent arguments, then directly confirms that TWF functions (8.2.8) and (8.2.8*), with $\lambda_o = 2s_o - 1$, are related to (8.2.2) via (8.1b) and (8.1b*), respectively.

## 9. Conclusions and further developments

From our perspective one of the most important results of this study is the ‘amateurish’ proof in Section 2 in support of the assertion that regular solutions of the Schrödinger equation with the generic IS@O radial potential do not vanish on the positive semi-axis within the LC range of the SI (the well-established fact for singular potentials in the LP region) . The author is not aware of any mathematical literature dealing in depth with this issue and was simply unable to proceed without this proof.

It was proven that DT converting the LC region onto itself turns any R@O solution into an irregular one (i..e., into the one with a smaller value of the ChExp). We found extremely useful Fulton’s representation [27] of a R@∞ solution as a superposition of Frobenius solutions at the origin. We extended his analysis of the *m*-function referred to as the TWF function in the paper) to SUSY partners of the given reflective IS@O potential using nodeless Frobenius solutions at the origin as FFs. It was directly shown that the partner *m*-functions are interrelated in an anomalous way if the Darboux transformation keeps the potential within the LC region.

The established transformation properties of the TWF functions under Darboux deformations of reflective centrifugal potentials were explicitly confirmed by analyzing their closed-form representations for the radial *r*- and *c*-GRef potentials. In particular, we explicitly demonstrate existence of *non-isospectral* partners of both radial potentials in the LC region and obtained algebraic formulas for their discrete energy spectra by analyzing poles of TWF functions (6.1.22) and (6.2.20).

We also took advantage of our revelation that both *r*-GRef and *c*-GRef potentials retain their form under action of double-step DTs using the basic SSs. In addition, we verified our general results for three (h-PT, E/MR, and KC) shape-invariant potentials -- the particular representatives of radial GRef potentials which retain their form under action of DTs with basic FFs.

One of interesting directions for future studies would be re-examination of the mathematical grounds of the transformations of the given SLE

$$\left\{\frac{d^2}{d\xi^2} + I^o[\xi;Q_o] + \varepsilon\,\wp[\xi]\right\}\Phi[\xi;\varepsilon\,|\,Q_o] = 0 \qquad (8.1)$$

under action of Darboux deformations of the ‘Liouville’ potential constructed by means of the appropriate Liouville transformation (see, i.e., [5]). These canonical Liouville-Darboux transformations (as we refer to them [15]) can be obviously introduced with no reference to the Liouville transformation used to convert the SLE with rational coefficients into the Schrödinger equation and back. It thus seems more informative to directly extend Fulton’s theory [27] to the M-function for the singular eigenvalue problem [3]

$$(\hat{H}_o - \varepsilon)\,\Theta[\xi;\varepsilon\,|\,Q_o] = 0 \qquad (9.2)$$

with the self-adjoint operator:

$$\hat{H}_o \equiv \wp^{1/2}[\xi]\left\{-\frac{d^2}{d\xi^2} + I^o[\xi;Q_o]\right\}\wp^{-1/2}[\xi] \qquad (9.3)$$

$$= -\frac{d}{d\xi}\wp^{-1}[\xi]\frac{d}{d\xi} + V_o[\xi\,|\,Q_o], \qquad (9.3^*)$$

where

$$V_o[\xi;Q_o] \equiv I^o[\xi;Q_o] + \frac{d}{d\xi}\left(ld\,\wp[\xi]\right). \tag{9.4}$$

One can directly verify that solutions of second-order differential equations (9.1) and (9.2) are related via the gauge transformation

$$\Theta[\xi;\varepsilon\,|\,Q_o] = \wp^{1/2}[\xi]\,\Phi[\xi;\varepsilon\,|\,Q_o]. \tag{9.5}$$

In this paper we only discussed the general case of a non-integer $M = 2s_o - 1$. The TWF function for the generic Kramer potential with integer values of M is required a special consideration (Case IC in [27] or Cases IIA and IIB for odd and even M, respectively, in [48]). The case of odd M (integer $s_o$) is of special importance because one comes to an infinite sequence of SUSY partners by applying multi-step DTs with SSs of type **a** to an arbitrary non-singular potential on the half-line. A natural extension of this study would be derive transformation properties of the TWF function for this chain of DTs and then illustrate the results by applying multi-step DTs with Gauss SS **a**,m to the non-singular potential $V[z\,|\,_1\mathcal{G}^{211}]$ obtained by setting the SI $\lozenge_o$ to $1/2$ ($M = s_o = 1$). It is worth mentioning that all the potential in this series do not have a Coulomb term ($q_1 \neq 0$). On other hand, the sequence of E/MR potentials with $M \equiv \lambda_o = 1, 3, ..., 2s_{max} - 1 < [\mu_o + 1]$, starting from the Hulthén potential, gives the unique exactly-solvable example of Case IIA for $q_1 \neq 0$.

As pointed out in Appendix C, the reciprocal transformation of z formally converts (5.1.5) into another rational SL equation with three regular singular points. The Liouville transformation of the resultant equation restricted to the finite interval (0,1) leads to the so-called [35] 'LTP *r*-GRef potential' on the line, $V[z\,|\,_1\mathcal{G}^{110}]$. One of common remarkable features of the potentials $V[z\,|\,_1\mathcal{G}^{211}]$ and $V[z\,|\,_1\mathcal{G}^{110}]$ is that they are both form-invariant under double-step DTs using basic SSs. In addition, the singe-step DTs with basic FFs convert each of these potentials into a pair of rational double-step SUSY partners quantized via Heun polynomials. It can be then shown that two sets of Heun polynomials are related via the reciprocal transformations of their arguments. The same is true for Lambe and Ward's [84] 'quasi-algebraic' solutions (the specific case of 'AEH' solutions in our terms). As a result there must be also a very interesting interrelationship between integral representations of two Heun equations using Lambe-Ward kernels.

## Appendix A

### Jost function for the radial *c*-GRef potential

It seems useful to clarify some details of the derivation presented in [17] for the Jost function of radial potential (5.2.2), making the final formula consistent with our current notation and also correcting a few misprints in its original version. Let us start from the generic case b > 0. First we need to introduce the Whittaker functions with the complex index

$$\mu(t) \equiv \mu(t; g_o; b) = -\frac{g_o + bt^2}{4t} = -\tfrac{1}{4}(g_o / t + bt) \tag{A.1}$$

which coincides at t = *i*k with $\kappa(k^2)$ on the upper imaginary axis of the plane of complex $k = i\sqrt{-\varepsilon}$, namely,

$$\mu(-ik) \equiv i\eta(k) - \tfrac{1}{4} ibk = \kappa(k^2; g_o; b) \quad \text{for } Im\ k = 0, Re\ k > 0. \tag{A.2}$$

Note, that, compared with [17], we changed the definition of the k-dependent phase

$$\eta(k) = \frac{g_o}{4k} = \frac{q_1}{2k} \tag{A.3}$$

to make it real at positive energies. We can then express the conventional asymptotic formula

$$W_{-\mu, \frac{1}{2}\lozenge_o}(e^{-i\pi}\varsigma) \sim e^{-\frac{1}{2}\varsigma} e^{i\mu\pi} \varsigma^{-\mu} \quad \text{for } |\varsigma| >> 1 \text{ and } |\arg(e^{-i\pi}\varsigma)| < \pi \tag{A.4}$$

in terms of r by noting that

$$\varsigma(r) = e^{\frac{1}{2} i\pi} k\, \zeta(r) \tag{A.5}$$

where the variable $\zeta(r)$ behaves at large r as

$$\zeta(r) \sim 2[r + r_o(b)] - \tfrac{1}{2} b\, \ln r \quad \text{at} \ \ r >> 1, \tag{A.6}$$

with the parameter $r_o \equiv r_o\,(b)$ standing for $x^+$ in [17]. Substituting (A.5) and (A.6) into (A.4) gives

$$W_{-\mu(-ik),\frac{1}{2}\Diamond_o}(-2i\,kr) \sim e^{ik(r+r_o)-i\eta(k)\ln r}\, e^{\frac{1}{2}i\pi\,\mu(-ik)}(2k)^{\frac{1}{4}ik\,b} \tag{A.7}$$

$$\text{for } r >> 1 \ \text{ and } \ -\tfrac{1}{2}\pi < \arg k < \tfrac{1}{2}\pi.$$

(In [17] the same notation b was accidentally used for the four-times smaller parameter.) Comparing the asymptotic behavior of the Whittaker function with (3.22) thus come to (22) in [17], where we are only interested in the upper index, namely,

$$_0f^+(r;k;\Diamond_o,g_o;0) = \sqrt[4]{1 + b/\zeta(r)}\; e^{-ikr_o - \frac{1}{2}i\pi\mu(-ik)}(2k)^{\frac{1}{4}ik\,b}$$
$$\times W_{-\mu(k),\frac{1}{2}\Diamond_o}\left(e^{-i\pi}\varsigma(r)\right). \tag{A.8}$$

in our current terms.

It directly follows from the asymptotic behavior of Whittaker function (5.2.10) at small $\varsigma$ that

$$\lim_{\varsigma\to 0}\left[\varsigma^{\frac{1}{2}(\Diamond_o - 1)} W_{-\mu,\frac{1}{2}\Diamond_o}(e^{-i\pi}\varsigma)\right] = \frac{\Gamma(\Diamond_o)}{\Gamma\left(\frac{1}{2}(1+\Diamond_o) + \mu(-ik)\right)} e^{\frac{1}{2}i\pi(\Diamond_o - 1)} \tag{A.9}$$

Defining the Jost function defined according to (3.9) and taking into account

$$\varsigma(r) \sim e^{\frac{1}{2}i\pi} kr^2 / b \ \ \text{at } r<<1 \tag{A.10}$$

one finds

$$_0f(k;\Diamond_o,g_o;b) = 2\sqrt[4]{k}\; e^{-ikr_o(b) - \frac{1}{2}i\pi\,\mu(-ik;g_o;b)}(2k)^{\frac{1}{4}ik\,b}$$
$$\times \left(e^{\frac{1}{2}i\pi} b/k\right)^{\frac{1}{2}(\Diamond_o - 1)} \frac{\Gamma(\Diamond_o + 1)}{\Gamma\left(\frac{1}{2}(1+\Diamond_o) + \mu(-ik;g_o;b)\right)} \tag{A.11}$$

The Jost function for the centrifugal Kepler-Coulomb potential requires a special consideration. First remember that $\lambda_o = 2\Diamond_o$ in this case, by analogy with E/MR potential. Namely, by substituting $\zeta(r)$ for 2r at large r asymptotic formula (A.8) can be simplified as follows

$$. \ e^{i\eta(k) ln(2kr)} {}_0f^{+}(r;k;\Diamond_o,g_o;0) = e^{\frac{1}{2}\pi\eta(k)} W_{-i\eta(k),\Diamond_o}(-2ikr). \qquad \text{(A.12)}$$

As a result we come to the conventional formula for the Coulomb potential [39]:

$$ {}_0f(k;\Diamond_o,g_o;0) = \frac{e^{-\frac{1}{2} i\mu(-ik;g_o)\pi}}{(-2ik)^{\Diamond_o - \frac{1}{2}}} \times \frac{\Gamma(2\Diamond_o + 1)}{\Gamma(\frac{1}{2} + \Diamond_o + \mu(-ik;g_o))}, \qquad \text{(A.13)}$$

with $\Diamond_o - \frac{1}{2}$ used instead of the generally non-integer 'angular momentum' $\ell$.

## Appendix B

**Transformation properties of energy-dependent parameters of hypergeometric solutions under Darboux deformations of the h-PT potential**

The purpose of this appendix is to list the explicit formula for energy-dependent arguments (5.1.20a) and (5.1.20b) of gamma-functions in the right-hand side of (5.1.19), based on form-invariance of the h-PT potential under DTs with basic FFs.

*I. FFs regular at the origin*

i) $\mathbf{t}_+ = \mathbf{a}$

$${}_1\alpha_{++}(\varepsilon;\lambda_{o;\mathbf{a}},\mu_{o;\mathbf{a}}) = {}_1\alpha_{++}(\varepsilon;\lambda_o,\mu_o), \qquad \text{(B.1a)}$$

$${}_1\beta_{++}(\varepsilon;\lambda_{o;\mathbf{a}},\mu_{o;\mathbf{a}}) = {}_1\beta_{++}(\varepsilon;\lambda_o,\mu_o) + 1, \qquad \text{(B*.1a)}$$

$${}_1\alpha_{-+}(\varepsilon;\lambda_{o;\mathbf{a}},\mu_{o;\mathbf{a}}) = {}_1\alpha_{-+}(\varepsilon;\lambda_o,\mu_o) - 1, \qquad \text{(B.2a}'\text{)}$$

$${}_1\beta_{-+}(\varepsilon;\lambda_{o;\mathbf{a}},\mu_{o;\mathbf{a}}) = {}_1\beta^0_{-+}(\varepsilon;\lambda_o,\mu_o); \qquad \text{(B*.2a)}$$

ii) $\mathbf{t}_+ = \mathbf{a}'$

$${}_1\alpha_{++}(\varepsilon;\lambda_{o;\mathbf{a}'},\mu_{o;\mathbf{a}'}) = \begin{cases} {}_1\alpha_{++}(\varepsilon;\lambda_o,\mu_o) + 1 & \text{for } \mu_o > 1 \\ \text{or} & \\ {}_1\beta_{++}(\varepsilon;\lambda_o,\mu_o) & \text{for } \mu_o < 1; \end{cases} \qquad \text{(B.1a}'\text{)}$$

$$ {}_1\beta_{++}(\varepsilon;\lambda_{o;a'},\mu_{o;a'}) = \begin{cases} {}_1\beta_{++}(\varepsilon;\lambda_o,\mu_o) & \text{for } \mu_o > 1 \\ \text{or} & \\ {}_1\alpha_{++}(\varepsilon;\lambda_o,\mu_o)+1 & \text{for } \mu_o < 1; \end{cases} \qquad \text{(B*.1a}'\text{)} $$

$$ {}_1\alpha_{-+}(\varepsilon;\lambda_{o;a'},\mu_{o;a'}) = \begin{cases} {}_1\alpha_{-+}(\varepsilon;\lambda_o,\mu_o) & \text{for } \mu_o > 1 \\ \text{or} & \\ \beta_{-+}(\varepsilon;\lambda_o,\mu_o)+1 & \text{for } \mu_o < 1; \end{cases} \qquad \text{(B*.2a}'\text{)} $$

$$ {}_1\beta_{-+}(\varepsilon;\lambda_{o;a'},\mu_{o;a'}) = \begin{cases} {}_1\beta_{-+}(\varepsilon;\lambda_o,\mu_o)-1 & \text{for } \mu_o > 1 \\ \text{or} & \\ {}_1\alpha_{-+}(\varepsilon;\lambda_o,\mu_o) & \text{for } \mu_o < 1; \end{cases} \qquad \text{(B*.2a}'\text{)} $$

iii) $t_+ = c$

$$ {}_1\alpha_{++}(\varepsilon;\lambda_{o;c},\mu_{o;c}) = {}_1\alpha_{++}(\varepsilon;\lambda_o,\mu_o)+1, \qquad \text{(B.1c)} $$

$$ {}_1\beta_{++}(\varepsilon;\lambda_{o;c},\mu_{o;c}) = {}_1\beta_{++}(\varepsilon;\lambda_o,\mu_o), \qquad \text{(B*.1c)} $$

$$ {}_1\alpha_{-+}(\varepsilon;\lambda_{o;c},\mu_{o;c}) = {}_1\alpha_{-+}(\varepsilon;\lambda_o,\mu_o), \qquad \text{(B.2c)} $$

$$ {}_1\beta_{-+}(\varepsilon;_{o;c},\mu_{o;c}) = {}_1\alpha_{-+}(\varepsilon;\lambda_o,\mu_o)-1. \qquad \text{(B*.2c)} $$

*II. FFs irregular at the origin*

**The LP range**

i) $t_- = b$

$$ {}_1\alpha_{++}(\varepsilon;\lambda_{o;b},\mu_{o;b}) = \begin{cases} {}_1\alpha_{++}(\varepsilon;\lambda_o,\mu_o) & \text{for } \mu_o > 1 \\ \text{or} & \\ {}_1\beta_{++}(\varepsilon;\lambda_o,\mu_o)-1 & \text{for } \mu_o < 1; \end{cases} \qquad \text{(B.1b)} $$

$$ {}_1\beta_{++}(\varepsilon;\lambda_{o;b},\mu_{o;b}) = \begin{cases} {}_1\beta_{++}(\varepsilon;\lambda_o,\mu_o)-1 & \text{for } \mu_o > 1 \\ \text{or} & \\ {}_1\alpha_{++}(\varepsilon;\lambda_o,\mu_o) & \text{for } \mu_o < 1; \end{cases} \qquad \text{(B*.1b)} $$

$$ {}_1\alpha_{-+}(\varepsilon;\lambda_{o;b},\mu_{o;b}) = \begin{cases} {}_1\alpha_{-+}(\varepsilon;\lambda_o,\mu_o)-1 & \text{for } \mu_o > 1 \\ \text{or} & \\ {}_1\beta_{-+}(\varepsilon;\lambda_o,\mu_o) & \text{for } \mu_o < 1; \end{cases} \qquad \text{(B.2b)} $$

$$ {}_1\beta_{-+}(\varepsilon;\lambda_{o;b},\mu_{o;b}) = \begin{cases} {}_1\alpha_{-+}(\varepsilon;\lambda_o,\mu_o) & \text{for } \mu_o > 1 \\ \text{or} & \\ {}_1\beta_{-+}(\varepsilon;\lambda_o,\mu_o)+1 & \text{for } \mu_o < 1; \end{cases} \qquad \text{(B*.2b)} $$

ii) $\mathsf{t}_- = \mathsf{b}'$ or $\mathsf{d}$

$$ {}_1\alpha_{++}(\varepsilon;\lambda_{o;\mathsf{t}_-},\mu_{o;\mathsf{t}_-}) = {}_1\alpha_{++}(\varepsilon;\lambda_o,\mu_o) - 1, \qquad \text{(B.1d)} $$

$$ {}_1\beta_{++}(\varepsilon;\lambda_{o;\mathsf{t}_-},\mu_{o;\mathsf{t}_-}) = {}_1\beta_{++}(\varepsilon;\lambda_o,\mu_o); \qquad \text{(B*.1d)} $$

$$ {}_1\alpha_{-+}(\varepsilon;\lambda_{o;\mathsf{t}_-},\mu_{o;\mathsf{t}_-}) = {}_1\alpha_{-+}(\varepsilon;\lambda_o,\mu_o), \qquad \text{(B.2d)} $$

$$ {}_1\alpha_{-+}(\varepsilon;\lambda_{o;\mathsf{t}_-},\mu_{o;\mathsf{t}_-}) = {}_1\alpha_{-+}(\varepsilon;\lambda_o,\mu_o) + 1. \qquad \text{(B*.2d)} $$

(When selecting different cases, we took into account that basic solutions $\mathsf{b}',0$ and $\mathsf{d}',0$ exist only within the LP range and for $\lambda_o,\mu_o < 1$, respectively.)

**The LC range**

i) $\mathsf{t}_- = \mathsf{b}$

$$ {}_1\alpha^0_{++}(\varepsilon;\lambda_{o;\mathsf{b}},\mu_{o;\mathsf{b}}) = \begin{cases} {}_1\alpha^0_{-+}(\varepsilon;\lambda_o,\mu_o) & \text{for } \mu_o > 1 \\ \text{or} & \\ {}_1\beta^0_{-+}(\varepsilon;\lambda_o,\mu_o) & \text{for } \mu_o < 1; \end{cases} \qquad \text{(B.3b)} $$

$$ {}_1\beta^0_{++}(\varepsilon;\lambda_{o;\mathsf{b}},\mu_{o;\mathsf{b}}) = \begin{cases} {}_1\beta^0_{-+}(\varepsilon;\lambda_o,\mu_o) + 1 & \text{for } \mu_o > 1 \\ \text{or} & \\ {}_1\alpha^0_{-+}(\varepsilon;\lambda_o,\mu_o) + 1 & \text{for } \mu_o < 1; \end{cases} \qquad \text{(B*.3b)} $$

$$ {}_1\alpha^0_{-+}(\varepsilon;\lambda_{o;\mathsf{b}},\mu_{o;\mathsf{b}}) = \begin{cases} {}_1\alpha^0_{++}(\varepsilon;\lambda_o,\mu_o) - 1 & \text{for } \mu_o > 1 \\ \text{or} & \\ {}_1\alpha^0_{++}(\varepsilon;\lambda_o,\mu_o) - 1 & \text{for } \mu_o < 1; \end{cases} \qquad \text{(B.4b)} $$

$$ {}_1\beta^0_{-+}(\varepsilon;\lambda_{o;\mathsf{b}},\mu_{o;\mathsf{b}}) = \begin{cases} {}_1\beta^0_{++}(\varepsilon;\lambda_o,\mu_o) & \text{for } \mu_o > 1 \\ \text{or} & \\ {}_1\alpha^0_{++}(\varepsilon;\lambda_o,\mu_o) & \text{for } \mu_o < 1. \end{cases} \qquad \text{(B*.4b)} $$

ii) $\mathsf{t}_- = \mathsf{d}$

$$ {}_1\alpha^0_{++}(\varepsilon;\lambda_{o;\mathsf{d}},\mu_{o;\mathsf{d}}) = {}_1\alpha^0_{-+}(\varepsilon;\lambda_o,\mu_o), \qquad \text{(B.3d)} $$

$$ {}_1\beta^0_{++}(\varepsilon;\lambda_{o;\mathsf{d}},\mu_{o;\mathsf{d}}) = {}_1\beta^0_{-+}(\varepsilon;\lambda_o,\mu_o) + 1; \qquad \text{(B*.3d)} $$

$$ {}_1\alpha^0_{-+}(\varepsilon;\lambda_{o;\mathsf{d}},\mu_{o;\mathsf{d}}) = {}_1\alpha^0_{++}(\varepsilon;\lambda_o,\mu_o) - 1 \qquad \text{(B.4d)} $$

$$ {}_1\beta^0_{-+}(\varepsilon;\lambda_{o;\mathsf{d}},\mu_{o;\mathsf{d}}) = {}_1\beta^0_{-+}(\varepsilon;\lambda_o,\mu_o). \qquad \text{(B*.4d)} $$

iii) $\mathsf{t}_- = \mathsf{d}'$ ($\mu_o < 1$)

$$ {}_1\alpha^0_{\pm+}(\varepsilon;\lambda_{0;\mathsf{d}'},\mu_{0;\mathsf{d}'}) = {}_1\alpha^0_{\pm+}(\varepsilon;\lambda_{0;\mathsf{b}},\mu_{0;\mathsf{b}}), \qquad \text{(B.5d}')$$

$$ {}_1\beta^0_{\pm+}(\varepsilon;\lambda_{0;\mathsf{d}'},\mu_{0;\mathsf{d}'}) = {}_1\beta^0_{\pm+}(\varepsilon;\lambda_{0;\mathsf{b}},\mu_{0;\mathsf{b}}). \qquad \text{(B.3d}')$$

**Appendix C**

**Reciprocal transformation of the GRef Bose invariant and related sequences of GS Heine polynomials in mutually reciprocal arguments**

The main purpose of this appendix is to relate GS Heine polynomials appearing in the right-hand side of (8.1.29) to Quesne's polynomials $y^{(A,B)}_{m+v-1}(\tilde{\eta})$ in [34] for type-I rational SUSY partners of the E/MR potential. The reason for giving a special attention to these potentials is that the appropriate sequence of GS Heine polynomials starts from a polynomial of order m–1, not m, contrary to the general case.

During a preliminary analysis of this problem we revealed a very interesting dualism existent between the radial *r*-GRef potential of our current interest and the so-called [15] 'LTP' *r*-GRef potential on the line which explains some common features associated with their Darboux deformations using basic FFs. In the particular case of their shape-invariant limits we come to the dualism between the E/MR and RM potentials recently disclosed by Quesne [34].

First, by applying the reciprocal transformation $\tilde{z} = z^{-1}$ to rational SL equation (5.1.5) one can easily verify that the equation does not change its generic form:

$$\left\{\frac{d^2}{d\tilde{z}^2} + {}_1\tilde{I}^o_G[\tilde{z};\tilde{\lambda}_o,\tilde{\mu}_o] + \varepsilon\, {}_1\tilde{\wp}_G[\tilde{z};a,b]\right\}\tilde{\Phi}[\tilde{z};\varepsilon\,|\,\tilde{z};\tilde{\lambda}_o,\tilde{\mu}_o] = 0. \qquad \text{(C.1)}$$

Keeping in mind that we deal with the particular case of a linear fractional transformation the Schwartz derivative is equal to zero [57] so that the reference PFs are related via the simple formula

$$z^4\, {}_1I^o_G[z;\lambda_o,\mu_o] = {}_1I^o_G[z^{-1};\mu_o,\lambda_o] = {}_1\tilde{I}^o_G[\tilde{z};\tilde{\lambda}_o,\tilde{\mu}_o], \qquad \text{(C2)}$$

where $\tilde{\lambda}_o = \mu_o,\ \tilde{\mu}_o = \mu_o$. It is remarkable that the density function in (C.1) is constructed by means of the polynomial of the first order:

$$ {}_1T_1[\tilde{z};a,b] = a + b\tilde{z}, \tag{C3} $$

i.e.,

$$ {}_1\tilde{\wp}_G[z;a,b] = \frac{{}_1T_1[\tilde{z};a,b]}{4\tilde{z}^2(\tilde{z}-1)^2}. \tag{C.4} $$

We thus demonstrated that two families of the potentials $V[z\,|\,{}_1\boldsymbol{\mathcal{G}}^{211}]$ and $V[\tilde{z}\,|\,{}_1\boldsymbol{\mathcal{G}}^{110}]$ can be described by the same SL equation, with the only difference that quantization is done on two different intervals. [We have already pointed [15] to the same dualism for the *t*-and *h*-versions of the PT potential.] The most important consequence of this observation is that AEH solutions of the SL equation for rational SUSY partners of the mentioned pair of the *r*-GRef potentials are described by the same sequences of GS Heine polynomials. The same is obviously true for their shape-invariant limiting cases.

Comparing AEH solutions expressed in terms of z and $\tilde{z}$

$$ \sqrt{\left|\frac{d\tilde{z}}{dz}\right|}\, z^{\frac{1}{2}(1\pm\lambda_o)}(1-z)^{\frac{1}{2}(1+\lambda_{1;\dagger_\pm,m})}\, P_m^{(\lambda_{1;\dagger_\pm,m},\pm\lambda_o)}(2z-1) \tag{C.5} $$

$$ = \tilde{k}_{\dagger_\pm,m}\, \tilde{z}^{\frac{1}{2}(1\pm\tilde{\lambda}_{0;\dagger_\pm,m})}(\tilde{z}-1)^{\frac{1}{2}(1+\tilde{\lambda}_{1;\dagger_\pm,m})}\, P_m^{(\tilde{\lambda}_{1;\dagger_\pm,m},\tilde{\lambda}_{0;\dagger_\pm,m})}(2\tilde{z}-1), $$

one finds

$$ \tilde{\lambda}_{1;\dagger_\pm,m} = \lambda_{1;\dagger_\pm,m},\ \tilde{\lambda}_{0;\dagger_\pm,m} = \mp\lambda_o - \lambda_{1;\dagger_\pm,m} - 2m - 1. \tag{C.6} $$

Here we set

$$ \tilde{k}_{\dagger_\pm,m} \equiv \frac{P_m^{(\lambda_{1;\dagger_\pm,m},\lambda_{0;\dagger_\pm,m})}(1)}{P_m^{(\tilde{\lambda}_{1;\dagger_\pm,m},\tilde{\lambda}_{0;\dagger_\pm,m})}(1)}\ . \tag{C.7} $$

As expected, the exponent difference at the singular point $\tilde{z}=0$ is equal to

$$|\tilde{\lambda}_{0;\dagger_{\pm},m}| = \mu(-a\lambda^2_{1;\dagger_{\pm},m};\mu_o)\,, \qquad \text{(C.8)}$$

where the function $\mu(a\varepsilon;\mu_o)$ is defined via (5.1.23).

Since the Jacobi polynomials in the numerator and denominator of this fraction differ only by the second index the fraction, according to (22.2.1) in [55], is equal to 1 and therefore

$$P_m^{(\lambda_{1;\dagger_{\pm},m},\pm\lambda_o)}(2z-1) = \tilde{z}^{-m}\, P_m^{(\lambda_{1;\dagger_{\pm},m},\tilde{\lambda}_{0;\dagger_{\pm},m})}(2\tilde{z}-1)\,. \qquad \text{(C.9)}$$

In Quesne's terms

$$\lambda_{1;\dagger_+,m} = \frac{B}{A_{\dagger_+}-1+m} - A_{\dagger_+} + 1 - m = \alpha_m\,. \qquad \text{(C.10a)}$$

$$\tilde{\lambda}_{0;\dagger_+,m} = -\frac{B}{A_{\dagger_+}-1+m} - A_{\dagger_+} + 1 - m = \beta_m, \qquad \text{(C.10b)}$$

where we use symbol $A_{\dagger_+}$ = A +1, instead of A, to stress that $A_{\dagger_+} = s_{o;\dagger_+} = s_o + 1$, not $s_o$. The GS Heine polynomial appearing in the right-hand side of (8.1.29) is thus nothing but

$$Hi_{m+v-1}[z\,|\,{}^1_1\mathcal{G}^{202}_{\dagger_+,m};\mathbf{c},v] = \upsilon_{m+v-1;v(s_o,B)}\, \tilde{z}^{m+v-1}\, y^{(s_o+1,B)}_{m+v-1}(2\tilde{z}-1) \qquad \text{(C.11)}$$

where $\mu_o = 2\sqrt{B}$ and the scale factor is found from the requirement that the leading coefficient of polynomial (C.11) is equal to 1.

The only thing left to prove is that GS Heine polynomial (8.1.29) does have the zero root as claimed in Section 8. This is the very feature which makes special this series of rational SUSY partners of the E/MR potential, compared with its general analogue $V[z\,|\,{}^1_1\mathcal{G}^{211}_{\mathbf{a}',m}]$.

Making use of (83) in [85]

$$(\eta-1)\dot{P}_n^{(\nu,\lambda)}(\eta)+\nu P_n^{(\nu,\lambda)}(\eta)=(\nu+n)P_n^{(\nu-1,\lambda+1)}(\eta) \tag{C.12}$$

we represent GS Heine polynomials (6.1.17) as

$$\upsilon_{\mathbf{a}',m;\mathbf{c},v}(I)\,\mathrm{Hi}_{m+v}[z\,|\,{}_1^1\mathcal{G}^{211}_{\mathbf{a}',m};\mathbf{c},v] \tag{C.13}$$

$$=(\lambda_{1;\mathbf{a}',m}+m)P_v^{(\lambda_{1;\mathbf{c},v},\lambda_o)}(2z-1)P_m^{(\lambda_{1;\mathbf{a}',m}-1,\lambda_o+1)}(2z-1)$$

$$-(\lambda_{1;\mathbf{c},v}+v)P_m^{(\lambda_{1;\mathbf{a}',m},\lambda_o)}(2z-1)P_v^{(\lambda_{1;\mathbf{c},v}-1,\lambda_o+1)}(2z-1)$$

$$+\tfrac{1}{2}(\lambda_{1;\mathbf{c},v}-\lambda_{1;\mathbf{a}',m})P_m^{(\lambda_{1;\mathbf{a}',m},\lambda_o)}(2z-1)\,P_v^{(\lambda_{1;\mathbf{c},v},\lambda_o)}(2z-1),$$

where

$$\upsilon_{\mathbf{a}',m;\mathbf{c},v}(I)=2^{m+v}k_v(\lambda_{1;\mathbf{c},v},\lambda_o)\,k_m(\lambda_{1;\mathbf{t},m},\lambda_o)[m-v-\tfrac{1}{2}(\lambda_{1;\mathbf{c},v}-\lambda_{1;\mathbf{a}',m})] \tag{C.14}$$

with $k_m(\nu,\lambda)$ standing for the leading coefficient of the polynomial $P_n^{(\nu,\lambda)}(\eta)$ in $\eta$. Taking into account that [55]

$$P_n^{(\nu-1,\lambda+1)}(-1)=\frac{\lambda+n+1}{\lambda+1}P_n^{(\nu,\lambda)}(-1) \tag{C.15}$$

we thus need to prove that:

$$(\lambda_{1;\mathbf{a}',m}+m)\frac{\lambda_o+m+1}{\lambda_o+1}-(\lambda_{1;\mathbf{c},v}+v)\frac{\lambda_o+v+1}{\lambda_o+1}=\tfrac{1}{2}(\lambda_{1;\mathbf{a}',m}-\lambda_{1;\mathbf{c},v}) \tag{C.16}$$

if the exponents $\lambda_{1;\mathbf{t}_+,m}$ are defined via (8.1.25). Substituting (8.1.25) for $\mathbf{t}_+=\mathbf{a}'_+,\mathbf{c}$ into the left-hand side of (C.16) thus gives

$$(v-m)\left[\frac{\mu_o^2}{(2m+1+\lambda_o)(2v+1+\lambda_o)}+1\right]=\lambda_{1;\mathbf{a}',m}-\lambda_{1;\mathbf{c},v} \tag{C.17}$$

which completes the proof.